\documentclass[fleqn,usenatbib]{mnras}

\usepackage{newtxtext,newtxmath}
\usepackage{color}
\definecolor{grey}{rgb}{0.4,0.5,0.6}
\definecolor{trolleygrey}{rgb}{0.5, 0.5, 0.5}

\newcommand*\diff{\mathop{}\!\mathrm{d}}

\usepackage[T1]{fontenc}

\DeclareRobustCommand{\VAN}[3]{#2}
\let\VANthebibliography\thebibliography
\def\thebibliography{\DeclareRobustCommand{\VAN}[3]{##3}\VANthebibliography}

\usepackage{graphicx}	
\usepackage{amsmath}	

\definecolor{darkred}{rgb}{0.8,0.0,0.0}

\title[Globular clusters and dwarf spheroidal galaxies]{BALRoGO: Bayesian Astrometric Likelihood Recovery of Galactic Objects -- Global properties of over one hundred globular clusters with \textit{Gaia} EDR3}

\author[Eduardo Vitral]{
Eduardo Vitral$^{1}$\thanks{E-mail: vitral@iap.fr}
\\
$^{1}$Sorbonne Universit\'e, CNRS, UMR 7095, Institut d’Astrophysique de Paris, 98 bis bd Arago, 75014 Paris, France \\
}

\date{Accepted XXX. Received YYY; in original form ZZZ}

\pubyear{2021}

\begin{document}
\label{firstpage}
\pagerange{\pageref{firstpage}--\pageref{lastpage}}
\maketitle

\begin{abstract}
We present BALRoGO: Bayesian Astrometric Likelihood Recovery of Galactic Objects, a public code to measure the centers, effective radii, and bulk proper motions of Milky Way globular clusters and Local Group dwarf spheroidals, whose data are mixed with Milky Way field stars.
Our approach presents innovative methods such as surface density fits allowing for strong interloper contamination and proper motion fits using a Pearson VII distribution for interlopers, instead of classic Gaussian-mixture recipes. We also use non-parametric approaches to represent the color-magnitude diagram of such stellar systems based in their membership probabilities, previously derived from surface density and proper motion fits. 
The robustness of our method is verified by comparing its results with previous estimates from the literature as well as by testing it on mock data from $N-$body simulations. We applied BALRoGO to \textsc{Gaia EDR3} data for over one hundred Milky Way globular clusters and nine Local Group dwarf spheroidals, and we provide positions, effective radii, and bulk proper motions. Finally, we make our algorithm available as an open source software.
\end{abstract}

\begin{keywords}
methods: data analysis -- astrometry -- proper motions -- (Galaxy:) globular clusters: general -- galaxies: dwarf -- stars: kinematics and dynamics
\end{keywords}



\section{Introduction}

The galactic neighborhood is a replete garden of important objects, such as globular clusters and dwarf spheroidal galaxies, that help to explain the key ingredients of galactic evolution and other unsolved mysteries in astrophysics. 
Globular clusters (GCs) are among the oldest relics of the Universe, sometimes reaching ages close to 13 Gyr (\citealt{MarinFranch+09}), which are spherically shaped, compact collections of stars that orbit the Milky Way (MW), and whose origin is still uncertain (\citealt*{Peebles&Dicke68,Peebles84,Searle&Zinn78,Bullock&Johnston05,Abadi+06,Penarrubia+09}). They have recently been discovered to harbor multiple stellar populations (e.g. \citealt{Carretta+09}) and are frequently served as test laboratories for intermediate-mass black holes search (e.g. \citealt*{vanderMarel&Anderson&Anderson10,Noyola+08}, and more recently \citealt{Vitral&Mamon21}). Dwarf spheroidal galaxies (dSphs), as well, are composed of old stellar systems and are located farther away than GCs, but still in the Local Group. They are much fainter group of stars and are often studied as tracers of dark matter (e.g., \citealt*{Battaglia+13}, \citealt{Walker13} and \citealt{Boldrini+20c}).

The study of both GCs and dSphs made a major leap as data from the \textsc{Gaia} astrometric mission became available, specially with its second and early third release (hereafter \textsc{Gaia DR2} and \textsc{Gaia EDR3}, respectively). This mission provided an overall astrometric coverage of stellar velocities, positions and magnitudes of more than $10^{9}$ stars in the MW and beyond and therefore allowed many deeper studies of spherical systems such as GCs and dSphs, which can be well separated from MW field stars (i.e., interlopers) when combining both position, distance, and most importantly, proper motions. New bulk proper motions were derived in \cite{GaiaHelmi+18}, \cite{Baumgardt+19} and \cite{Vasiliev19b} with \textsc{Gaia DR2} for more than 150 GCs, and different methods were used to do so, relying whether on Gaussian mixtures or iterative routines to differentiate tracers from interlopers. Similarly, \cite{McConnachie&Venn20a} derived bulk proper motions for dSphs using \textsc{Gaia DR2}, and later with \textsc{Gaia EDR3} (\citealt{McConnachie&Venn20}) by using a Bayesian routine also employing Gaussian mixtures, inspired by the work of \cite{Pace&Li19}. Recently, by analyzing the GC NGC~6397, \cite{Vitral&Mamon21} showed that the separation between GC stars and interlopers was much more robust by assigning distribution functions to the proper motion space which were not based on a Gaussian mixture, as usual, but rather considering a GC Gaussian component plus a Pearson VII (\citealt{Pearson16}) distribution for the interlopers. An alternative to such problem is to consider not one single Gaussian for interlopers, but rather a multiple Gaussian component, such as in \cite{Vasiliev&Baumgardt&Baumgardt21}, who recently derived GC astrometric parameters with \textsc{Gaia EDR3}.

In fact, with the arrival of \textsc{Gaia} data, extracting GC members, as well as dSphs became a very important topic, for which a robust approach is required in order not to let interloper contamination bias one's results. For example, if the selection of GC stars at its outskirts is not well made, one could misinterpret the presence of tidal tails, and thus derive wrong conclusions on the cluster evolution. 

Given the importance of a separation algorithm between such galactic objects and interlopers, \cite{BustosFierro&Calderon19} proposed a method based on Gaussian mixtures and clustering algorithms, with only one step not performed by a machine and applied it to eight GCs. Their work joined previous attempts to separate cluster members from interlopers through parametric and non-parametric methods (e.g., \citealt{Vasilevskis+65} and \citealt{GaladiEnriquez+98}, respectively). In this paper, we also present an algorithm to extract GCs and dSphs members embedded in MW interlopers, which employs a combination of Bayesian fits, giving the membership probability of each star in the field, and non-parametric cuts on the color-magnitude diagram. Our method takes as an input a cone search delivered by \textsc{Gaia} and, from that data, estimates the object center in right ascension and declination (i.e., $\alpha$ and $\delta$), the bulk proper motions, $\mu_{\alpha,*}$ (i.e., $[\diff \alpha / \diff t] \, \cos{\delta}$) and $\mu_{\delta}$, and the effective radius (half projected number radius) by fitting discrete data. To validate our method, we compare bulk proper motions provided in the literature using \textsc{Gaia EDR3}, and we compare our effective radii estimates for over a hundred sources with other estimates. We use \textsc{Gaia EDR3} data from the two GCs NGC~6752 and NGC~6205 (M~13) along with the Draco dSph galaxy as illustrative examples in our Figures, throughout the paper.

In addition, we derive effective radii, bulk proper motions and centers for over a hundred GCs from the \textsc{New Galactic Catalog} (NGC) using \textsc{Gaia EDR3} and make it available in Table format. Our algorithm developed with \textsc{Python} is also provided as an open source code named \textsc{BALRoGO: Bayesian Astrometric Likelihood Recover of Galactic Objects}\footnote{\url{https://gitlab.com/eduardo-vitral/balrogo}}, and allows the user to adjust the inputs of our algorithm to better suit any particular source.

\section{Methodology}

Our algorithm is specially designed to deal with data from the \textsc{Gaia} astrometric mission, and therefore we begin by acquiring all the stars in a two degrees cone search for each of the GCs and dSphs in Table~\ref{tab: results}. The first two steps of our approach deal with all the stars in the two degrees field of view, and use only positional ($\alpha$, $\delta$) information. As it is going to be shown, they allow for interlopers in their fitting routines.

\subsection{Center estimation}
\label{ssec: center}

\begin{figure}
\centering
\includegraphics[width=0.8\hsize]{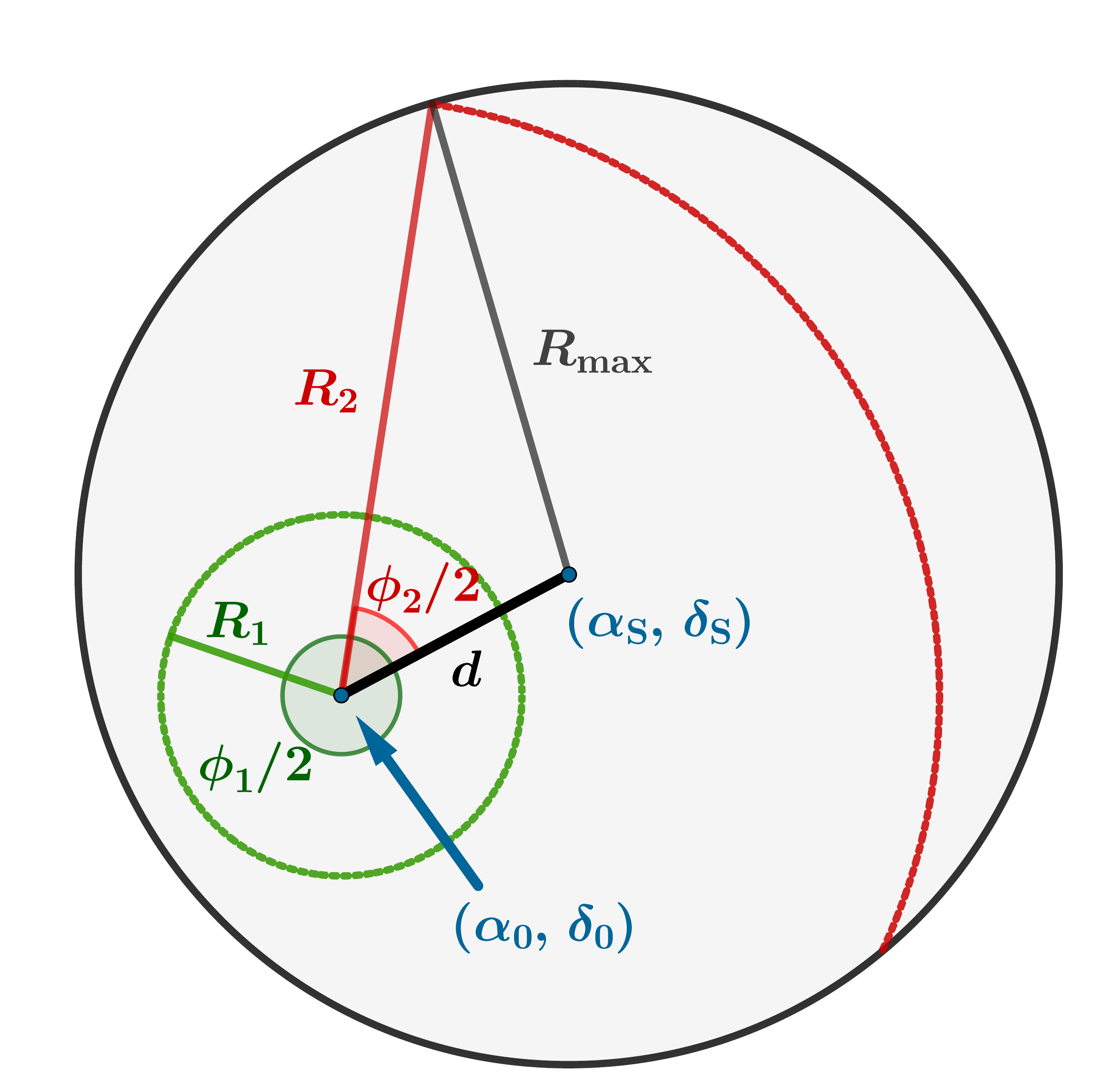}
\caption{{\it Geometry of the center estimation:} The original cone search, centered in ($\alpha_{\rm S}, \, \delta_{\rm S}$) is shown as the gray circle of radius $R_{\rm max}$, while the new fitted center is ($\alpha_{0}, \, \delta_{0}$). The angles $\phi_1$ and $\phi_2$ represent the effective circular sections (of respective radii $R_1$ and $R_2$) where we consider circular symmetry. The distance between the two centers is labeled as $d$, and we consider this value to approach zero for the cone searches around the centers from the SIMBAD data base.}
\label{fig: geometry}
\end{figure}

The first step of our analysis is the estimation of the GC or dSph (hereafter, galactic object for short) center. Although \textsc{BALRoGO} allows for an iterative frequentist approach, this can be dangerous when dealing with objects that suffer from crowding issues (\citealt{Arenou+18}), since the amount of stars decreases when one approaches the center too much. Therefore, as for many other steps, we opt for a Bayesian fit of the galactic object center.

This is done by fitting a Plummer profile (\citealt{Plummer1911}) plus a constant contribution of interlopers to the surface density of the stars inside the two degrees cone (in some cases, we selected smaller radii due to high MW contamination, or due to parasite satellites). Such approach aims for nothing more than finding a center of mass of the stellar distribution, by fitting the center that best suits an spherical shaped collection of stars. Even if the galactic object does not follow precisely a Plummer model (e.g., core-collapsed GCs), other profiles, such as S\'ersic \citep{Sersic63,Sersic68}, Hernquist \citep{Hernquist90} or \citep{Kazantzidis+04} for example also follow a circular symmetry, and therefore should have their centers well determined by a Plummer fit. In addition, allowing for the contribution of interlopers in the fit enables the user not to lose many galactic object stars and thus have a better knowledge of the radial extent of the source, as well as the radial membership probability of a star with respect to the galactic object. 

Following the normalization of \citeauthor{Vitral&Mamon20} (\citeyear{Vitral&Mamon20}, equations~15--17):

\begin{subequations} 
\label{eq: tilderho}
\begin{equation}
    \Sigma(R) = \frac{N_\infty}{\pi R_{\rm scale}^2} \,
    \widetilde \Sigma \left (\frac{R}{R_{\rm scale}}\right) \ , 
\end{equation}
\begin{equation}
    N(R) = N_{\infty} \, \widetilde N \left(\frac{R}{R_{\rm scale}}\right) \ ,
\end{equation}
\end{subequations}
where $N_{\infty}$ is the projected number of tracers at infinity, one can write the global surface density as:

\begin{equation}
    \widetilde{\Sigma}(X) = \widetilde{\Sigma}_{\rm sys}(X) + \mathcal{R} \, \frac{\widetilde{N}_{\rm sys,tot}}{X^2_{\rm max} - X^2_{\rm min}}  \  ,
    \label{eq: total-sd}
\end{equation}
where $\mathcal{R} = N_{\rm ilop, tot} / N_{\rm sys, tot}$ is the ratio between the number of interlopers and system (galactic object) stars, $\widetilde{N}_{\rm sys,tot} = \widetilde{N}_{\rm sys}(X_{\rm max}) - \widetilde{N}_{\rm sys}(X_{\rm min})$ is the total number of system stars, and $X=R/a$ is the normalized projected radius, with $a$ being the Plummer scale radius, or the Plummer projected, two-dimensional, half number radius. The normalized surface density of the analyzed system, for a Plummer profile reads:

\begin{equation}
    \widetilde{\Sigma}_{\rm sys}(X) = \frac{1}{\left(1 + X^2\right)^2}  \ ,
    \label{eq: plummer-sd}
\end{equation}
while the normalized projected number of stars for the same model is:
\begin{equation}
    \widetilde{N}_{\rm sys}(X) = \frac{1}{\pi} \, \int_{X_{\rm min}}^{X_{\rm max}} \phi(X) \, X \, \widetilde{\Sigma}_{\rm sys}(X) \diff X  \ ,
    \label{eq: plummer-n}
\end{equation}
where $\phi(X)$ is the angle corresponding to the effective circular section where we analyze the data (see Fig~\ref{fig: geometry}), and whose analytical expression for small cone apertures (i.e. $R_{\rm max} \ll 1$ radian) is:

\begin{equation}
\phi(X = R/a) =
    \begin{cases}
        2 \, \pi
        & , \ \mathrm{if} \ R \leq R_{\rm max} - d \ , 
        \\
        \displaystyle
        2 \, \arccos{\left[\frac{R^2 + d^2 - R_{\rm max}^2}{2 \, R \, d}\right]}
        & , \ \mathrm{if} \ R > R_{\rm max} - d \ ,
    \end{cases}
\label{eq: casephifull}
\end{equation}
where $R_{\rm max}$ is the maximum radius of the original cone search (usually two degrees) and $d$ is the distance between the fitted center and the center from the original cone search ($\alpha_{\rm S}, \, \delta_{\rm S}$), set as the source center on SIMBAD\footnote{\url{http://simbad.u-strasbg.fr/simbad/}} by the automatic \textsc{Gaia} advanced query. The projected radius $R$ is defined, in spherical trigonometry, as:
\begin{equation}
    R = \arccos[\sin{\delta} \, \sin{\delta_{0}} + \cos{\delta} \, \cos{\delta_{0}} \, \cos{(\alpha - \alpha_{0})}]  \ ,
    \label{eq: Rproj}
\end{equation}
where ($\alpha_{0}, \, \delta_{0}$) is the center of the galactic object. The likelihood function is therefore written as:

\begin{equation}
    \mathcal{L} = \prod_i \, \frac{\phi(X)}{\pi} \, \frac{X}{a} \frac{\widetilde{\Sigma}(X)}{\widetilde{N}_{\rm sys,tot} \, (1 + \mathcal{R})} \  .
    \label{eq: plummer-lik}
\end{equation}

We minimize $- \log{\mathcal{L}}$ with the \textsc{differential\_evolution} routine from the \textsc{scipy.optimize} method and find the centers that best fit the data, later displaying them in Table~\ref{tab: results}. For simplicity, since we expect the quantity $d$ to approach zero, given the reliable previous measurements of the centers in SIMBAD, one can assume the case where $\phi(X) = 2 \, \pi$, and thus derive:

\begin{equation}
    \widetilde{N}_{\rm sys}(X) = \frac{X^2}{1 + X^2}  \ .
    \label{eq: plummer-n-simple}
\end{equation}

Notably, for more general cases, the \textsc{BALRoGO} routine allows to use the more general representation of $\phi(X)$, and thus to account for a more complicated expression for $\widetilde{N}_{\rm sys}(X)$, presented in appendix~\ref{app: general-n}.

\subsection{Surface density}
\label{ssec: sd}

\begin{figure*}
\centering
\includegraphics[width=0.33\textwidth]{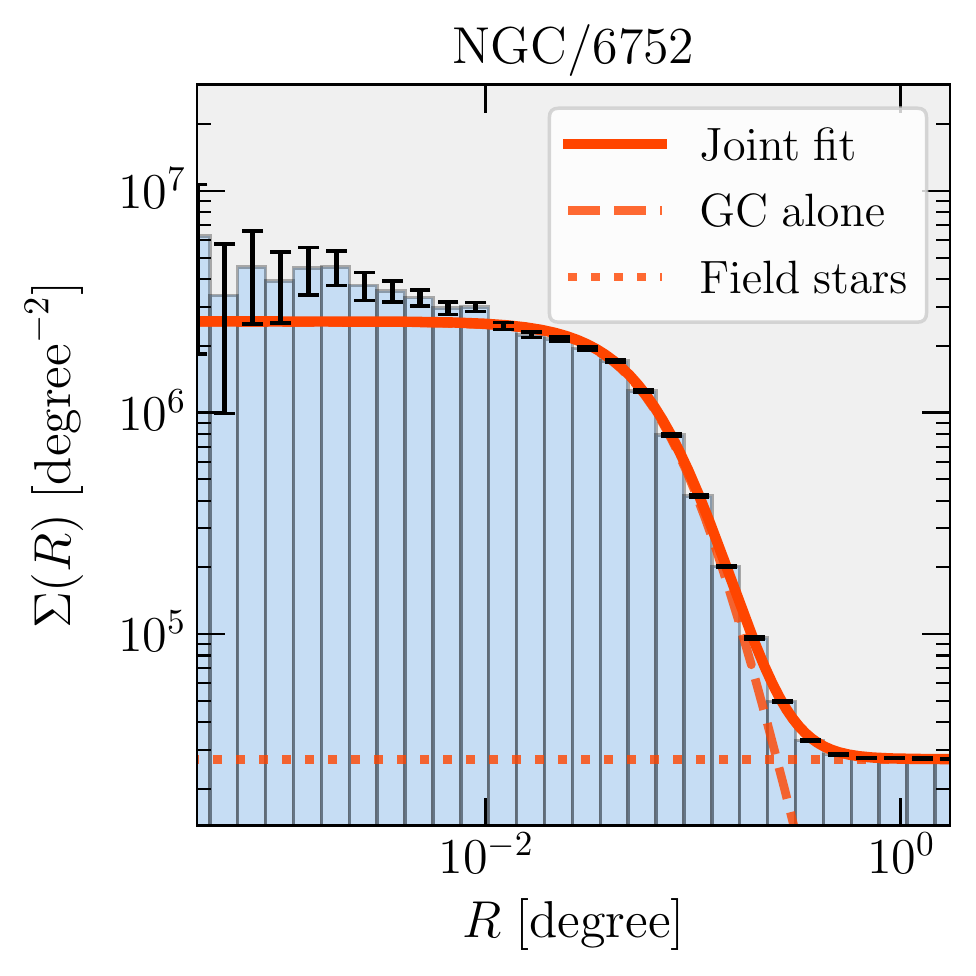}
\includegraphics[width=0.33\textwidth]{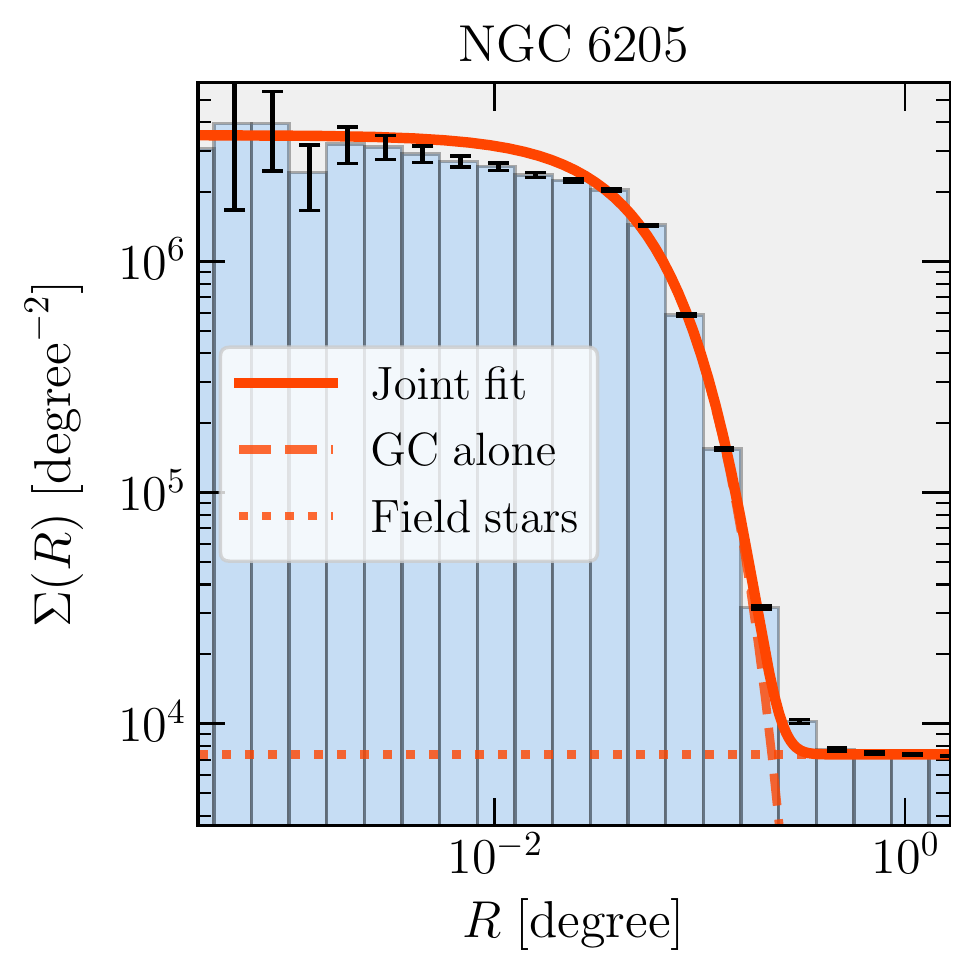}
\includegraphics[width=0.33\textwidth]{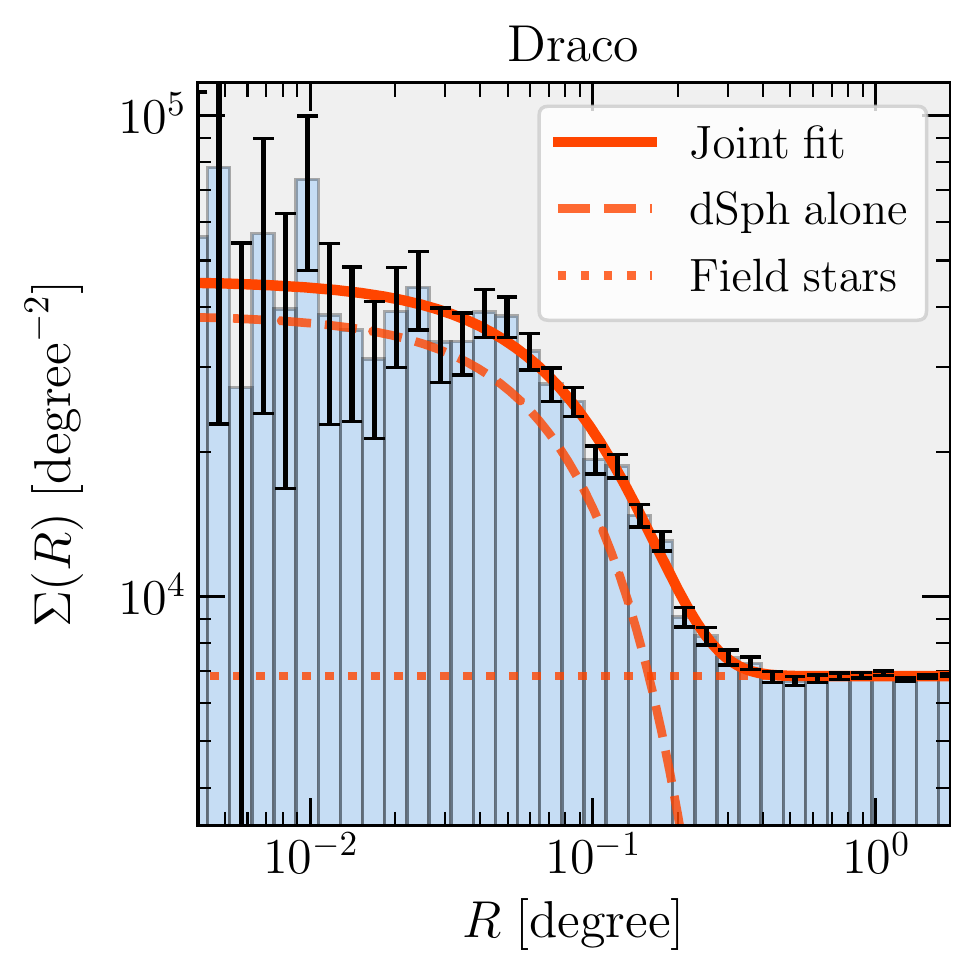}
\caption{{\it Surface density fits:} Fits for the surface density of the galactic object (NGC~6752, NGC~6205 or M~13 and the Draco dSph, respectively) plus a constant contribution of interlopers, according to section~\ref{ssec: sd}. The \emph{histogram} shows the empirical profile, using logarithmic radial bins extending from the innermost bin point to 2 degrees. 
The \emph{curves} show different models: our MLE fit (\emph{red}) 
of a Plummer, S\'ersic and S\'ersic models respectively (\emph{dashed}) plus constant field stars surface density (\emph{dotted}), as well as the total (\emph{solid}) to compare with the data. The error bars were calculated considering only Poisson noise.}
\label{fig: sd-fits}
\end{figure*}

The next step of our algorithm is to fit the surface density of the galactic object plus interlopers once again, using a maximum likelihood estimation (MLE), but this time with fixed centers from the previous step. It may seem as an unnecessary step given the previous one, but it actually allows the fitting routine to better explore the other parameters ranges like the scale radius and the interloper fraction, and thus derive more robust results. In this step, we also reduced the maximum projected radius whenever there was a parasite galactic object on the field or when the MW contamination was too intense, disturbing the fits.

In this step, we not only tested Plummer profiles, but also S\'ersic models \citep{Sersic63,Sersic68}, which can be useful for core-collapsed clusters such as NGC~6397, and the Kazantzidis model, which is motivated from dynamical simulations of repeated tidal encounters \citep{Kazantzidis+04}. Equations~\ref{eq: total-sd}, \ref{eq: Rproj} and \ref{eq: plummer-lik} remain valid for these two other models, while the normalized surface density and projected number of stars become:

\begin{subequations}
\begin{equation}
    \widetilde{\Sigma}_{\rm sys}(X) = \frac{b^{2n}(n)}{2n \, \Gamma{(2n)}} \exp{\left[ -b(n) \, X^{1/n} \right]} \  ,
    \label{eq: sersic-sd}
\end{equation}
\begin{equation}
    \widetilde{N}_{\rm sys}(X) = \frac{\gamma{(2n, \, b(n) \, X^{1/n})}}{\Gamma{(2n)}} \  .
    \label{eq: sersic-n}
\end{equation}
\end{subequations}
for S\'ersic, where $\gamma(a,x) = \int_{0}^{x} t^{a-1} e^{-t} \mathrm{d} t$ is the lower incomplete gamma function and $X = R/R_{\rm e}$, $R_{\rm e}$ being the effective projected radius. And for Kazantzidis:
\begin{subequations}
\begin{equation}
    \widetilde{\Sigma}_{\rm sys}(X) = K_{0}(X) / 2 \  ,
    \label{eq: kaz-sd}
\end{equation}
\begin{equation}
    \widetilde{N}_{\rm sys}(X) = 1 - X \, K_{1}(X) \  .
    \label{eq: kaz-n}
\end{equation}
\end{subequations}
Where $K_{0}$ and $K_{1}$ are modified Bessel functions of the second kind and $X = R/a_{\rm K}$, $a_{\rm K}$ being the radius of density slope $-2$.

We compared the likelihood of these three models by means of Bayesian inference, using the corrected Akaike Information Criterion (AICc,
\citealt{Sugiyara78}):
\begin{equation}
      \mathrm{AICc} =\mathrm{AIC} + 2 \, \frac{N_{\mathrm{free}} \,  (1 + N_{\mathrm{free}})}{N_{\mathrm{data}} - N_{\mathrm{free}} - 1} \ ,
\end{equation}
where AIC is the original 
{Akaike Information Criterion}
\citep{akaike1973information}:
\begin{equation}
 \label{eq: AIC}
    \mathrm{AIC} = - 2 \, \ln \mathcal{L_{\mathrm{MLE}}} + 2 \, N_{\mathrm{free}} \ ,
\end{equation}
Our motivation for this criterion is the same one portrayed in \cite{Vitral&Mamon21}, section 8.2. Whenever the difference between two models (i.e., $\Delta \rm AICc = AICc_{i} - AICc_{j}$) had an absolute value smaller than two (i.e., no particular preference), priority was given to Plummer, Kazantzidis and S\'ersic profiles, respectively.
The results of our fits of surface density model plus constant interloper contribution are displayed in Figure~\ref{fig: sd-fits} for NGC~6752, NGC~6205 (M 13) and the Draco dSph, respectively.

\subsection{Data cleaning}
\label{ssec: clean}

The two previous steps required no data cleaning, since the filtering could bias the results, for example, by forcing radial incompleteness towards the center of the galactic object (mostly because of crowdness issues) and cutting of fainter stars at the tracer outskirts, thus damaging the fitting of the center and other surface density parameters. However, the next step, which is aimed to look into the proper motion space is, in its turn, much more sensible to poor astrometry. This is a consequence of the much more subtle measurements of proper motions than sky coordinates, given that the \textsc{Gaia} astrometric mission has not yet a long timeline data base to fully trust its main velocity values without any filtering.
Another reason why a conservative data cleaning of proper motions is necessary is that our proper motion fits themselves do not take into account the convolution of the interloper component with \textsc{Gaia} errors, which will be latter justified. Therefore, our data cleaning follows a similar approach as done in \cite{Vitral&Mamon21} for the GC NGC~6397, with some few changes, detailed in the following.

\subsubsection{Maximum projected radius}

We first filtered the {\sc Gaia} stars inside a maximum projected radius: we selected stars inside a cone of radius $r_{\rm cut}$, where $r_{\rm cut} = 10 \, R_{\rm scale}$. $R_{\rm scale}$ was $R_{\rm e}$, $a$, and $a_{\rm K}$ for S\'ersic, Plummer and Kazantzidis, respectively. 
For some sources, this limit was changed in order to account for a more representative subset, but never out-passed the range between 0.03 and two degrees.

\subsubsection{Low error stars}

The next step was to retain only stars with $\epsilon_\mu < \sigma_{\rm limit}$, where $\sigma_{\rm limit}$ is an error threshold and $\epsilon_\mu$ is the proper motion error (semi-major axis of the error ellipse), defined in \citeauthor{Lindegren+18} (\citeyear{Lindegren+18}, eq. B.2):
\begin{eqnarray} \label{eq: err_lind18}
    \epsilon_\mu &=& \sqrt{\frac{1}{2}(C_{33}+C_{44}) + \frac{1}{2}\sqrt{(C_{44}-C_{33})^2 + 4 C_{34}^2}}
    \label{errLindegren} \ , \\
C_{33} &=& \epsilon_{\mu_{\alpha,*}}^2 \ ,\\
C_{34} &=&  \epsilon_{\mu_{\alpha,*}}\, \epsilon_{\mu_\delta}\,\rho \ ,\\
C_{44} &=& \epsilon_{\mu_\delta}^2 \ ,
\end{eqnarray}
where $\epsilon$ denotes the error or uncertainty and where $\rho$ is the correlation coefficient between $\mu_{\alpha,*}$ and $\mu_\delta$.\footnote{We use the standard notation $\mu_{\alpha*} = \cos \delta\,\left[{\rm d}\alpha/{\rm d}t\right]$,
$\mu_\delta={\rm d}\delta/{\rm d}t$.}
The error threshold was derived by default, as the following routine:

\begin{enumerate}
    \item First, we selected the stars inside the maximum radius mentioned above.
    \item With this subset, we generated a two-dimensional histogram in proper motion space and assigned the two highest local peaks of the distribution with \textsc{Python}'s routine \textsc{peak\_local\_max} from the \textsc{skimage.feature} method, which would correspond to the tracer and interloper clumps of stars. The binning of the histogram was chosen such that each dimension $x_{i}$ had a number of bins computed as:
    \begin{equation}
        N_{\rm bin} = \left\lfloor \frac{4 \times (\mathrm{max}[x_{i}] - \mathrm{min}[x_{i}])}{\mathrm{min}[\eta_{.84}(x_{i})-\eta_{.50}(x_{i}),\eta_{.50}(x_{i})-\eta_{.16}(x_{i})]} \right\rceil  \ ,
    \end{equation}
    where $\eta_{.n}(x_{i})$ is the $n$-th percentile of the $x_{i}$ data. The typical values of $N_{\rm bin}$ were around 300.
    \item We naively estimated the Gaussian dispersion of each clump by taking the distance between the peaks found above and the closest point where the histogram reached $\rm e^{-1/2}$ times the value at the peak.
    \item We assigned the interloper peak as the one with great dispersion, while the narrow dispersion clump (calculated as above) was considered as the tracer velocity dispersion $\sigma_{\rm tracer}$. The few cases where this relation was not satisfied (e.g., for 47 Tuc) could be corrected by selecting a narrow field of view around the source center, allowing for very few interlopers.
    \item The error threshold $\sigma_{\rm limit}$ was taken as $\sigma_{\rm limit} = k \, \sigma_{\rm tracer}$, where the default $k$ was set as 0.5. For many clusters this value was changed in order to select a more representative subset. The GCs had all $k$ values between 0.2 and two, while the dSphs had $k$ up to five, given their greater distances, and consequently, much higher errors (otherwise, we would have too few tracers for dSphs in order to fit them properly).
\end{enumerate}

\subsubsection{Quality flags}

We only kept stars whose astrometric solution presented a sufficiently low uncertainty:
\begin{equation} \label{eq: Gaia_flag1}
    \sqrt{\frac{\chi^2}{\nu' - 5}} < 1.2 \, \mathrm{max}\left\{1,\exp\left[-0.2 \, (G-19.5)\right]\right\} \  ,
\end{equation}
where $\nu'$ is the number of points (epochs) in the astrometric fit of a given star and 5 is the number of free parameters of the astrometric fit (2 for the position, 1 for the parallax and two for the proper motions). 
Eq.~(\ref{eq: Gaia_flag1}) gives a sharper HR diagram, removing artifacts such as double stars, calibration problems, and astrometric effects from binaries. It is more optimized than the (\url{astrometric_excess_noise}$<1$) criterion, used in \cite{Baumgardt+19} and \cite{Vasiliev19b}, especially for brighter stars ($G\lesssim15$), according to \cite{Lindegren+18}. We consider this step to substitute possible filters based on the {\sc Gaia} \url{RUWE} parameter.

Second, we only kept stars with good photometry, by selecting the ones that satisfied:

\begin{equation} \label{eq: Gaia_flag2}
    C^{*}(r) < N \, \sigma_{C} \ ,
\end{equation}
where we select $N = 3$, such as in \cite{Vasiliev&Baumgardt&Baumgardt21} and \cite{McConnachie&Venn20}, while $\sigma_{C}$ follows equation~(18) from \cite{Riello+20}. $C^{*}(r)$ is defined such as:

\begin{equation}
    C^{*}(r) = C(r) - f(G_{\mathrm{BP}} - G_{\mathrm{RP}}) \ , 
\end{equation}
with $C(r)$ being the corrected excess factor and $f(x) = \sum_i a_i \, x^i$,  $a_i$ reported in Table~2 from \cite{Riello+20}.
Eq.~(\ref{eq: Gaia_flag2}) performs an additional filter for unreliable astrometric solutions (mainly in the cases of blended stars), affecting mainly faint sources in crowded areas.
Those variables correspond to the following quantities in the {\sc Gaia DR2} and {\sc EDR3} archive:
\begin{itemize}
    \item $\chi^{2}$: \url{astrometric_chi2_al} 
    \item $\nu'$: \url{astrometric_n_good_obs_al} 
    \item $C(r)$: \url{phot_bp_rp_excess_factor} 
    \item $G_{\mathrm{BP}} - G_{\mathrm{RP}}$: \url{bp_rp}
\end{itemize}

\subsection{Proper motions}
\label{ssec: pm}

\begin{figure*}
\centering
\includegraphics[width=0.33\textwidth]{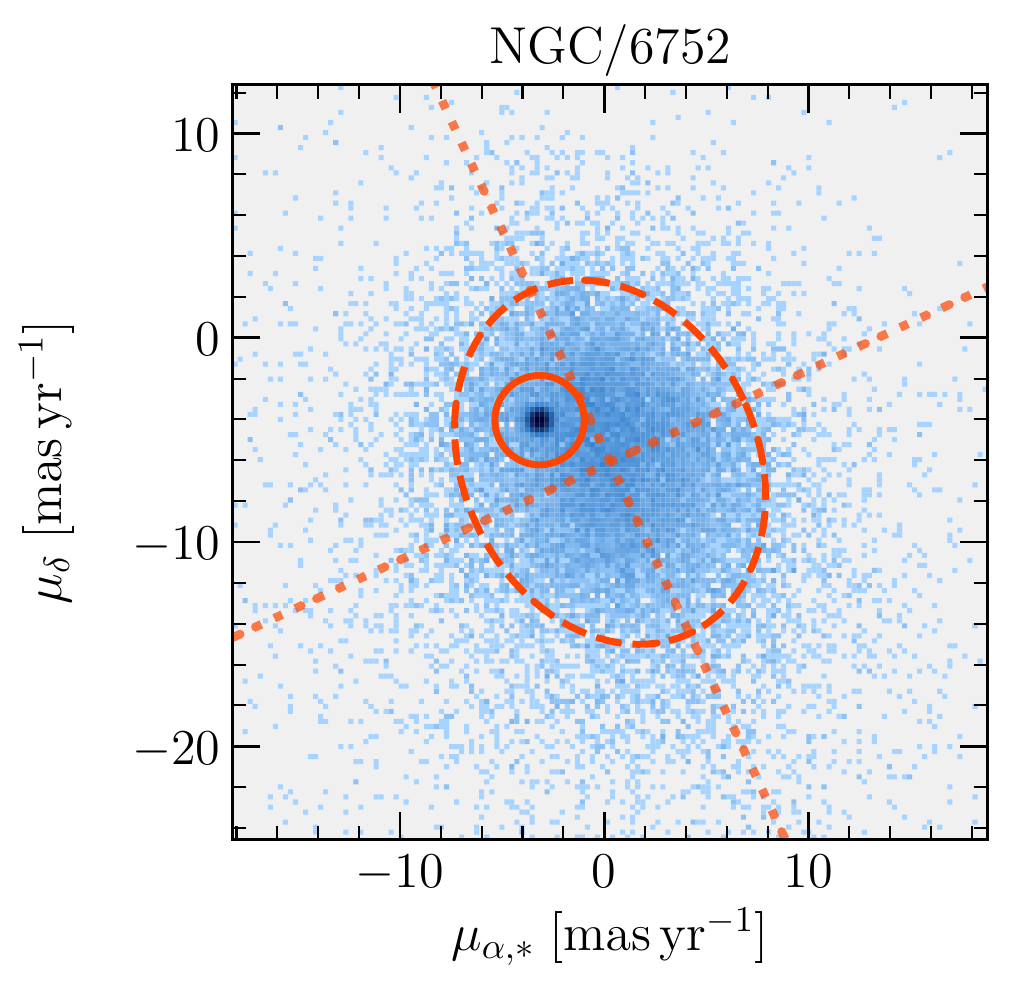}
\includegraphics[width=0.33\textwidth]{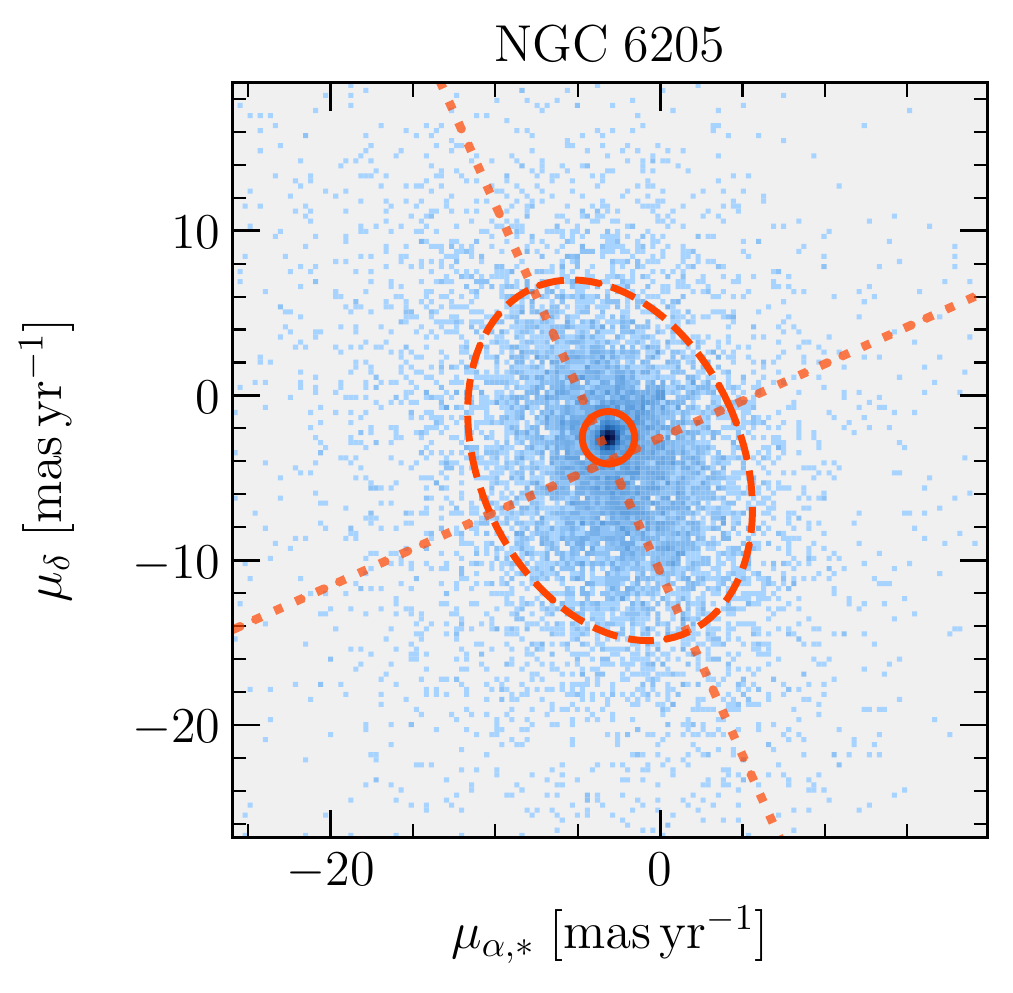}
\includegraphics[width=0.33\textwidth]{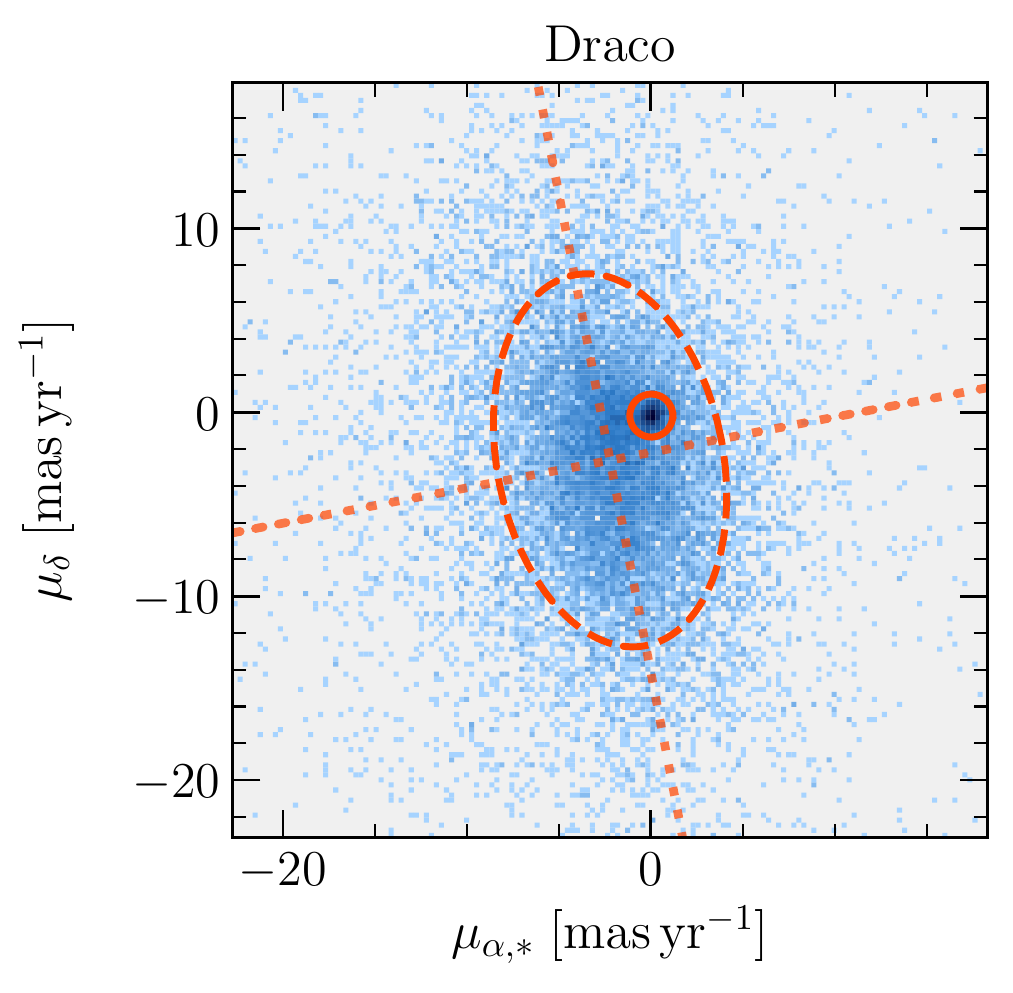}
\includegraphics[width=0.33\textwidth]{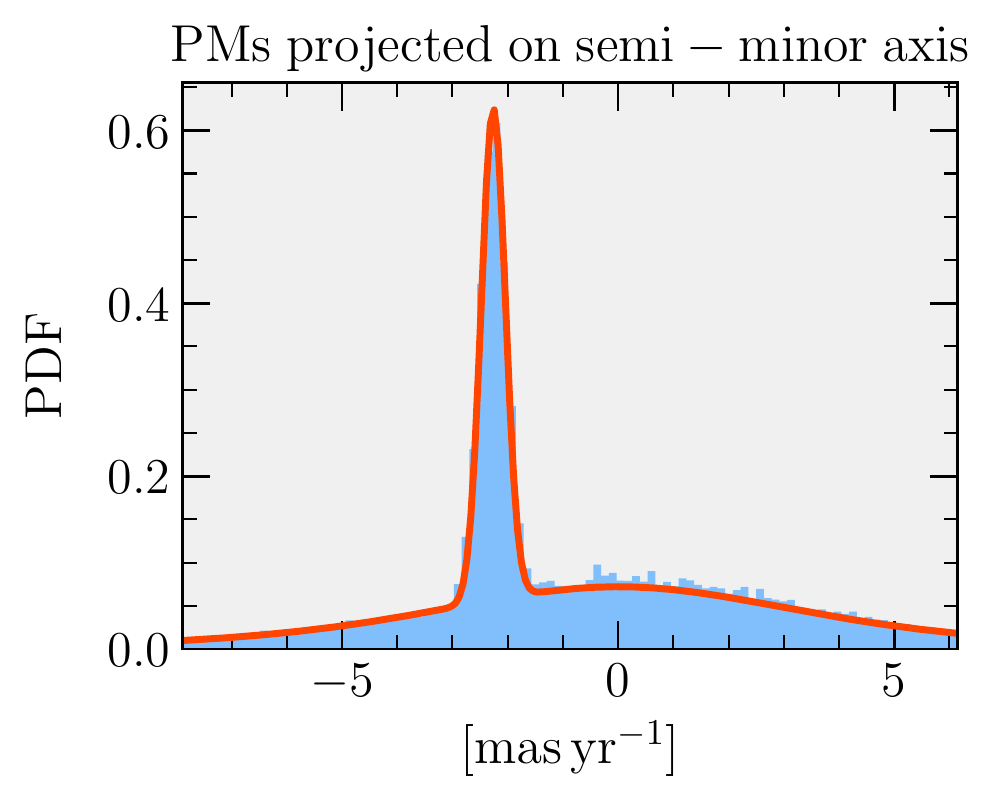}
\includegraphics[width=0.33\textwidth]{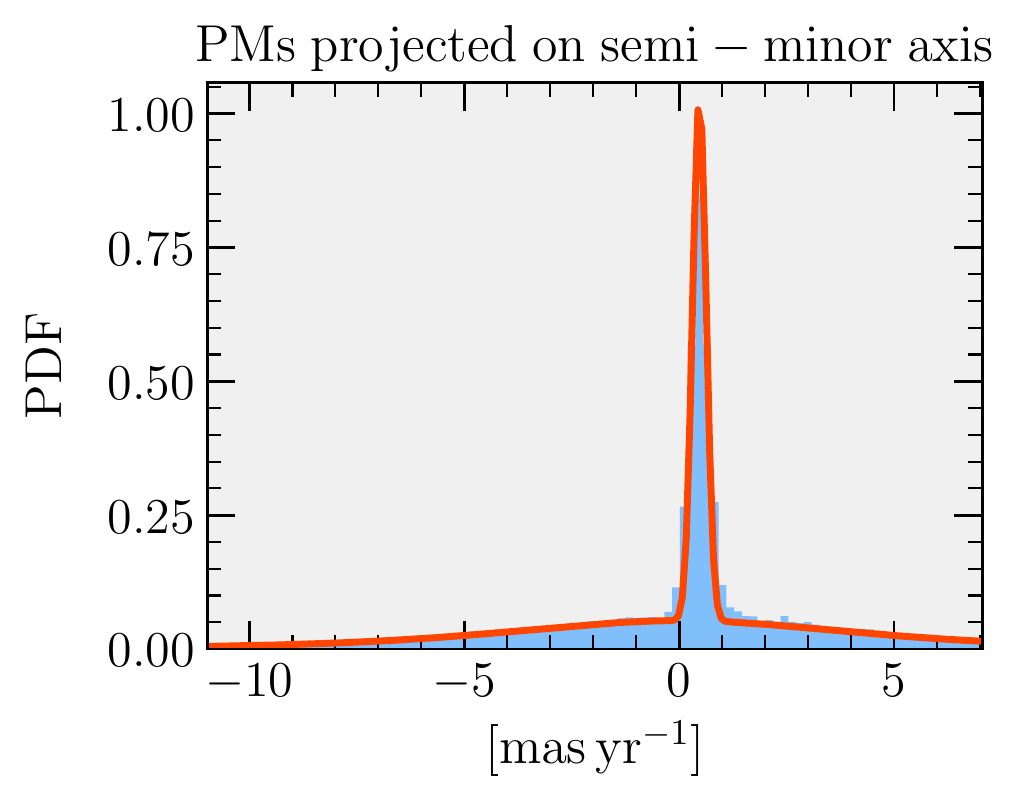}
\includegraphics[width=0.33\textwidth]{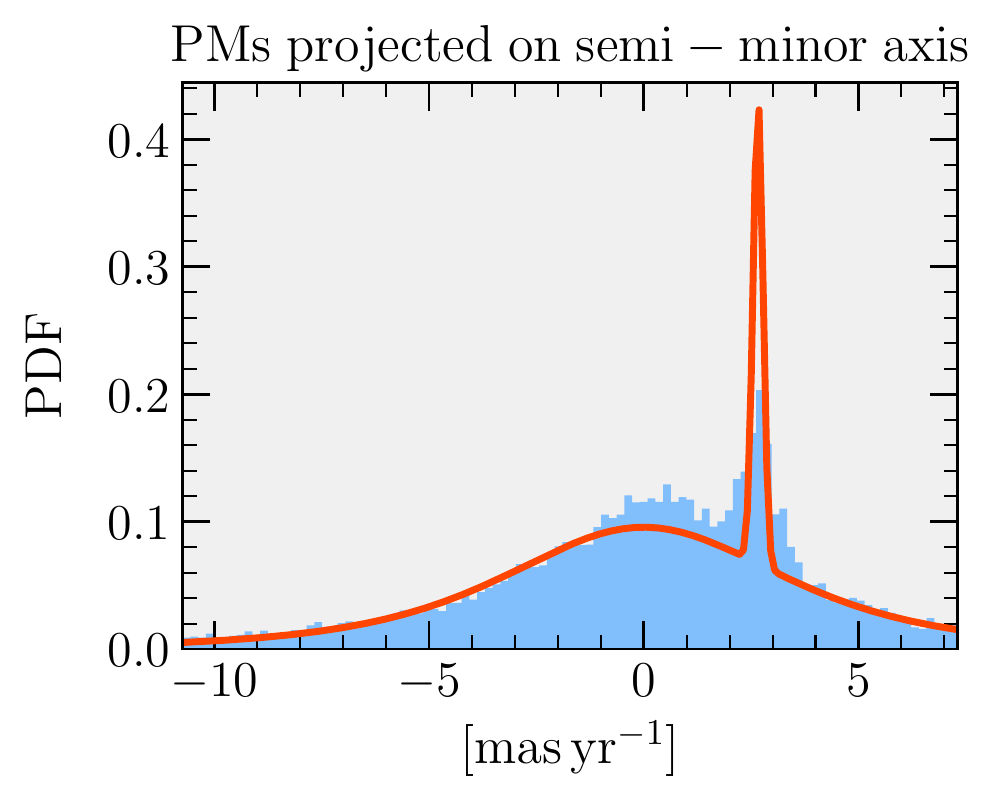}
\includegraphics[width=0.33\textwidth]{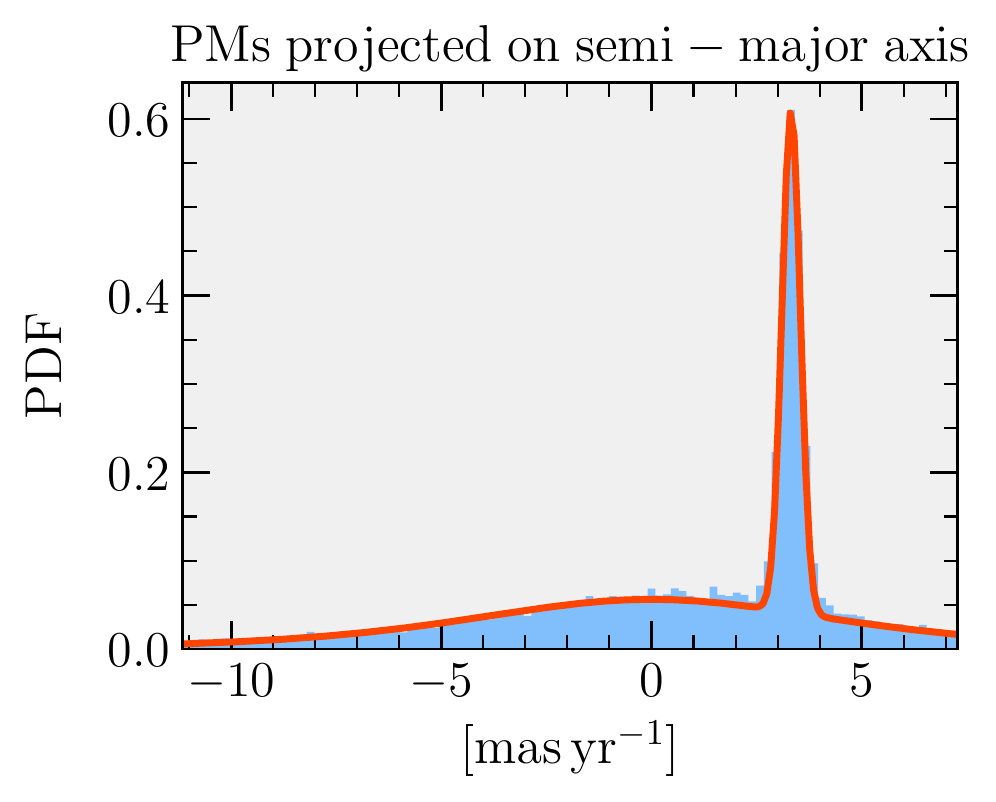}
\includegraphics[width=0.33\textwidth]{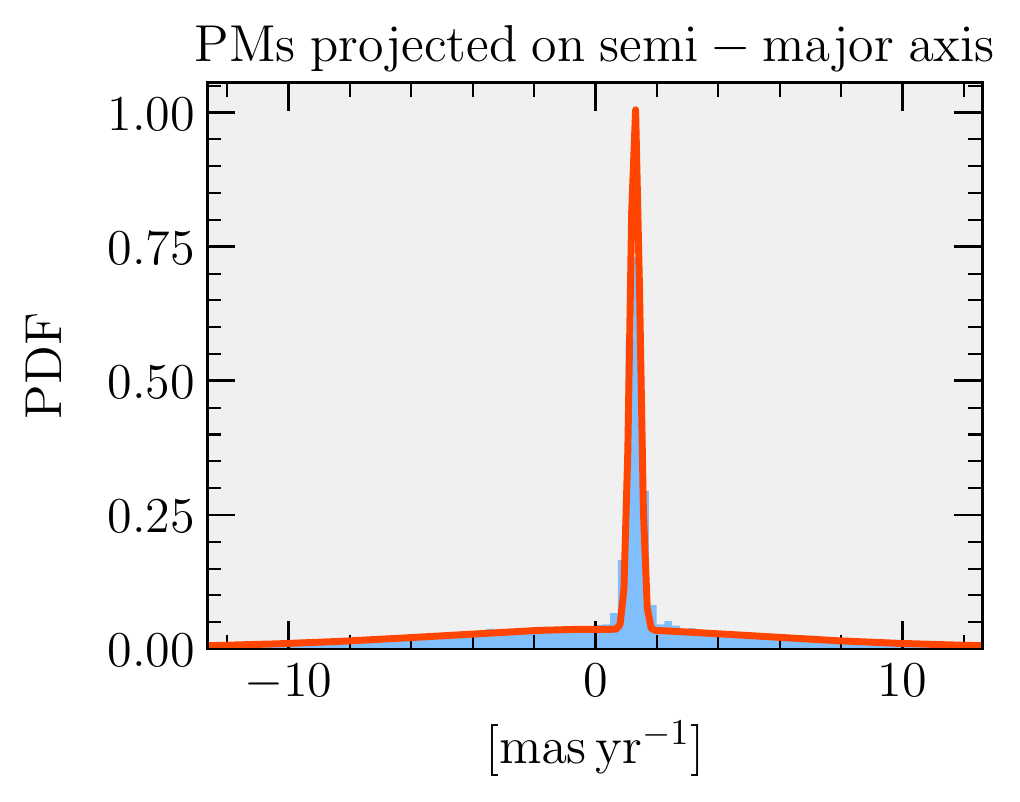}
\includegraphics[width=0.33\textwidth]{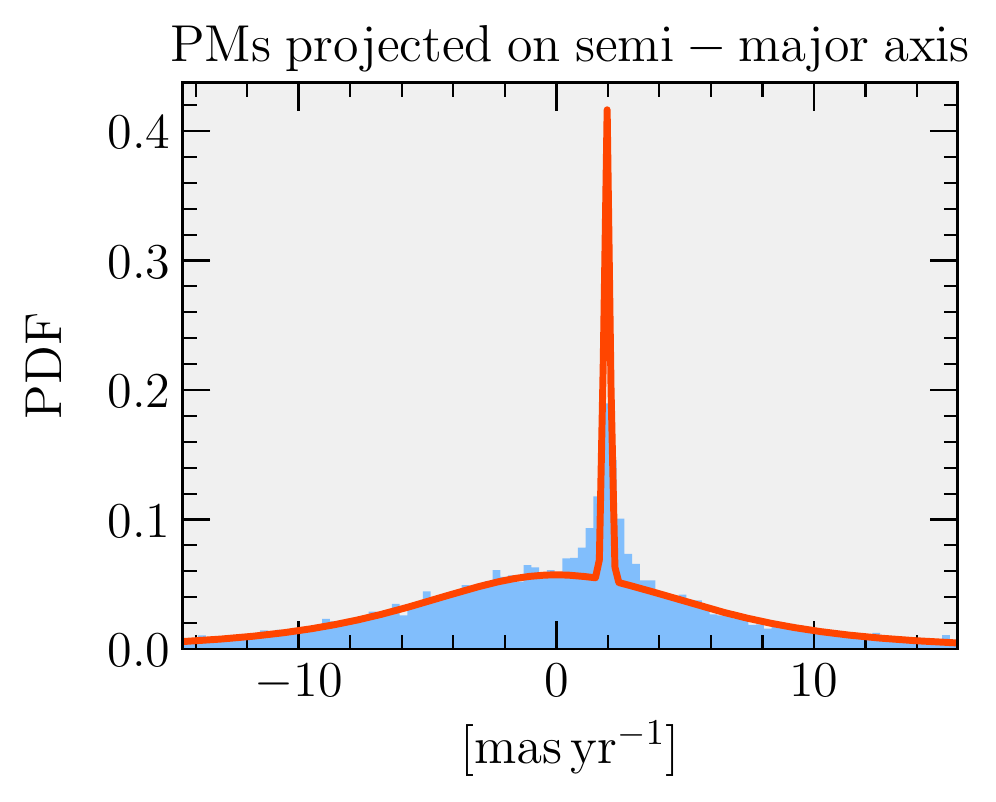}
\caption{{\it Proper motion fits:} We display here the results of the proper motion fits of NGC~6752, NGC~6205 (M~13) and the Draco dSph in the first, second and third columns, respectively. The first row displays the entire proper motion subset color-coded by stellar counts, from light blue to dark blue. The dashed ellipse displays the Pearson VII asymmetric distribution, with its semi-minor and semi-major directions as two perpendicular dotted lines, while the continuous circle represents the galactic object (globular cluster or dwarf spheroidal) proper motion mean with a radius equals ten times its intrinsic dispersion, for better visualization. The second and third rows display the fits (solid red) projected on the semi-minor and semi-major axis respectively, with the data in blue.}
\label{fig: pm-fits}
\end{figure*}

After having cleaned the data set according to the previous section, we proceed to fit the proper motion joint distribution of galactic objects plus interlopers. These two components present two clumps of data in proper motion space which allow one to assign membership probabilities to each clump.

\cite{Vasiliev19b} and \cite{McConnachie&Venn20} used Bayesian approaches based on Gaussian mixture models (a Gaussian distribution for both cluster and interloper clumps), while the original work from \cite{GaiaHelmi+18} used an iterative method based on clustering approaches and \cite{Baumgardt+19} performed an iterative cleaning routine based on an $n-\sigma$ outlier rejection. Even though Bayesian approaches tend to be more reliable, the use of Gaussian mixtures must, however, be taken with caution, since the distribution of field stars has often much wider wings than a Gaussian distribution (it can be cleverly bypassed by assigning multiple Gaussians to the interloper component, such as in \citealt{Vasiliev&Baumgardt&Baumgardt21}). This trend on the interloper proper motion distribution was first noticed in \citeauthor{Vitral&Mamon21} (\citeyear{Vitral&Mamon21}, see appendix B), where the authors opted for a symmetric Pearson VII distribution for the interlopers and the results were much more robust. The trend is again confirmed in this work, where we employ a refined version of their Bayesian method, by assigning a symmetric Gaussian distribution for the galactic object plus a non-symmetric Pearson VII distribution for interlopers. The choice of a non-symmetric distribution allows for a much better adjustment of the interlopers, which improves the membership probability of the stars in the subset, while the impact on the bulk proper motions of the galactic object varies from cluster to cluster\footnote{For example, there is no strong difference when using symmetric and non-symmetric interloper distributions for some GCs such as NGC~6397 and NGC~6121 (M4)}.

The main drawback of such an approach is that in order to introduce a non-symmetric Pearson VII distribution, which is analytically more complicated than a symmetric distribution, we abandon the convolution of the interloper distribution with Gaussian errors. This was not an issue for \cite{Vitral&Mamon21}, who used a polynomial approximation to this convolution for the symmetric case, which was simpler, in order to avoid extra integrals. This becomes much more computationally costly whenever accepting the non symmetricity of the data. We therefore try to counter this issue by trusting in our data cleaning described in the previous section, which applied conservative cuts on proper motion errors. Moreover, ignoring the error convolution for the non-Gaussian field population should not impact significantly the fits, since the interlopers show a much broader distribution, which is less affected than the narrow galactic object component (Gaussian). We further test this assumption in section~\ref{ssec: mock-analysis}.

\subsubsection{Probability distribution function}

To construct the probability distribution function (PDF) of the analyzed subset, we consider the PDF from these tracers and from the MW contaminants:

\begin{equation}
\label{eq: global_pdf}
    \mathrm{PDF} = f_{\rm GO} \, \mathrm{PDF}_{\rm GO} + (1 - f_{\rm GO}) \, \mathrm{PDF}_{\rm MW} \ ,
\end{equation}
where $f_{\rm GO}$ is the fraction of galactic object stars. The PDF of galactic objects is a straightforward Gaussian:

\begin{subequations} \label{eq: pdf_go}
\begin{equation}
    \mathrm{PDF}_{\rm GO}(\mu_{\alpha,*},\mu_{\delta}) = \frac{\exp{\left(- \zeta^{2} \right)}}{2 \, \pi \, \sigma^{2}_{\rm GO}} \ ,
\end{equation}
\begin{equation}
    \zeta^{2} = \frac{(\mu_{\alpha,*} - \mu_{\alpha,* \, \rm GO})^2 \, + \, (\mu_{\delta} - \mu_{\delta, \, \rm GO})^2}{2 \, \sigma^{2}_{\rm GO}}
\end{equation}
\end{subequations}
where $\sigma_{\rm GO}$ is the convolved proper motion dispersion of the galactic object stars, and $\mu_{\alpha,* \, \rm GO}$ and $\mu_{\delta, \, \rm GO}$ are their bulk proper motions in ($\alpha$, $\delta$). The convolution of the Gaussian component is done by considering $\sigma_{\rm GO}^{2} = \sigma_{\rm GO, int}^{2} + \epsilon^{2}$, where $\epsilon$ is the proper motion error (see equation~[20] from \citealt{Vitral&Mamon21}) and $\sigma_{\rm GO, int}$ is the intrinsic dispersion of the source. For the PDF of interlopers, we first shift the origin to the bulk proper motions of the interlopers clump, and then rotate the reference frame into the main axis of the interlopers proper motion ellipsoidal distribution:

\begin{subequations}
\label{eq: ref_frame}
\begin{equation}
    \mu_{x} = (\mu_{\alpha,*} - \mu_{\alpha,* \, \rm MW}) \, \cos{\theta} + (\mu_{\delta} - \mu_{\delta, \, \rm MW}) \, \sin{\theta} \ ,
\end{equation}
\begin{equation}
    \mu_{y} = - (\mu_{\alpha,*} - \mu_{\alpha,* \, \rm MW}) \, \sin{\theta} + (\mu_{\delta} - \mu_{\delta, \, \rm MW}) \, \cos{\theta} \ ,
\end{equation}
\end{subequations}
where $\mu_{\alpha,* \, \rm MW}$ and $\mu_{\delta, \, \rm MW}$ are the contaminants bulk proper motions in ($\alpha$, $\delta$) and $\theta$ is the angle between the original ($\mu_{\alpha,*}$, $\mu_{\delta}$) frame and the new one. 
Then, if we call the semi-major and semi-minor axis of the Pearson VII ellipsoidal distribution $a_{x}$ and $a_{y}$, we can write the interlopers PDF as:

\begin{equation}
    \mathrm{PDF_{MW}} = \displaystyle{ \left[ \frac{\Gamma{\left( - \frac{1}{2} - \frac{\tau}{2} \right)}}{\Gamma{\left(-1 - \frac{\tau}{2}\right)}} \right]^{2} \, \frac{ \left\{ \left[1 + \left(\frac{\mu_{x}}{a_{x}}\right)^{2} \right] \, \left[1 + \left(\frac{\mu_{y}}{a_{y}}\right)^{2} \right] \,  \right\}^{(1 + \tau)/2}}{ \pi \, a_{x} \, a_{y}} } \ ,
    \label{eq: pdf-mw}
\end{equation}
where $\Gamma{(x)}$ is the gamma function of $x$ and $\tau$ is an intrinsic slope of the distribution (with $\tau < -2$). Thus, we had ten free parameters, which were fitted by an MLE routine using the \textsc{differential\_evolution} method in \textsc{Python}. 

In Figure~\ref{fig: pm-fits}, we present the outcome of those fits for NGC~6752, NGC~6205 (M~13) and the Draco dSph, respectively. The first row shows the fit over the entire proper motion space, where one can verify the asymmetry of the interlopers distribution, while the second and third rows display the fits projected on the semi-minor and semi-major axis respectively.

\subsubsection{Handling errors}

\begin{figure*}
\centering
\includegraphics[width=0.33\textwidth]{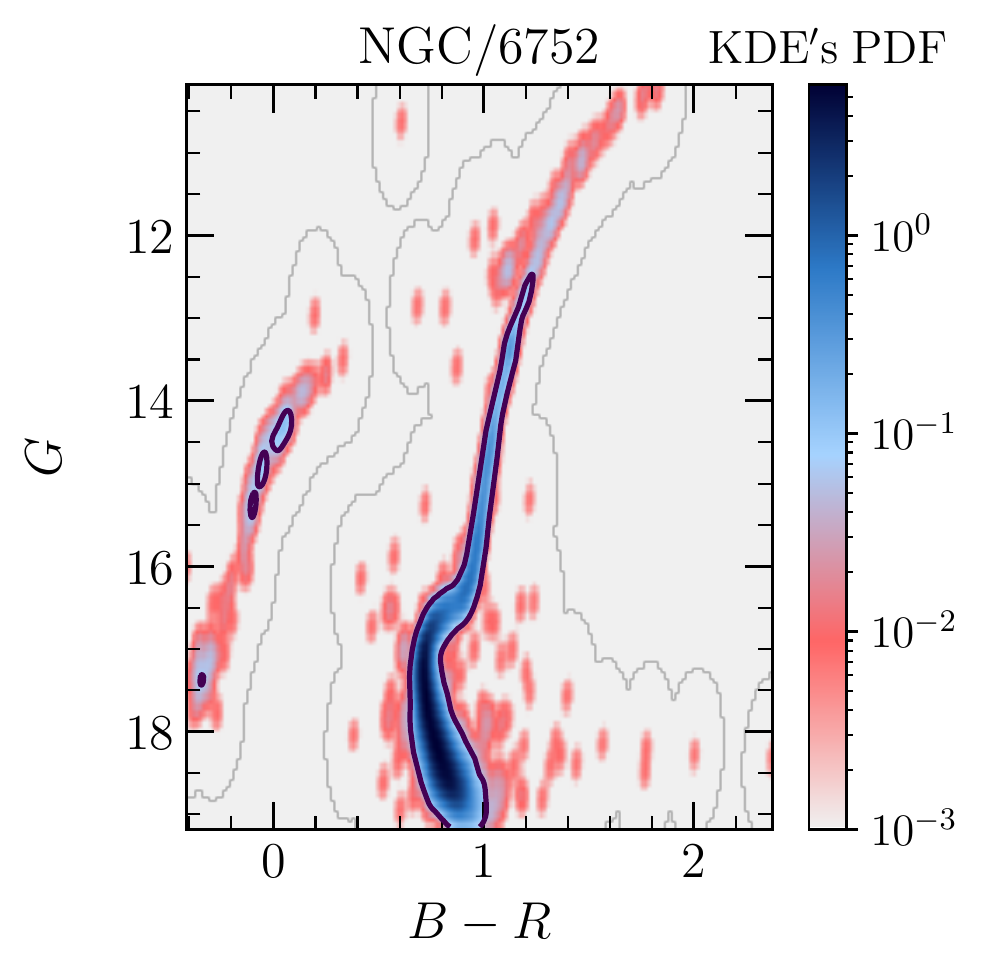}
\includegraphics[width=0.33\textwidth]{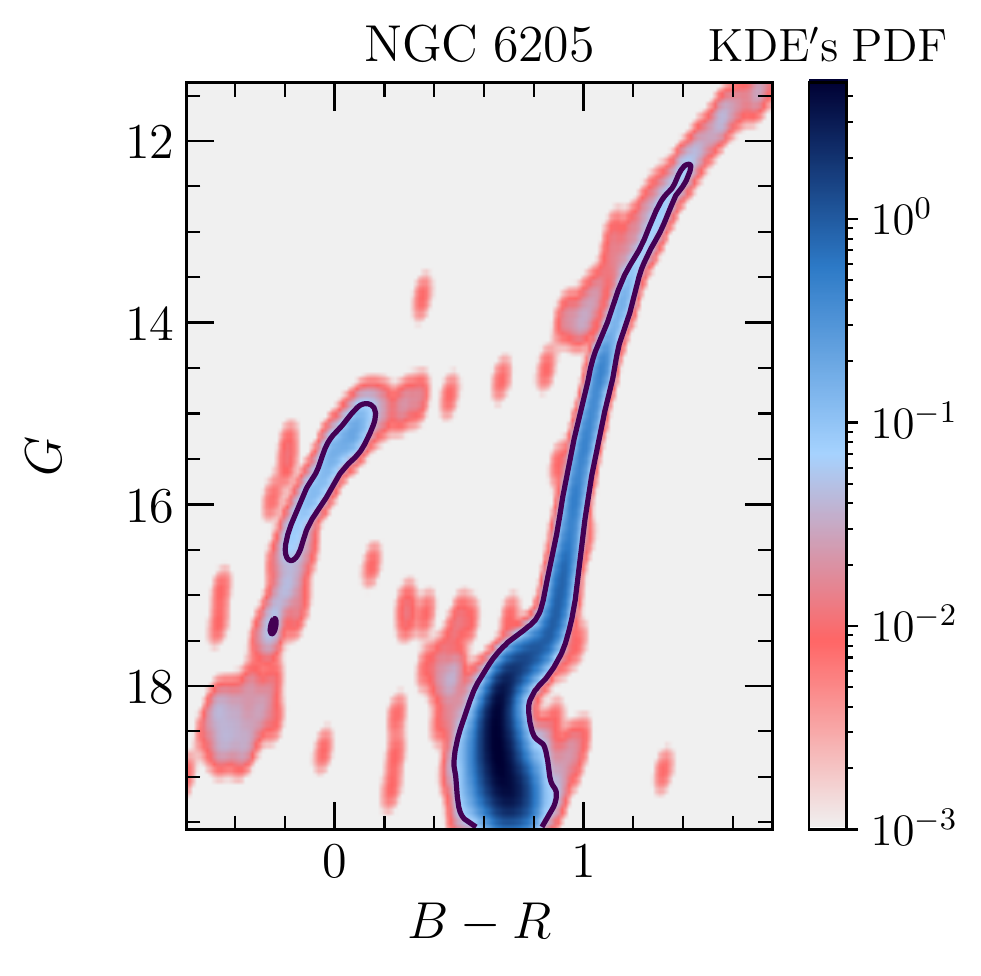}
\includegraphics[width=0.33\textwidth]{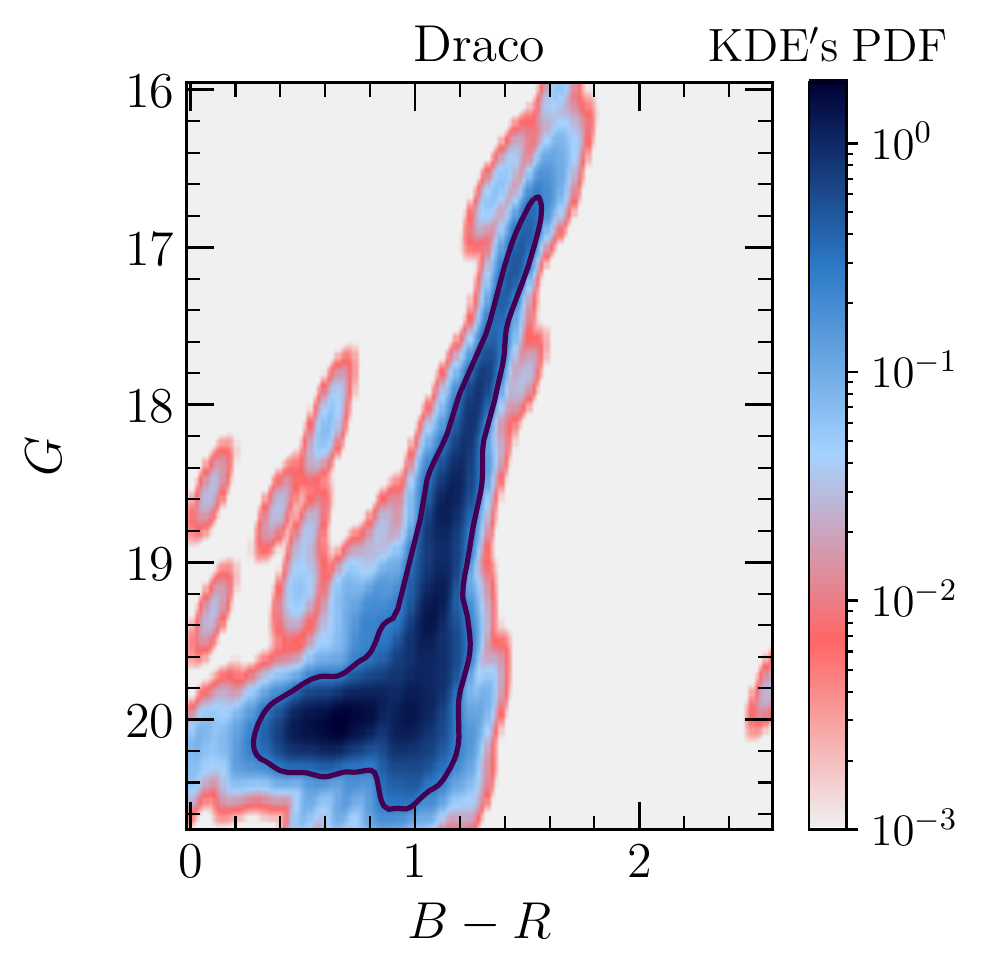}
\caption{{\it Color-magnitude diagram fits:} We display the color-magnitude diagram (CMD) of NGC~6752, NGC~6205 (M~13) and the Draco dSph color-coded by the Kernel Density Estimation (KDE) non-parametric PDF. The black line contours indicate 3$-\sigma$ (2.5$-\sigma$, for Draco) confidence regions, which we used to filter out interlopers and binaries that lie away from the CMD.}
\label{fig: cmd-fits}
\end{figure*}

In order to derive statistical errors of our Bayesian estimates, such as the bulk proper motion uncertainties shown in Table~\ref{tab: results}, we used \textsc{Python}'s \textsc{numdifftools.Hessian} method to compute the Hessian matrix of the proper motion PDF (i.e., eq~[\ref{eq: global_pdf}]). After, we assigned the uncertainties of each parameter as the square root of the respective diagonal position of the inverted Hessian matrix. To these statistical uncertainties, one should expect to incorporate a systematic error at the level of $\sim 0.025$ mas yr$^{-1}$ for \textsc{Gaia EDR3} (as estimated by \citealt{Lindegren+20} and \citealt{Vasiliev&Baumgardt&Baumgardt21}) and of $\sim 0.06$ mas yr$^{-1}$ for \textsc{Gaia DR2} (\citealt{Vasiliev19c}).

\subsection{Color magnitude diagram}


Once one has a precise analytical description of both the surface density and the proper motion distribution of the ensemble of field stars plus the analyzed galactic object, it is much easier to extract tracer members by means of membership probabilities. However, the incredible amount of astrophysical information from the \textsc{Gaia} releases allows to go even further, by analyzing how likely it is for a star to be part of the color-magnitude diagram of the tracer. It also allows to spot particular groups of stars such as binaries and blue stragglers (e.g., \citealt{Leonard89}) that lie away from the CMD.

The last step of our default filtering routine aims at constraining the region covered by the galactic object on the color-magnitude diagram (CMD). Since we do not have an analytical form to correctly describe the CMD, we opt, for the first time until now, to not use a Bayesian method, but rather a non-parametric Kernel Density Estimation (KDE) approach. We do so with the \textsc{Python} method \textsc{scipy.stats.gaussian\_kde}, similarly to \cite{Vitral&Mamon21}. Nevertheless, our approach has some few differences:

\begin{itemize}
    \item We first select stars with a membership probability greater than 0.8 (this limit is a modifiable parameter in \textsc{BALRoGO}).
    \item We used a KDE bandwidth of half the one derived by the Silverman's rule \citep{Silverman86}.
    \item We used the membership probabilities of each star as weights to the KDE routine.
    \item We selected stars inside a 3$-\sigma$ density contour in the CMD (this $n-\sigma$ limit is a modifiable parameter in \textsc{BALRoGO}, and set as 2.5 for dSphs in this article).
\end{itemize}

The CMD of NGC~6752, NGC~6205 (M~13) and the Draco dSph are presented in Figure~\ref{fig: cmd-fits}, with the 3$-\sigma$ (2.5$-\sigma$, for the dSph) contour region displayed as a black thick line.




\subsection{Mock data}
\label{ssec: mock}

\begin{figure*}
\centering
\includegraphics[width=0.33\textwidth]{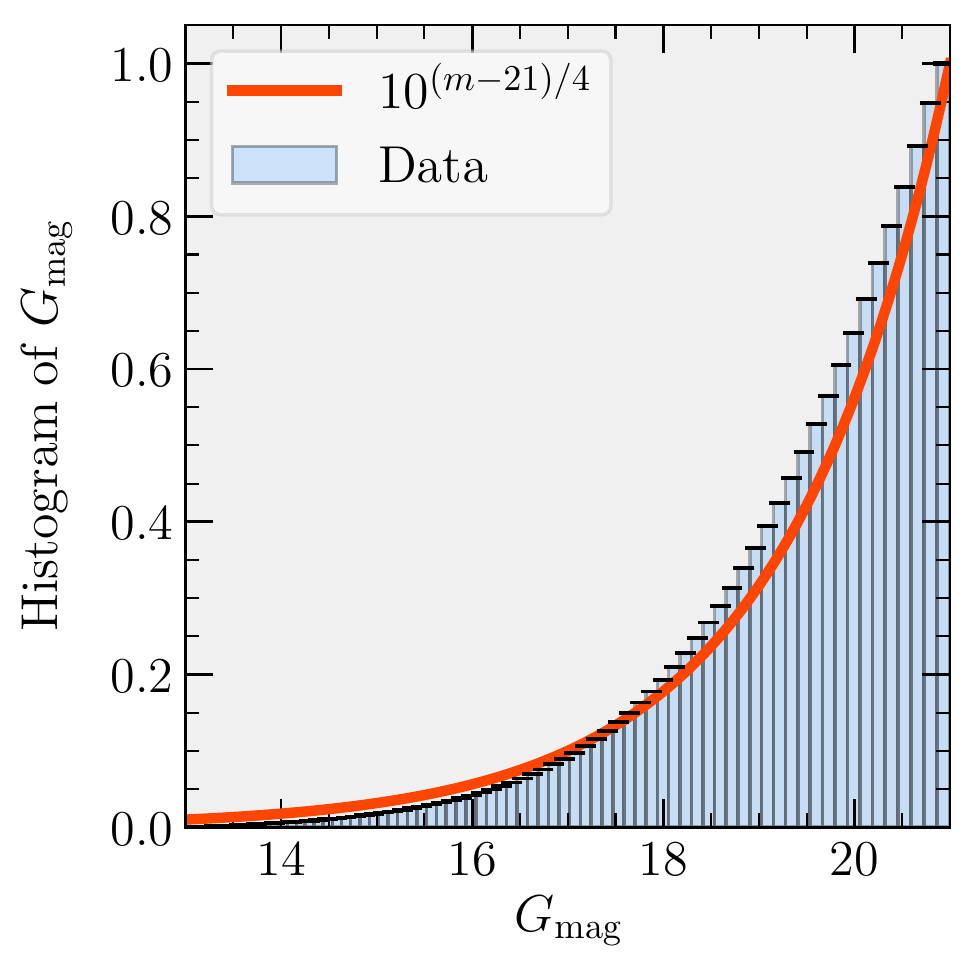}
\includegraphics[width=0.33\textwidth]{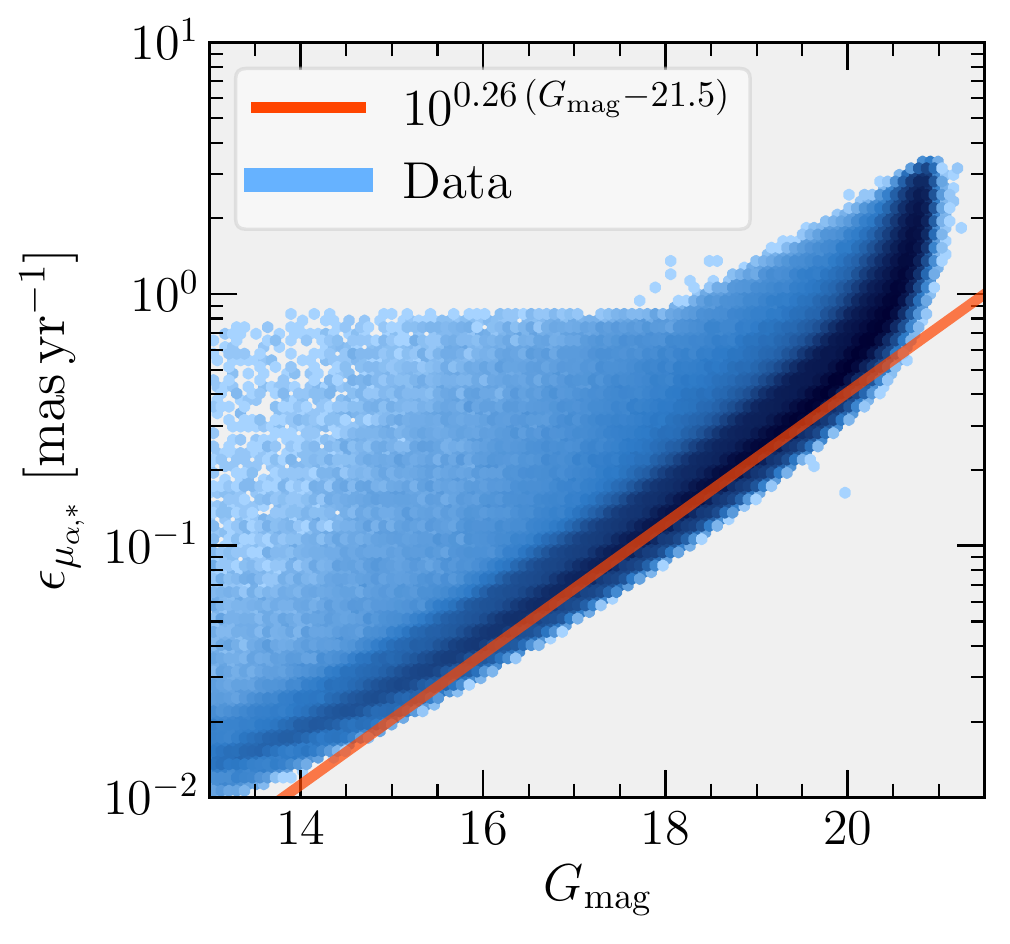}
\includegraphics[width=0.33\textwidth]{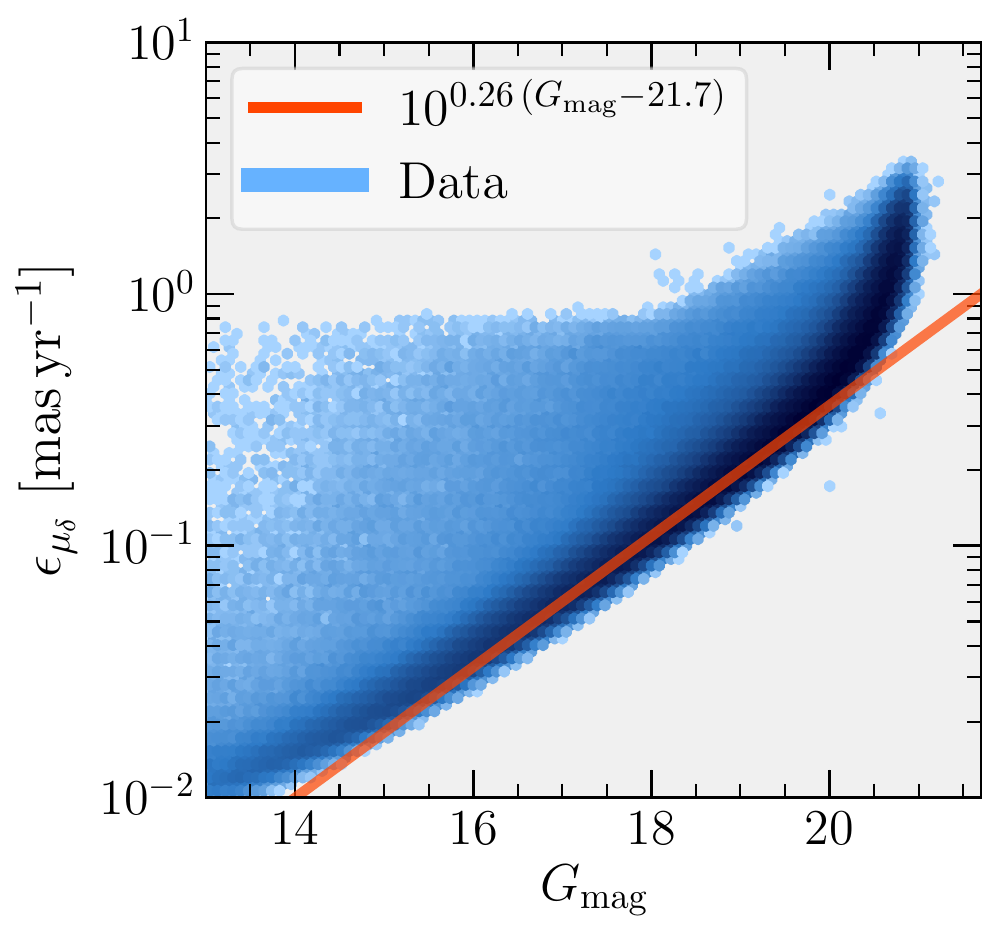}
\caption{{\it \textsc{Gaia EDR3} uncertainties:} On the \textbf{left}, we show the cumulative histogram of \textsc{Gaia EDR3} $G$ magnitudes in blue, along with the function $10^{(m-21)/4}$ in red. In the \textbf{middle}, we display the \textsc{Gaia EDR3} $\mu_{\alpha,*}$ uncertainties blue color-coded by stellar counts, with the function $10^{0.26 \, (m-21.5)}$ in red. In the \textbf{right}, we display the \textsc{Gaia EDR3} $\mu_{\delta}$ uncertainties blue color-coded by stellar counts, with the function $10^{0.26 \, (m-21.7)}$ in red. The \textsc{Gaia EDR3} data used for those plots is the stack of all stars in a two degrees cone search around the nearby globular clusters NGC~6121 (M~4), NGC~5139 ($\omega$ Cen), NGC~6397 and NGC~6752. }
\label{fig: pm-error}
\end{figure*}

For astronomers interested in generating mock data sets, \textsc{BALRoGO} has a \textsc{mock} method equipped with many capabilities such as (1) Projecting cartesian data into sky coordinates\footnote{$\alpha$, $\delta$, $\mu_{\alpha,*}$ and $\mu_{\delta}$}, (2) Adding field stars uniformly distributed in an spherical cap and following a Pearson VII distribution in proper motions space, and (3) Adding realistic \textsc{Gaia EDR3} errors to velocities. We describe below these three functionalities.

\subsubsection{Coordinate transformation}

\textsc{BALRoGO}'s method \textsc{cart6d\_to\_gaia} is able to convert cartesian coordinates into plane of sky coordinates by making use of \textsc{Astropy} \citep{AstropyCollaboration+13} \textsc{coordinates} method, allowing the users to decide weather they want a realistic data set with proper motion uncertainties of not. Moreover, if the users want to shift a certain source from one ($\alpha$, $\delta$) position in the sky to another one, this can be promptly done by calling the method \textsc{angle.transrot\_source}, which takes into account the spherical symmetry of the sky projection.

\subsubsection{Field stars}

When observing regions of the sky narrow enough (such as the two degrees cone searches we made in this work), one can expect to find uniformly distributed field stars in the field of view. By taking into account spherical trigonometry, one can generate such random distribution by inverting the probabilities $\Pr \{ r<R \}$ and $\Pr \{ \theta < \Theta \}$, where $R$ and $\Theta$ are the angular distance of a tracer from the source's center and the angle between this tracer and the source's center with respect to the increasing declination axis. We have, thus:

\begin{subequations}
\begin{equation}
\label{eq: random-radius}
    R = \arccos{[(1 - \mathrm{U}) + \mathrm{U} \, \cos{(R_{\rm lim})}]} \ ,
\end{equation}
\begin{equation}
\label{eq: random-theta}
    \Theta = 2 \, \pi \, \mathrm{U} \ ,
\end{equation}
\label{eq: random-positions}
\end{subequations}
where $R_{\rm lim}$ is the maximum distance from the source's center and U is a random variable uniformly distributed between zero and one. Similarly, one is able to generate random proper motions from a symmetric Pearson VII distribution ($a_{y} = a_{x}$ in equation~[\ref{eq: pdf-mw}]) by inverting the probability $\Pr \{ |\mu| < \mathrm{M} \}$:

\begin{equation}
    \mathrm{M} = a  \, \sqrt{\mathrm{U}^{1 / (1 + \tau / 2)} - 1} \ ,
    \label{eq: random-pms}
\end{equation}
where $a$ and $\tau$ are the scale radius and slope from the Pearson VII distribution, and once again U is a random variable uniformly distributed between zero and one. We distribute those proper motion moduli azimuthally by choosing angles from a distribution such as the one from equation~(\ref{eq: random-theta}). The precise derivation of the probabilities above is presented in appendix~\ref{app: random-fs}. 

\subsubsection{Realistic \textsc{Gaia EDR3} uncertainties}

The proper motion uncertainties in \textsc{Gaia EDR3} present a clear dependence on the apparent magnitude: In Figure~\ref{fig: pm-error}, we stacked all \textsc{Gaia EDR3} stars in a two degrees cone search around the nearby GCs NGC~6121 (M~4), NGC~5139 ($\omega$ Cen), NGC~6397 and NGC~6752, in order to show that dependence. For that reason, to generate realistic \textsc{Gaia EDR3} proper motion uncertainties, \textsc{BALRoGO} first generates random magnitudes, again, by inverting the probability $\Pr \{ G_{\rm mag} < m \}$, which can be easily derived from the cumulative distribution of $G_{\rm mag}$ magnitudes from \textsc{Gaia EDR3} (i.e., the \url{phot\_g\_mean\_mag} parameter), by imposing a threshold magnitude $m_{\rm lim}$. The respective equations for a distribution of magnitude $m$ are:

\begin{subequations}
\begin{equation}
    \Pr \{ G_{\rm mag} < m \} = 10^{(m - m_{\rm lim}) / f} \ ,
\end{equation}
\begin{equation}
    m = f \, \log_{10}{\rm U} + m_{\rm lim} \ ,
    \label{eq: random-mag}
\end{equation}
\end{subequations}
where U is a random variable uniformly distributed between zero and one, and $m_{\rm lim} = 21$ and $f = 4$, according to the fits displayed in Figure~\ref{fig: pm-error}. Once one has a random set of magnitudes from equation~(\ref{eq: random-mag}), we can again make use of the fits displayed in Figure~\ref{fig: pm-error} to assign:

\begin{subequations}
\begin{equation}
    \epsilon_{\mu_{\alpha,*}} = 10^{0.26 \, (m - 21.5)} \ ,
\end{equation}
\begin{equation}
    \epsilon_{\mu_{\delta}} = 10^{0.26 \, (m - 21.7)} \ .
\end{equation}
\end{subequations}

Those uncertainties are provided to the user, after adding Gaussian errors to $\mu_{\alpha,*}$ and $\mu_{\delta}$, with respective standard deviation of $\epsilon_{\mu_{\alpha,*}}$ and $\epsilon_{\mu_{\delta}}$.

\section{Results and discussion}

\begin{figure}
\centering
\includegraphics[width=0.9\hsize]{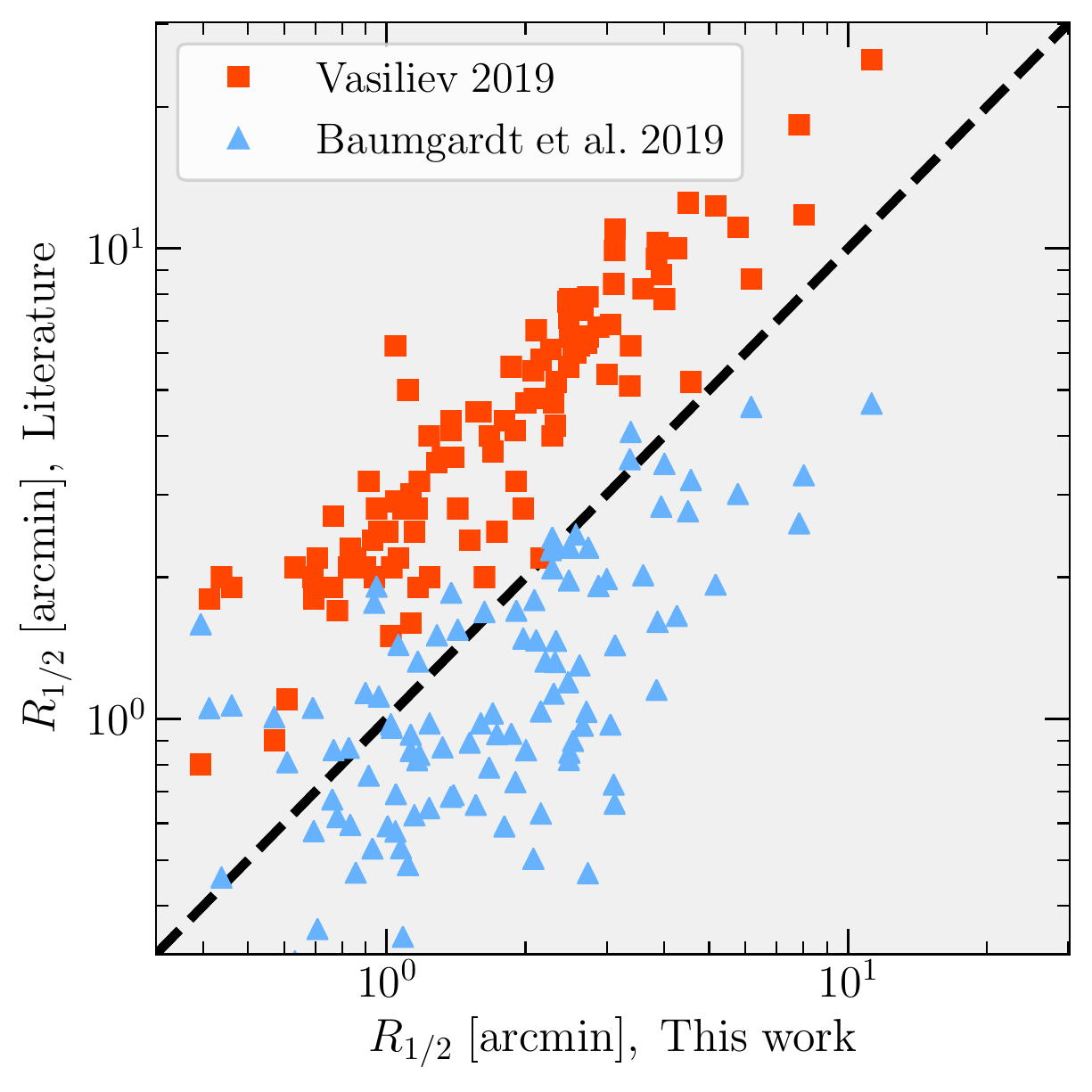}
\caption{{\it Effective radii:} Comparison between the half number radii in arcmin derived by \textsc{BALRoGO} from \textsc{Gaia EDR3} data in the x-axis, and the same quantity derived by \protect\citeauthor{Baumgardt+19} (\protect\citeyear{Baumgardt+19}, blue triangles) and \protect\citeauthor{Vasiliev19b} (\protect\citeyear{Vasiliev19b}, red squares) from \textsc{Gaia DR2} data in the y-axis. We display the $x = y$ line in dashed black.}
\label{fig: sd-comparison}
\end{figure}

\begin{figure}
\centering
\includegraphics[width=\hsize]{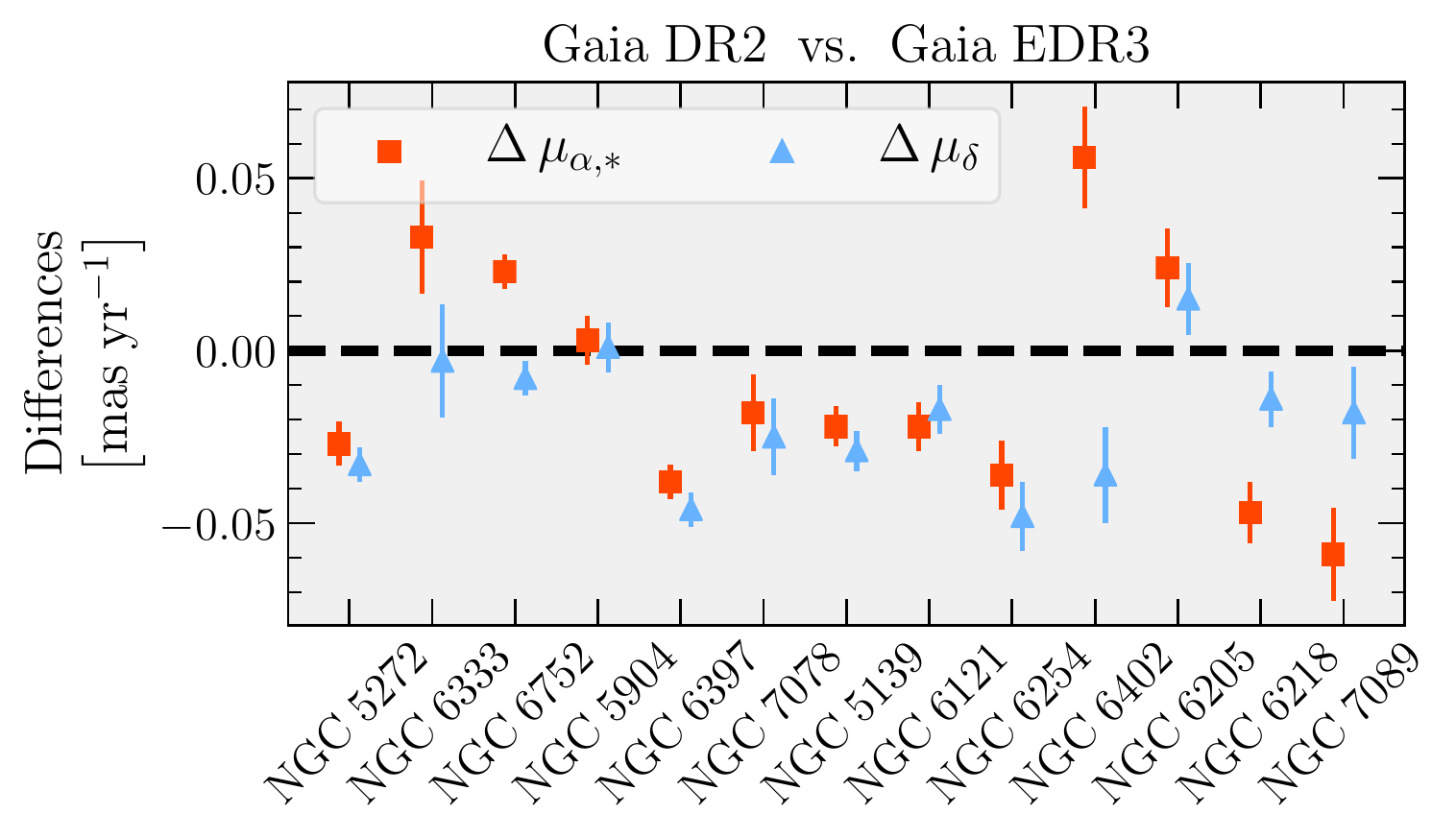}
\caption{{\it Improvement of the Gaia catalog:} Comparison of proper motion means derived by \textsc{BALRoGO} from \textsc{Gaia DR2} and \textsc{Gaia EDR3} data for the first ten globular clusters from the Messier catalogue plus NGC~6397, NGC~6752 and NGC~5139 ($\omega$ Cen). We display, in the y-axis, the differences $\Delta \, \mu_{\alpha,*} = \mu_{\alpha,* \, \rm DR2} - \mu_{\alpha,* \, \rm EDR3}$ and $\Delta \, \mu_{\delta} = \mu_{\delta \, \rm DR2} - \mu_{\delta \, \rm EDR3}$ as red squares and blue triangles, respectively, for the 13 globular clusters mentioned above, distributed along the x-axis. The errors bars were calculated as explained in section~\ref{ssec: pm-results}, and we plot a dashed black horizontal line at the value of zero as a reference.}
\label{fig: pm-gaias}
\end{figure}

\begin{figure}
\centering
\includegraphics[width=\hsize]{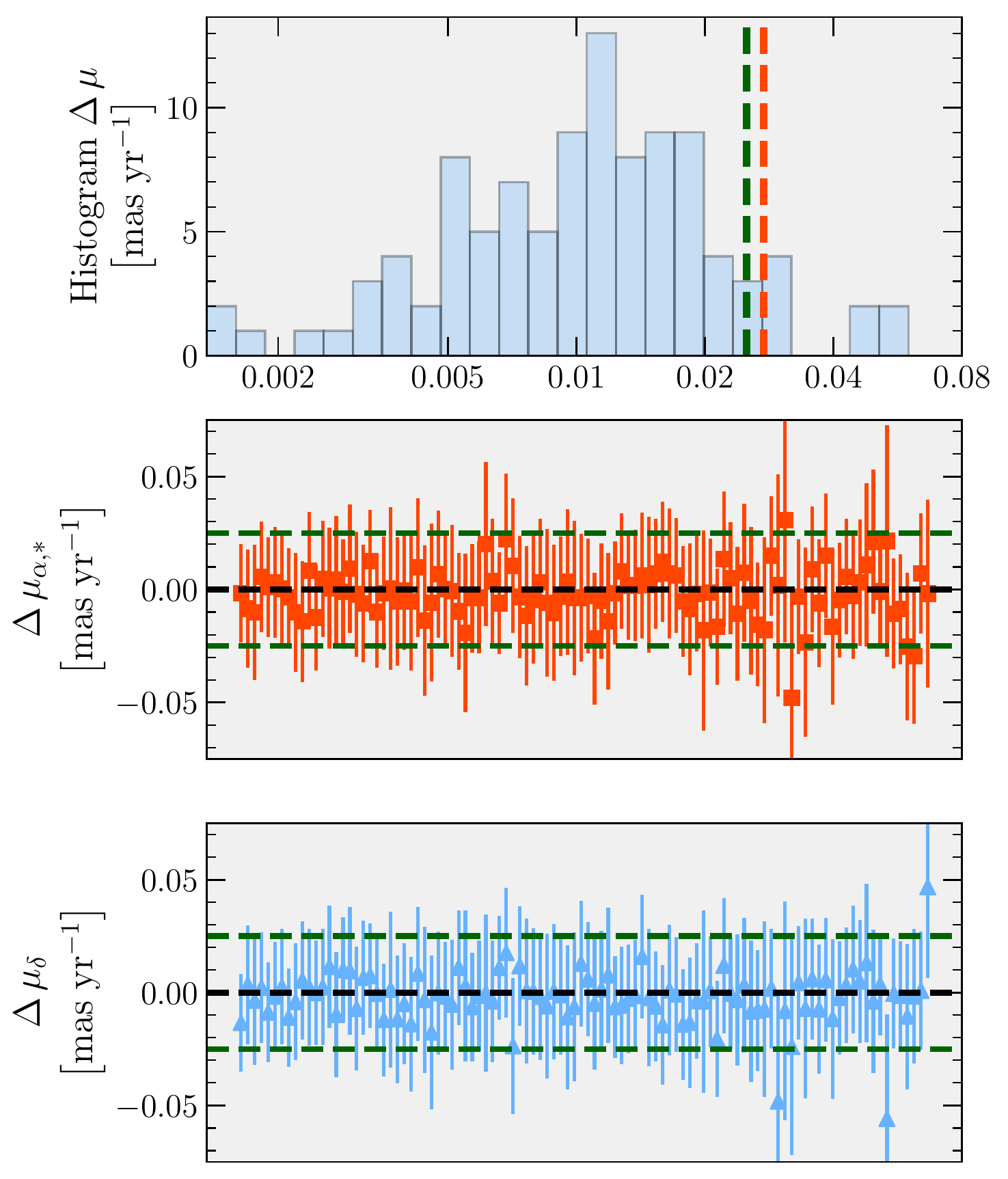}
\caption{{\it Bulk proper motions:} Comparison of proper motion means derived from \textsc{Gaia EDR3} data by \textsc{BALRoGO} and \protect\cite{Vasiliev&Baumgardt&Baumgardt21} for all the globular clusters in Table~\ref{tab: results}. The two \textit{bottom plots} display, in the y-axis, the differences $\Delta \, \mu_{\alpha,*} = \mu_{\alpha,* \, \rm this \ work} - \mu_{\alpha,* \, \rm other \ work}$ and $\Delta \, \mu_{\delta} = \mu_{\delta \, \rm this \ work} - \mu_{\delta \, \rm other \ work}$ as red squares and blue triangles, respectively, for the globular clusters in Table~\ref{tab: results}, distributed along the x-axis in a similar fashion than Figure~\ref{fig: pm-gaias}. In the \textit{top plot}, we display the histogram of $\Delta \mu = \sqrt{(\mu_{\alpha,* \, \rm other \ work} - \mu_{\alpha,* \,  \rm this \ work})^2 + (\mu_{\delta \, \rm other \ work} - \mu_{\delta \,  \rm this \ work})^2}$ in mas yr$^{-1}$, with the uncertainty floor of the \textsc{Gaia EDR3} data reported in \protect\cite{Vasiliev&Baumgardt&Baumgardt21}, of $0.025$ mas yr$^{-1}$ as a dashed green line and the median uncertainty in these differences, calculated with uncertainty propagation, as a dashed red line.}
\label{fig: pm-comparison}
\end{figure}

In Table~\ref{tab: results}, we display the main results of our fits, including our Bayesian estimates of the center, bulk proper motion and half number radii for over a hundred GCs and a few dSph galaxies measured by \textsc{Gaia EDR3}. This section aims to compare our results with previous estimates and discuss the implications of our new methods.

\subsection{Centers}

Our GC centers, derived according to section~\ref{ssec: center}, were compared to the estimates from \cite{Goldsbury+10} for a robustness check: The median separation between their centers and ours is of $0.12$ arcsec, while the median of their reported uncertainties is of $0.2$ arcsec.
Similarly, we compared our estimates with the sources from the \cite{Harris10} catalog, and obtained a median separation of $0.67$ arcsec.

This strengthens our initial assumption of small separations (i.e., $d \approx 0$, in Figure~\ref{fig: geometry}), and gives us confidence in our Bayesian center estimation. 
The maximum separation between our measurements and those from \cite{Goldsbury+10} and \cite{Harris10} was of $26.5$ arcsec, for the GC NGC~4147, followed by NGC~6553, with a separation of $26.1$ arcsec and then by NGC~6558, with a separation of $9.5$ arcsec.

\subsection{Effective radii}
\label{ssec: rh-results}

Figure~\ref{fig: sd-comparison} displays the effective (two-dimensional) radii derived from \textsc{Gaia EDR3} by \textsc{BALRoGO} in the x-axis along with the same quantity derived from \textsc{Gaia DR2} by \cite{Vasiliev19b} and \cite{Baumgardt+19}, in red and blue respectively, in the y-axis. The conversion to projected half number radii was straightforward for both S\'ersic and Plummer fits, since the scale radius used in both models was already the half number radius. For the Kazantizidis model, we multiplied the Kazantizidis scale radius $a_{\rm K}$ from equations~\ref{eq: kaz-sd} and \ref{eq: kaz-n} by $1.257$ in order to retrieve the equivalent half number radius.

The values from \cite{Vasiliev19b} are in general higher, which he mentions in his Figure~C.1 to be a consequence of the incompleteness in the central regions of his filtered \textsc{Gaia} catalog, which may lead his derived scale radius to be much larger than the actual half number radius of the cluster computed from all stars. In contrast, the measurements from \cite{Baumgardt+19} seem slightly smaller than ours, which in turn may indicate that our measurements are also slightly overestimated due to the intrinsic incompleteness of \textsc{Gaia}. 

In any case, it is impressive that \textsc{BALRoGO} stands right in the middle of both estimates, which can be considered as a reliable indicator of an adequate goodness of fit. This highlights one of the strengths of our surface density fit method, which is taking into account a constant distribution of MW field stars, and neglecting any data filtering in this first step: It avoids a forced incompleteness towards both the cluster center and outskirts, due to crowdness and fainter stars respectively, which are associated with worse astrometric solutions that would be filtered out in most filtering routines.

\subsection{Proper motions}
\label{ssec: pm-results}

\subsubsection{Gaia DR2 vs. Gaia EDR3}

In this subsection we stress the important difference between the bulk proper motions derived with \textsc{Gaia DR2} and \textsc{Gaia EDR3}. In order to make such a comparison possible, we decided to perform the same analysis, but this time using \textsc{Gaia DR2}, for 13 GCs: the first ten GCs of the Messier catalogue, plus NGC~6397, NGC~6752 and NGC~5139 ($\omega$ Cen). We show, in Figure~\ref{fig: pm-gaias}, the differences between the bulk proper motions ($\mu_{\alpha,*}$, $\mu_{\delta}$) estimated by \textsc{BALRoGO} from \textsc{Gaia DR2} and from \textsc{Gaia EDR3}. 

The uncertainty of those mean values was calculated as $\epsilon = \sqrt{ \epsilon_{\rm DR2}^2 + \epsilon_{\rm EDR3}^2 }$, where $\epsilon_{i}$ stands for the uncertainties on the estimated values from the catalog $i$. One can notice that the disagreements, dominated by DR2 errors, lie in-between the uncertainty floor of the \textsc{Gaia DR2} mission reported in \cite{Vasiliev19c}, of $\sim  0.06$ mas yr$^{-1}$. This is an important indicator of the improvement of \textsc{Gaia EDR3}, with more reliable measurements and a longer baseline, which in turns leads to smaller systematic uncertainties of the order of $\sim 0.025$ mas yr$^{-1}$ (e.g., \citealt{Lindegren+20} and \citealt{Vasiliev&Baumgardt&Baumgardt21}).

\subsubsection{Comparison with the literature}

\cite{GaiaHelmi+18}, \cite{Baumgardt+19} and \cite{Vasiliev19b} measured bulk proper motions for over a hundred GCs with \textsc{Gaia DR2} by using different methods previously mentioned. In this work, we update those information with the new \textsc{Gaia EDR3}, which has more precise and robust measurements of proper motions given its longer baseline. Due to this fact, the values of proper motions means for nearly all GCs changed by more than their respective errors published in the works mentioned above. Therefore, it would be unfair to compare our results using \textsc{Gaia EDR3} with their results obtained from \textsc{Gaia DR2} modelling.

Fortunately, \cite{Vasiliev&Baumgardt&Baumgardt21} recently provided bulk proper motion fits for 170 GCs, which allow for such comparison. We show, in Figure~\ref{fig: pm-comparison}, the differences between the bulk proper motions ($\mu_{\alpha,*}$, $\mu_{\delta}$) estimated from \textsc{Gaia EDR3} by \textsc{BALRoGO} and by their work in the two bottom panels, in a similar fashion than displayed in Figure~\ref{fig: pm-gaias}, but for all GCs in Table~\ref{tab: results}. The upper plot displays the histogram of $\Delta \mu = \sqrt{(\mu_{\alpha,* \, \rm VB21} - \mu_{\alpha,* \,  \rm V21})^2 + (\mu_{\delta \, \rm VB21} - \mu_{\delta \,  \rm V21})^2}$ in mas yr$^{-1}$ (VB21 stands for \citealt{Vasiliev&Baumgardt&Baumgardt21} and V21 for this work) with the uncertainty floor of the \textsc{Gaia EDR3} data reported in \cite{Lindegren+20} and \cite{Vasiliev&Baumgardt&Baumgardt21}, of $0.025$ mas yr$^{-1}$ as a dashed green line and the median uncertainty in these differences, calculated with uncertainty propagation, as a dashed red line.

We observe a very good agreement between the measurements using \textsc{BALRoGO} and the measurements from \cite{Vasiliev&Baumgardt&Baumgardt21}: Most of the sources lie below the median uncertainty of $\approx 0.03$ mas yr$^{-1}$, with the exception of NGC~6440, NGC~6453, NGC~6522, NGC~6528 and NGC~6540, which present a $\Delta \mu \approx 0.05$ mas yr$^{-1}$, with high statistical uncertainties, on the order of $0.05$ mas yr$^{-1}$. The reason for such disagreement is likely the small amount of tracers of NGC~6453, NGC~6522, NGC~6528, NGC~6540, which all have a small extension (i.e., $R_{1/2} \lesssim 0.6$ arcmin), and are therefore more affected by data cleaning and also the increased amount of interlopers in the proper motion space of NGC~6440. In this plot, it is important to mention that the differences are certainly smaller than what could be expected from formal error bars, since both studies use the same EDR3 data (different from Figure~\ref{fig: pm-gaias}), in which case the systematic uncertainty cancels out (it could be viewed as the calibration error on the proper motion zero-point, which varies across the sky at this level).

In addition to the comparison between GCs made above, our proper motion fits of dSphs provide a very good agreement with the estimates from \cite{McConnachie&Venn20}, with differences $\lesssim 0.05$ mas yr$^{-1}$ for all of our fits in Table~\ref{tab: results}. In the very late stages of this work, \cite{Li+21} also provided bulk proper motion fits\footnote{They apply a similar method than presented in \cite{Vasiliev19b}, therefore also relying on Gaussian mixtures.} for 46 dSphs, presenting a very good overall agreement with our measurements, a part from the Bootes I dSph, for which our measurements are closer to those from \cite{McConnachie&Venn20}.
This gives us confidence on the use of a non-Gaussian mixture in our Bayesian fit, along with the choice of a Pearson VII non-symmetric distribution of proper motions for interlopers, even though we neglected the convolution of the field stars component with Gaussian errors, after cleaning the data.

\subsection{Convolution with Gaussian errors}
\label{ssec: mock-analysis}

\begin{figure*}
\centering
\includegraphics[width=0.33\hsize]{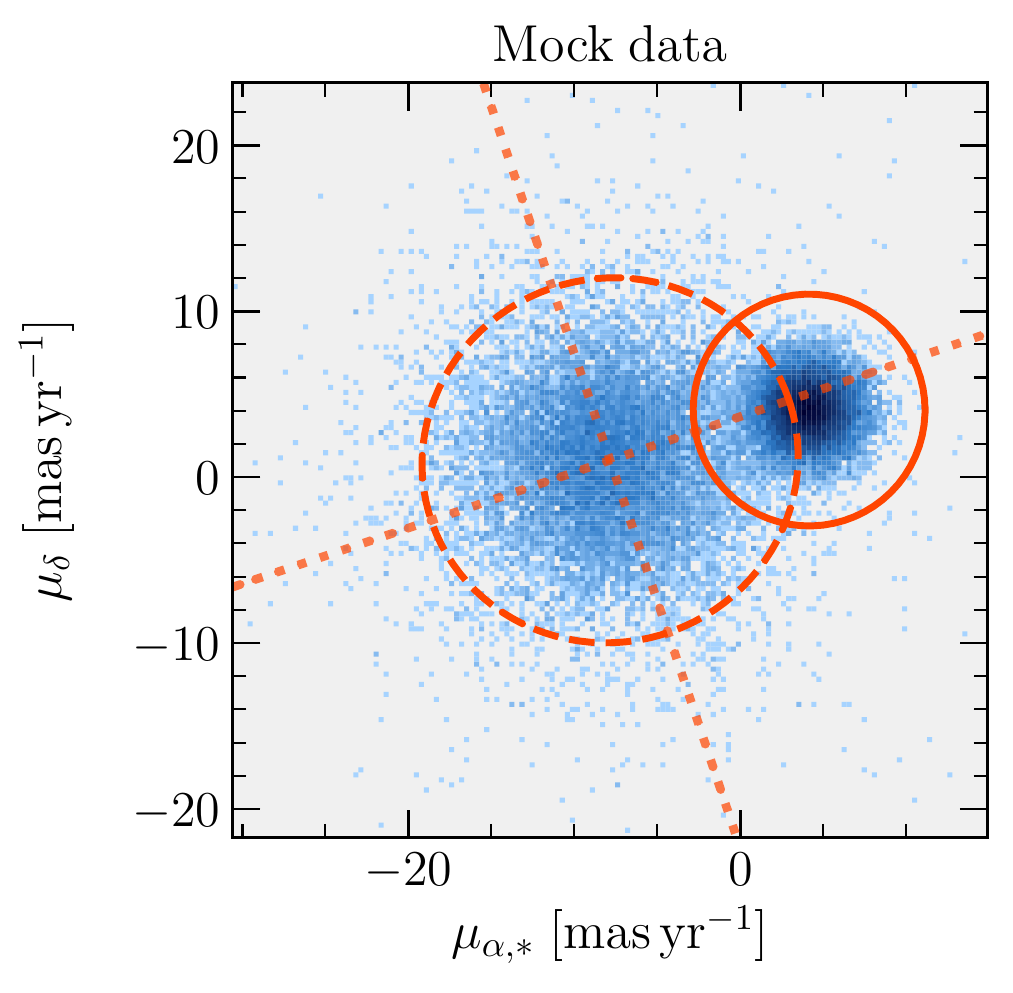}
\includegraphics[width=0.33\hsize]{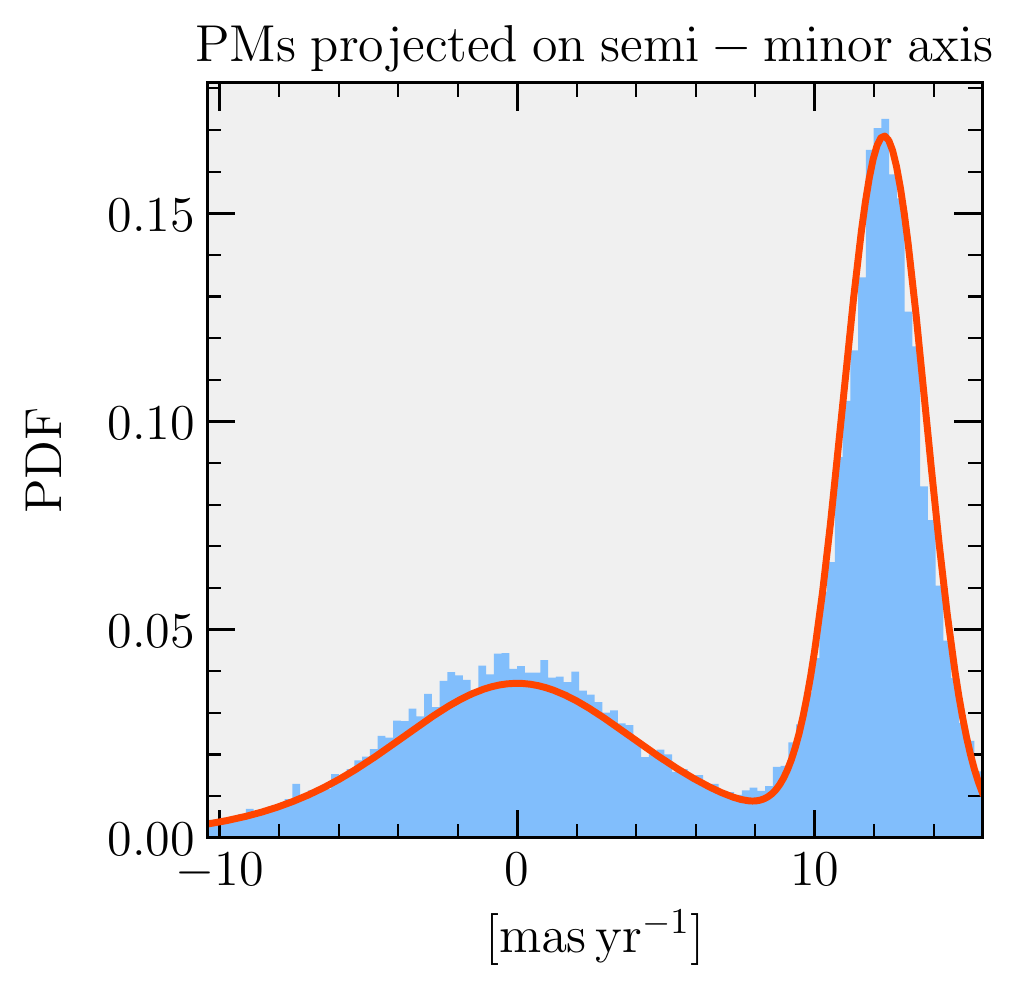}
\includegraphics[width=0.33\hsize]{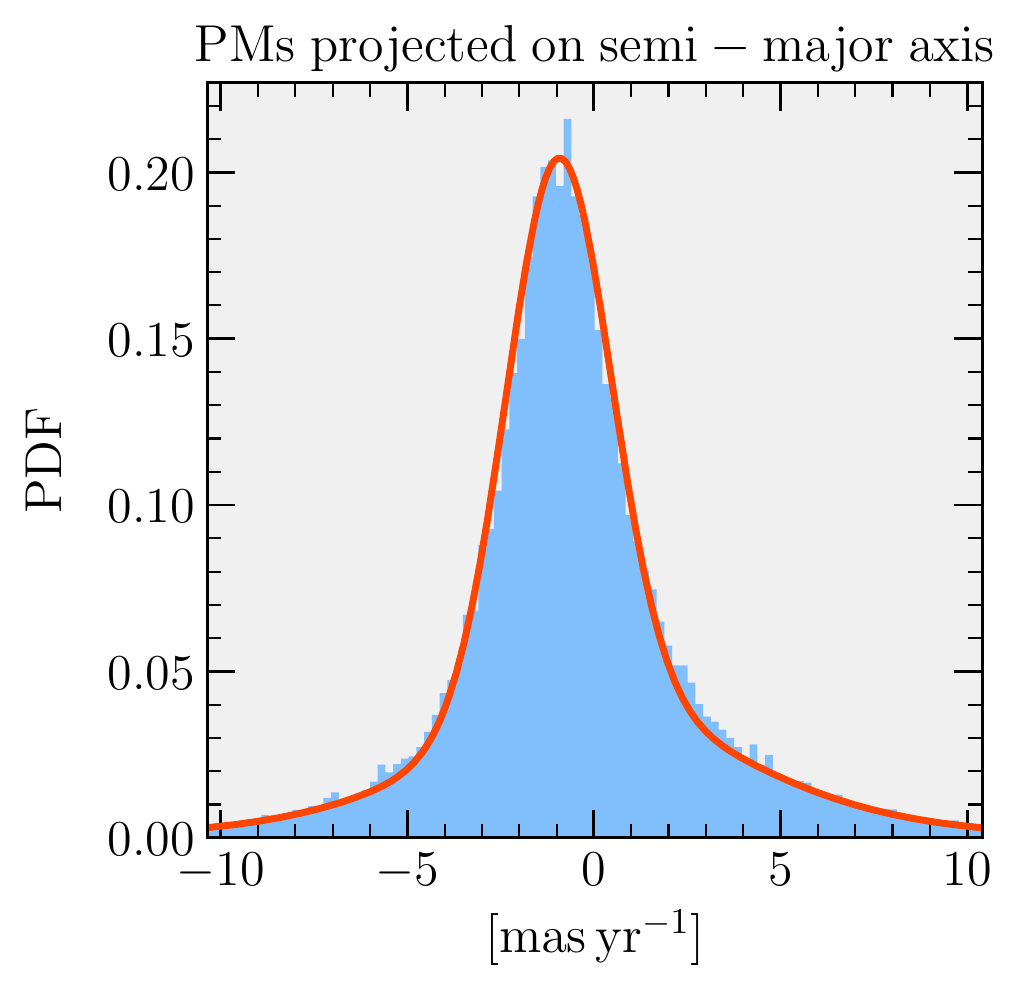}
\caption{{\it Mock data:} We display here the results of the proper motion fits of our mock data set with realistic \textsc{Gaia EDR3} errors. The image on the first column displays the entire proper motion subset color-coded by stellar counts, from light blue to dark blue. The dashed ellipse displays the fitted Pearson VII symmetric distribution, with its main axis directions as two perpendicular dotted lines, while the continuous circle represents the galactic object (mock globular cluster) proper motion mean with a radius equals five times its intrinsic dispersion, for better visualization. The second and third columns display the fits (solid red) projected on the semi-minor and semi-major axis respectively, with the data in blue.}
\label{fig: mock-pm}
\end{figure*}

\begin{table}

\centering
\tabcolsep=2.4pt
\caption{Comparison of estimates on the bulk $\mu_{\alpha, *}$ and $\mu_{\delta }$ from the clean and inaccurate mock data sets (see section~\ref{ssec: mock-analysis}).} 
\label{tab: mock-fits}
\footnotesize
\begin{tabular}{lcc}

\hline
Data set & $\mu_{\alpha *}$ & $\mu_{\delta }$ \\
[0.5ex] 
\hline
& [mas yr$^{-1}$] & [mas yr$^{-1}$] \\ [0.5ex] 
\hline \hline

Clean & $4.141 \pm 0.011$ & $4.040 \pm 0.011$  \\

Inaccurate & $4.146 \pm 0.012$  & $4.039 \pm 0.012$ \\
\hline

\end{tabular}
\normalsize
\end{table}

As previously pointed out, our conservative data cleaning from section~\ref{ssec: clean} should be enough to counter the fact that we do not convolve the Pearson VII distribution with Gaussian errors. In order to test this assumption, Pierre Boldrini kindly provided a GC mock using the initial condition $N-$body generator \textsc{magi} \citep{Miki&Umemura&Umemura18}. Adopting a distribution-function-based method, it ensures that the final realization of the cluster is in dynamical equilibrium \citep{Miki&Umemura&Umemura18}. This GC mock was inspired by the real cluster NGC~6397, and therefore followed a S\'ersic profile with stellar mass of $1.17 \times 10^{5}M_{\sun}$, Sérsic radius of $3.14$ pc, Sérsic index of $3.3$ \citep{Vitral&Mamon21}, along with orbital and tidal radius of $5.91$ kpc \citep{Vasiliev19b} and $59.9$ pc (using the relation from \citealt{Bertin2008}), respectively.

To this data, composed of $57 500$ stars, we added five times more stars, following spatial and velocity distributions of \textsc{Gaia EDR3} field stars, as described in section~\ref{ssec: mock}. From this new data set, we created an extra one with realistic \textsc{Gaia EDR3} proper motion uncertainties, constructed accordingly to section~\ref{ssec: mock}, in order to test if the lack of convolution with Gaussian errors could significantly impact our results. We randomly selected $10^4$ stars in both subsets (hereafter clean and inaccurate subsets, for simplicity) and ran \textsc{BALRoGO}'s routine on them. Figure~\ref{fig: mock-pm} displays the proper motion fits of the inaccurate subset, similar to Figure~\ref{fig: pm-fits}. 

The fitted bulk $\mu_{\alpha,*}$ and $\mu_{\delta}$ from the clean and inaccurate subsets, along with their respective uncertainties are displayed in Table~\ref{tab: mock-fits}. We can verify that even without convolving the global proper motion PDF with Gaussian errors, the fits on the bulk proper motion from the data sets with and without uncertainties agree within the 1$-\sigma$ error bars, and their disagreement lies below the \textsc{Gaia EDR3} uncertainty floor of 0.025 mas yr$^{-1}$. This strengthens our confidence in the proper motion cleaning routine from \textsc{BALRoGO}, as well as in its fits.

\subsection{Kurtosis of the interlopers proper motions}

This work confirms the tendency of field stars proper motions to follow a Pearson VII distribution, instead of a Gaussian, and we address now the interpretation of this result. Such an effect could be explained by the fact that the proper motions delivered by \textsc{Gaia} are not a direct measurement of the velocity, i.e. a measure of space variation per time, but rather a measurement of angular velocity, which neglects the distance of the stars.

According to the central limit theorem \citep{Laplace1810}, when one considers the ensemble of independent and identically distributed random variables sharing the same dispersion and mean, their properly normalized sum tends toward a normal distribution regardless of the variables original distribution. In the case of GCs and dSph, this can be considered as an adequate approximation, since these sources contain generally tens or hundreds of thousands of stars \citep{Binney&Tremaine08} which are located at distances that can be considered barely the same for a distant observer. This means that when converting the spatial velocities of its stars into angular velocities (i.e., dividing by their distance and turning them into proper motions), their originally quasi-independent and identically distributed velocities remain as such, as well as their similar dispersion, and therefore the variation around their mean is close to a Gaussian.

In the case of MW field stars however, since they have completely different distances, the distribution of proper motions is drawn away from an independent and identically distributed assumption, with each random measurement having a particular dispersion. In fact, their variation around their mean depends on the distance they lie from the observer, some of them much closer and others much farther than the galactic object analyzed. As a consequence, we expect to find more outliers, i.e., stars with proper motions that deviate more strongly from their mean, and thus a higher kurtosis (wider tails) in the distribution. That is why the Pearson VII distribution, with its wider tails, is better adapted to fit the proper motions of interlopers than a Gaussian distribution.

\section{Conclusions}

We present a new algorithm aimed at measuring some of the main structural parameters of galactic objects such as globular clusters and dwarf spheroidal galaxies. This algorithm, named \textsc{BALRoGO: Bayesian Astrometric Likelihood Recovery of Galactic Objects}, performs Bayesian and non-parametric fits in order to extract stars drowned in Milky Way interlopers, by accessing their membership probabilities (with respect to the galactic object analyzed). 

Our approach presents innovative points which have been previously used in \cite{Vitral&Mamon21}, but with some new improvements. Among the new approaches used in this algorithm, we highlight the following:

\begin{itemize}
    \item Bayesian surface density fits considering a constant contribution from Milky Way interlopers.
    \item We allow the surface density modelling of globular clusters with profiles such as S\'ersic and Kazantzidis, instead of the generally used King models \citep{King66}. Cored density profiles are well handled by a Plummer profile or low S\'ersic indexes.
    \item The proper motion Bayesian fit is not based on a Gaussian mixture, but rather a Pearson VII distribution for interlopers, given their distribution wider tails.
    \item We use the membership probabilities of the surface density and proper motion fits in order to derive a non-parametric representation of the color-magnitude diagram, and then select cluster members inside confidence regions from this representation.
\end{itemize}

Comparisons between our method and previous works such as \cite{Baumgardt+19}, \cite{Vasiliev19b}, \cite{McConnachie&Venn20} and \cite{Vasiliev&Baumgardt&Baumgardt21} indicate strong agreement of bulk proper motions for globular clusters and dwarf spheroidal galaxies, and of scale radii for globular clusters. In addition, we make available our measurements of center, bulk proper motions and scale radii for over a hundred globular clusters from the NGC catalog, along with a few dwarf spheroidal galaxies in Table~\ref{tab: results} (see data availability section).

The dynamics of such stellar systems is a fundamental key to understand some of the main aspects of galaxy evolution, as well as the astrophysical impacts of dark matter. With new releases of the \textsc{Gaia} astrometric mission, along with other astrometric data sets, the future of dynamical modeling is very enticing. For that reason, we believe it is important to make algorithms such as ours available, which can be easily used in order to derive important parameters of many stellar systems.

\section*{Acknowledgements}
First, I want to thank Alexandre Macedo for all his efforts towards making \textsc{BALRoGO} public, readable and accessible by anyone. I acknowledge the important comments of Gary Mamon, in particular for proposing equation~(\ref{eq: plummer-n}) and providing equation~(\ref{eq: casephifull}) in a more complicated form than presented here. I also thank Pierre Boldrini, who provided the globular cluster mock data presented in section~\ref{ssec: mock-analysis} and gave insightful comments on this work.
\\
I acknowledge the referee, Eugene Vasiliev, for his comprehensive analysis of this work and important suggestions that improved the data cleaning and the proper motion fitting procedure. I finally thank Zhao Su for noticing a bad center fit for NGC~1904 in the submitted version.
\\
Eduardo Vitral was funded by an AMX doctoral grant from \'Ecole Polytechnique.
\\
This work greatly benefited from public software
packages such as
{\sc Numpy} \citep{vanderWalt11},
{\sc Astropy} \citep{AstropyCollaboration+13},
{\sc emcee} \citep{Goodman&Weare10},
{\sc Matplotlib} \citep{Hunter07}, as well as the {\sc Spyder} Integrated Development Environment (Copyright (c) 2009- Spyder Project Contributors and others).

\section*{Data Availability}

The data that support the plots within this paper and other findings of this study are available from the corresponding author upon reasonable request. Table~\ref{tab: results} can be accessed in text format at \url{https://gitlab.com/eduardo-vitral/balrogo/-/raw/master/table\_gc\_dsph.dat}. The code repository link of \textsc{BALRoGO}, on GitLab, is \url{https://gitlab.com/eduardo-vitral/balrogo}.



\bibliographystyle{mnras}
\bibliography{src} 

\begin{thebibliography}{}
\makeatletter
\relax
\def\mn@urlcharsother{\let\do\@makeother \do\$\do\&\do\#\do\^\do\_\do\%\do\~}
\def\mn@doi{\begingroup\mn@urlcharsother \@ifnextchar [ {\mn@doi@}
  {\mn@doi@[]}}
\def\mn@doi@[#1]#2{\def\@tempa{#1}\ifx\@tempa\@empty \href
  {http://dx.doi.org/#2} {doi:#2}\else \href {http://dx.doi.org/#2} {#1}\fi
  \endgroup}
\def\mn@eprint#1#2{\mn@eprint@#1:#2::\@nil}
\def\mn@eprint@arXiv#1{\href {http://arxiv.org/abs/#1} {{\tt arXiv:#1}}}
\def\mn@eprint@dblp#1{\href {http://dblp.uni-trier.de/rec/bibtex/#1.xml}
  {dblp:#1}}
\def\mn@eprint@#1:#2:#3:#4\@nil{\def\@tempa {#1}\def\@tempb {#2}\def\@tempc
  {#3}\ifx \@tempc \@empty \let \@tempc \@tempb \let \@tempb \@tempa \fi \ifx
  \@tempb \@empty \def\@tempb {arXiv}\fi \@ifundefined
  {mn@eprint@\@tempb}{\@tempb:\@tempc}{\expandafter \expandafter \csname
  mn@eprint@\@tempb\endcsname \expandafter{\@tempc}}}

\bibitem[\protect\citeauthoryear{{Abadi}, {Navarro}  \& {Steinmetz}}{{Abadi}
  et~al.}{2006}]{Abadi+06}
{Abadi} M.~G.,  {Navarro} J.~F.,   {Steinmetz} M.,  2006, \mn@doi [\mnras]
  {10.1111/j.1365-2966.2005.09789.x}, \href
  {https://ui.adsabs.harvard.edu/abs/2006MNRAS.365..747A} {365, 747}

\bibitem[\protect\citeauthoryear{Akaike}{Akaike}{1973}]{akaike1973information}
Akaike H.,  1973, Information Theory and an Extension of the Maximum Likelihood
  Principle.
Springer New York, New York, NY, pp 199--213

\bibitem[\protect\citeauthoryear{{Arenou} et~al.,}{{Arenou}
  et~al.}{2018}]{Arenou+18}
{Arenou} F.,  et~al., 2018, \mn@doi [\aap] {10.1051/0004-6361/201833234}, \href
  {https://ui.adsabs.harvard.edu/abs/2018A&A...616A..17A} {616, A17}

\bibitem[\protect\citeauthoryear{{Astropy Collaboration} et~al.,}{{Astropy
  Collaboration} et~al.}{2013}]{AstropyCollaboration+13}
{Astropy Collaboration} et~al., 2013, \mn@doi [\aap]
  {10.1051/0004-6361/201322068}, \href
  {https://ui.adsabs.harvard.edu/abs/2013A&A...558A..33A} {558, A33}

\bibitem[\protect\citeauthoryear{{Battaglia}, {Helmi}  \&
  {Breddels}}{{Battaglia} et~al.}{2013}]{Battaglia+13}
{Battaglia} G.,  {Helmi} A.,   {Breddels} M.,  2013, \mn@doi [\nar]
  {10.1016/j.newar.2013.05.003}, \href
  {https://ui.adsabs.harvard.edu/abs/2013NewAR..57...52B} {57, 52}

\bibitem[\protect\citeauthoryear{{Baumgardt}, {Hilker}, {Sollima}  \&
  {Bellini}}{{Baumgardt} et~al.}{2019}]{Baumgardt+19}
{Baumgardt} H.,  {Hilker} M.,  {Sollima} A.,   {Bellini} A.,  2019, \mn@doi
  [\mnras] {10.1093/mnras/sty2997}, \href
  {https://ui.adsabs.harvard.edu/abs/2019MNRAS.482.5138B} {482, 5138}

\bibitem[\protect\citeauthoryear{{Bertin} \& {Varri}}{{Bertin} \&
  {Varri}}{2008}]{Bertin2008}
{Bertin} G.,  {Varri} A.~L.,  2008, \mn@doi [\apj] {10.1086/592684}, \href
  {https://ui.adsabs.harvard.edu/abs/2008ApJ...689.1005B} {689, 1005}

\bibitem[\protect\citeauthoryear{{Binney} \& {Tremaine}}{{Binney} \&
  {Tremaine}}{2008}]{Binney&Tremaine08}
{Binney} J.,  {Tremaine} S.,  2008, Galactic Dynamics: Second Edition.
Princeton University Press, Princeton, NJ, USA

\bibitem[\protect\citeauthoryear{{Boldrini}, {Miki}, {Wagner}, {Mohayaee},
  {Silk}  \& {Arbey}}{{Boldrini} et~al.}{2020}]{Boldrini+20c}
{Boldrini} P.,  {Miki} Y.,  {Wagner} A.~Y.,  {Mohayaee} R.,  {Silk} J.,
  {Arbey} A.,  2020, \mn@doi [\mnras] {10.1093/mnras/staa150}, \href
  {https://ui.adsabs.harvard.edu/abs/2020MNRAS.492.5218B} {492, 5218}

\bibitem[\protect\citeauthoryear{{Bullock} \& {Johnston}}{{Bullock} \&
  {Johnston}}{2005}]{Bullock&Johnston05}
{Bullock} J.~S.,  {Johnston} K.~V.,  2005, \mn@doi [\apj] {10.1086/497422},
  \href {https://ui.adsabs.harvard.edu/abs/2005ApJ...635..931B} {635, 931}

\bibitem[\protect\citeauthoryear{{Bustos Fierro} \& {Calder{\'o}n}}{{Bustos
  Fierro} \& {Calder{\'o}n}}{2019}]{BustosFierro&Calderon19}
{Bustos Fierro} I.~H.,  {Calder{\'o}n} J.~H.,  2019, \mn@doi [\mnras]
  {10.1093/mnras/stz1879}, \href
  {https://ui.adsabs.harvard.edu/abs/2019MNRAS.488.3024B} {488, 3024}

\bibitem[\protect\citeauthoryear{{Carretta} et~al.,}{{Carretta}
  et~al.}{2009}]{Carretta+09}
{Carretta} E.,  et~al., 2009, \mn@doi [\aap] {10.1051/0004-6361/200912096},
  \href {https://ui.adsabs.harvard.edu/abs/2009A&A...505..117C} {505, 117}

\bibitem[\protect\citeauthoryear{{Gaia Collaboration} et~al.,}{{Gaia
  Collaboration} et~al.}{2018}]{GaiaHelmi+18}
{Gaia Collaboration} et~al., 2018, \mn@doi [\aap]
  {10.1051/0004-6361/201832698}, \href
  {https://ui.adsabs.harvard.edu/abs/2018A&A...616A..12G} {616, A12}

\bibitem[\protect\citeauthoryear{{Galadi-Enriquez}, {Jordi}  \&
  {Trullols}}{{Galadi-Enriquez} et~al.}{1998}]{GaladiEnriquez+98}
{Galadi-Enriquez} D.,  {Jordi} C.,   {Trullols} E.,  1998, VizieR Online Data
  Catalog, \href {https://ui.adsabs.harvard.edu/abs/1998yCat..33370125G} {pp
  J/A+A/337/125}

\bibitem[\protect\citeauthoryear{{Goldsbury}, {Richer}, {Anderson}, {Dotter},
  {Sarajedini}  \& {Woodley}}{{Goldsbury} et~al.}{2010}]{Goldsbury+10}
{Goldsbury} R.,  {Richer} H.~B.,  {Anderson} J.,  {Dotter} A.,  {Sarajedini}
  A.,   {Woodley} K.,  2010, \mn@doi [\aj] {10.1088/0004-6256/140/6/1830},
  \href {https://ui.adsabs.harvard.edu/abs/2010AJ....140.1830G} {140, 1830}

\bibitem[\protect\citeauthoryear{{Goodman} \& {Weare}}{{Goodman} \&
  {Weare}}{2010}]{Goodman&Weare10}
{Goodman} J.,  {Weare} J.,  2010, \mn@doi [Communications in Applied
  Mathematics and Computational Science] {10.2140/camcos.2010.5.65}, \href
  {https://ui.adsabs.harvard.edu/abs/2010CAMCS...5...65G} {5, 65}

\bibitem[\protect\citeauthoryear{{Harris}}{{Harris}}{2010}]{Harris10}
{Harris} W.~E.,  2010, arXiv e-prints, \href
  {https://ui.adsabs.harvard.edu/abs/2010arXiv1012.3224H} {p. arXiv:1012.3224}

\bibitem[\protect\citeauthoryear{{Hernquist}}{{Hernquist}}{1990}]{Hernquist90}
{Hernquist} L.,  1990, \apj, 356, 359

\bibitem[\protect\citeauthoryear{{Hunter}}{{Hunter}}{2007}]{Hunter07}
{Hunter} J.~D.,  2007, \mn@doi [Computing in Science \& Engineering]
  {10.1109/MCSE.2007.55}, 9, 90

\bibitem[\protect\citeauthoryear{{Kazantzidis}, {Mayer}, {Mastropietro},
  {Diemand}, {Stadel}  \& {Moore}}{{Kazantzidis} et~al.}{2004}]{Kazantzidis+04}
{Kazantzidis} S.,  {Mayer} L.,  {Mastropietro} C.,  {Diemand} J.,  {Stadel} J.,
    {Moore} B.,  2004, \mn@doi [\apj] {10.1086/420840}, \href
  {https://ui.adsabs.harvard.edu/abs/2004ApJ...608..663K} {608, 663}

\bibitem[\protect\citeauthoryear{{King}}{{King}}{1966}]{King66}
{King} I.~R.,  1966, \mn@doi [\aj] {10.1086/109857}, \href
  {https://ui.adsabs.harvard.edu/abs/1966AJ.....71...64K} {71, 64}

\bibitem[\protect\citeauthoryear{{Laplace}}{{Laplace}}{1810}]{Laplace1810}
{Laplace} P.,  1810, M\'emoires de l'Acad\'emie Royale des Sciences de Paris,
  10, 301

\bibitem[\protect\citeauthoryear{{Leonard}}{{Leonard}}{1989}]{Leonard89}
{Leonard} P. J.~T.,  1989, \mn@doi [\aj] {10.1086/115138}, \href
  {https://ui.adsabs.harvard.edu/abs/1989AJ.....98..217L} {98, 217}

\bibitem[\protect\citeauthoryear{{Li}, {Hammer}, {Babusiaux}, {Pawlowski},
  {Yang}, {Arenou}, {Du}  \& {Wang}}{{Li} et~al.}{2021}]{Li+21}
{Li} H.,  {Hammer} F.,  {Babusiaux} C.,  {Pawlowski} M.~S.,  {Yang} Y.,
  {Arenou} F.,  {Du} C.,   {Wang} J.,  2021, arXiv e-prints, \href
  {https://ui.adsabs.harvard.edu/abs/2021arXiv210403974L} {p. arXiv:2104.03974}

\bibitem[\protect\citeauthoryear{{Lindegren} et~al.,}{{Lindegren}
  et~al.}{2018}]{Lindegren+18}
{Lindegren} L.,  et~al., 2018, \mn@doi [\aap] {10.1051/0004-6361/201832727},
  \href {https://ui.adsabs.harvard.edu/abs/2018A&A...616A...2L} {616, A2}

\bibitem[\protect\citeauthoryear{{Lindegren} et~al.,}{{Lindegren}
  et~al.}{2020}]{Lindegren+20}
{Lindegren} L.,  et~al., 2020, arXiv e-prints, \href
  {https://ui.adsabs.harvard.edu/abs/2020arXiv201203380L} {p. arXiv:2012.03380}

\bibitem[\protect\citeauthoryear{{Mar{\'\i}n-Franch}
  et~al.,}{{Mar{\'\i}n-Franch} et~al.}{2009}]{MarinFranch+09}
{Mar{\'\i}n-Franch} A.,  et~al., 2009, \mn@doi [\apj]
  {10.1088/0004-637X/694/2/1498}, \href
  {https://ui.adsabs.harvard.edu/abs/2009ApJ...694.1498M} {694, 1498}

\bibitem[\protect\citeauthoryear{{McConnachie} \& {Venn}}{{McConnachie} \&
  {Venn}}{2020a}]{McConnachie&Venn20}
{McConnachie} A.~W.,  {Venn} K.~A.,  2020a, \mn@doi [Research Notes of the
  American Astronomical Society] {10.3847/2515-5172/abd18b}, \href
  {https://ui.adsabs.harvard.edu/abs/2020RNAAS...4..229M} {4, 229}

\bibitem[\protect\citeauthoryear{{McConnachie} \& {Venn}}{{McConnachie} \&
  {Venn}}{2020b}]{McConnachie&Venn20a}
{McConnachie} A.~W.,  {Venn} K.~A.,  2020b, \mn@doi [\aj]
  {10.3847/1538-3881/aba4ab}, \href
  {https://ui.adsabs.harvard.edu/abs/2020AJ....160..124M} {160, 124}

\bibitem[\protect\citeauthoryear{{Miki} \& {Umemura}}{{Miki} \&
  {Umemura}}{2018}]{Miki&Umemura&Umemura18}
{Miki} Y.,  {Umemura} M.,  2018, \mn@doi [\mnras] {10.1093/mnras/stx3327},
  \href {https://ui.adsabs.harvard.edu/abs/2018MNRAS.475.2269M} {475, 2269}

\bibitem[\protect\citeauthoryear{{Noyola}, {Gebhardt}  \& {Bergmann}}{{Noyola}
  et~al.}{2008}]{Noyola+08}
{Noyola} E.,  {Gebhardt} K.,   {Bergmann} M.,  2008, \mn@doi [\apj]
  {10.1086/529002}, \href
  {https://ui.adsabs.harvard.edu/abs/2008ApJ...676.1008N} {676, 1008}

\bibitem[\protect\citeauthoryear{{Pace} \& {Li}}{{Pace} \&
  {Li}}{2019}]{Pace&Li19}
{Pace} A.~B.,  {Li} T.~S.,  2019, \mn@doi [\apj] {10.3847/1538-4357/ab0aee},
  \href {https://ui.adsabs.harvard.edu/abs/2019ApJ...875...77P} {875, 77}

\bibitem[\protect\citeauthoryear{{Pe{\~n}arrubia}, {Walker}  \&
  {Gilmore}}{{Pe{\~n}arrubia} et~al.}{2009}]{Penarrubia+09}
{Pe{\~n}arrubia} J.,  {Walker} M.~G.,   {Gilmore} G.,  2009, \mn@doi [\mnras]
  {10.1111/j.1365-2966.2009.15027.x}, \href
  {https://ui.adsabs.harvard.edu/abs/2009MNRAS.399.1275P} {399, 1275}

\bibitem[\protect\citeauthoryear{{Pearson}}{{Pearson}}{1916}]{Pearson16}
{Pearson} K.,  1916, \mn@doi [Philosophical Transactions of the Royal Society
  of London Series A] {10.1098/rsta.1916.0009}, \href
  {https://ui.adsabs.harvard.edu/abs/1916RSPTA.216..429P} {216, 429}

\bibitem[\protect\citeauthoryear{{Peebles}}{{Peebles}}{1984}]{Peebles84}
{Peebles} P.~J.~E.,  1984, \mn@doi [\apj] {10.1086/161714}, \href
  {https://ui.adsabs.harvard.edu/abs/1984ApJ...277..470P} {277, 470}

\bibitem[\protect\citeauthoryear{{Peebles} \& {Dicke}}{{Peebles} \&
  {Dicke}}{1968}]{Peebles&Dicke68}
{Peebles} P.~J.~E.,  {Dicke} R.~H.,  1968, \mn@doi [\apj] {10.1086/149811},
  \href {https://ui.adsabs.harvard.edu/abs/1968ApJ...154..891P} {154, 891}

\bibitem[\protect\citeauthoryear{{Plummer}}{{Plummer}}{1911}]{Plummer1911}
{Plummer} H.~C.,  1911, \mnras, 71, 460

\bibitem[\protect\citeauthoryear{{Riello} et~al.,}{{Riello}
  et~al.}{2020}]{Riello+20}
{Riello} M.,  et~al., 2020, arXiv e-prints, \href
  {https://ui.adsabs.harvard.edu/abs/2020arXiv201201916R} {p. arXiv:2012.01916}

\bibitem[\protect\citeauthoryear{{Searle} \& {Zinn}}{{Searle} \&
  {Zinn}}{1978}]{Searle&Zinn78}
{Searle} L.,  {Zinn} R.,  1978, \mn@doi [\apj] {10.1086/156499}, \href
  {https://ui.adsabs.harvard.edu/abs/1978ApJ...225..357S} {225, 357}

\bibitem[\protect\citeauthoryear{{S\'ersic}}{{S\'ersic}}{1963}]{Sersic63}
{S\'ersic} J.~L.,  1963, Bull. Assoc. Argentina de Astron., 6, 41

\bibitem[\protect\citeauthoryear{{Sersic}}{{Sersic}}{1968}]{Sersic68}
{Sersic} J.~L.,  1968, Atlas de galaxias australes.
Cordoba, Argentina: Observatorio Astronomico

\bibitem[\protect\citeauthoryear{{Silverman}}{{Silverman}}{1986}]{Silverman86}
{Silverman} B.~W.,  1986, Density estimation for statistics and data analysis

\bibitem[\protect\citeauthoryear{{Sugiura}}{{Sugiura}}{1978}]{Sugiyara78}
{Sugiura} N.,  1978, \mn@doi [Communications in Statistics - Theory and
  Methods] {10.1080/03610927808827599}, 7, 13

\bibitem[\protect\citeauthoryear{{Vasilevskis}, {Sanders}  \& {van
  Altena}}{{Vasilevskis} et~al.}{1965}]{Vasilevskis+65}
{Vasilevskis} S.,  {Sanders} W.~L.,   {van Altena} W.~F.,  1965, \mn@doi [\aj]
  {10.1086/109821}, \href
  {https://ui.adsabs.harvard.edu/abs/1965AJ.....70..806V} {70, 806}

\bibitem[\protect\citeauthoryear{{Vasiliev}}{{Vasiliev}}{2019a}]{Vasiliev19b}
{Vasiliev} E.,  2019a, \mn@doi [\mnras] {10.1093/mnras/stz171}, \href
  {https://ui.adsabs.harvard.edu/abs/2019MNRAS.484.2832V} {484, 2832}

\bibitem[\protect\citeauthoryear{{Vasiliev}}{{Vasiliev}}{2019b}]{Vasiliev19c}
{Vasiliev} E.,  2019b, \mn@doi [\mnras] {10.1093/mnras/stz2100}, \href
  {https://ui.adsabs.harvard.edu/abs/2019MNRAS.489..623V} {489, 623}

\bibitem[\protect\citeauthoryear{{Vasiliev} \& {Baumgardt}}{{Vasiliev} \&
  {Baumgardt}}{2021}]{Vasiliev&Baumgardt&Baumgardt21}
{Vasiliev} E.,  {Baumgardt} H.,  2021, arXiv e-prints, \href
  {https://ui.adsabs.harvard.edu/abs/2021arXiv210209568V} {p. arXiv:2102.09568}

\bibitem[\protect\citeauthoryear{{Vitral} \& {Mamon}}{{Vitral} \&
  {Mamon}}{2020}]{Vitral&Mamon20}
{Vitral} E.,  {Mamon} G.~A.,  2020, \mn@doi [\aap]
  {10.1051/0004-6361/201937202}, \href
  {https://ui.adsabs.harvard.edu/abs/2020A&A...635A..20V} {635, A20}

\bibitem[\protect\citeauthoryear{{Vitral} \& {Mamon}}{{Vitral} \&
  {Mamon}}{2021}]{Vitral&Mamon21}
{Vitral} E.,  {Mamon} G.~A.,  2021, \mn@doi [\aap]
  {10.1051/0004-6361/202039650}, \href
  {https://ui.adsabs.harvard.edu/abs/2021A&A...646A..63V} {646, A63}

\bibitem[\protect\citeauthoryear{{Walker}}{{Walker}}{2013}]{Walker13}
{Walker} M.,  2013, Dark Matter in the Galactic Dwarf Spheroidal Satellites.
p.~1039, \mn@doi{10.1007/978-94-007-5612-0_20}

\bibitem[\protect\citeauthoryear{{van der Marel} \& {Anderson}}{{van der Marel}
  \& {Anderson}}{2010}]{vanderMarel&Anderson&Anderson10}
{van der Marel} R.~P.,  {Anderson} J.,  2010, \mn@doi [\apj]
  {10.1088/0004-637X/710/2/1063}, \href
  {https://ui.adsabs.harvard.edu/abs/2010ApJ...710.1063V} {710, 1063}

\bibitem[\protect\citeauthoryear{{van der Walt}, {Colbert}  \&
  {Varoquaux}}{{van der Walt} et~al.}{2011}]{vanderWalt11}
{van der Walt} S.,  {Colbert} S.~C.,   {Varoquaux} G.,  2011, Computing in
  Science Engineering, 13, 22

\makeatother
\end{thebibliography}




\appendix

\section{Projected number}
\label{app: general-n}

We present here the solution of equation~\ref{eq: plummer-n}, for a distribution following the Plummer \citep{Plummer1911} profile, in the approximation of small cone apertures in the sky (i.e. $R_{\rm max} \ll 1$ radian, where $R_{\rm max}$ and $d$ can be seen in Figure~\ref{fig: geometry}). For the case where $R \leq R_{\rm max} - d$ ($d$ also defined in Figure~\ref{fig: geometry}), one has equation~\ref{eq: plummer-n-simple}.

However, whenever $R > R_{\rm max} - d$, the indefinite integral of equation~\ref{eq: plummer-n} yields:

\begin{eqnarray}
    \widetilde{N}(x) &\!\!\!\!=\!\!\!\!& - \frac{a^2 \, \arccos{\left[\frac{R^2 + d^2 - R_{\rm max}^2}{2 \, R \, d}\right]}}{\pi (a^2 + R^2)} + 
    \\
    &\!\!\!\!\mbox{}\!\!\!\!& \Bigg( \sqrt{\Xi_1} \, \Bigg( - \sqrt{\Xi_2} \ \times  \nonumber
    \\
    &\!\!\!\!\mbox{}\!\!\!\!& \arctan{\left[ \frac{(d^2 - R_{\rm max}^2)^2 - R^2 \, (d^2 + R_{\rm max}^2)}{(d^2 - R_{\rm max}^2) \, \sqrt{\Xi_1} } \right]} + \nonumber
    \\
    &\!\!\!\!\mbox{}\!\!\!\!& (a^2 + d^2 - R_{\rm max}^2) \, \arctan{\left[ \frac{\Xi_3}{\sqrt{\Xi_2} \, \sqrt{\Xi_1}} \right]} \Bigg) \Bigg) \ \div \nonumber
    \\
    &\!\!\!\!\mbox{}\!\!\!\!&  \left( 2 \, \pi \, \sqrt{\Xi_2} \, \sqrt{\Xi_1} \right) + \mathcal{C} \ . \nonumber
    \label{eq: general-n-plu}
\end{eqnarray}
where $\mathcal{C} = 1$ is an integration constant, $a$ is the Plummer effective radius, and $\Xi_1$, $\Xi_2$ and $\Xi_3$ are defined as:

\begin{subequations}
\begin{equation}
    \Xi_1 = - d^4 - (R^2 - R_{\rm max}^2)^2 + 2 \, d^2 \, (R^2 + R_{\rm max}^2)
\end{equation}
\begin{equation}
    \Xi_2 = a^4 + (d^2 - R_{\rm max}^2)^2 + 2 \, a^2 \, (d^2 + R_{\rm max}^2)
\end{equation}
\begin{equation}
    \Xi_3 = (d^2 - R_{\rm max}^2)^2 + a^2 \, (d^2 + R_{\rm max}^2) - R^2 \, (a^2 + d^2 + R_{\rm max}^2)
\end{equation}
\end{subequations}

This treatment can be chosen in the \textsc{BALRoGO} method \textsc{position.find\_center()}, by providing the argument \url{method="mle\_robust"}.

\section{Field stars mock data}
\label{app: random-fs}

In this section, we derive equations~(\ref{eq: random-positions}) and (\ref{eq: random-pms}). 

\subsection{Random positions}

\begin{figure}
\centering
\includegraphics[width=0.8\hsize]{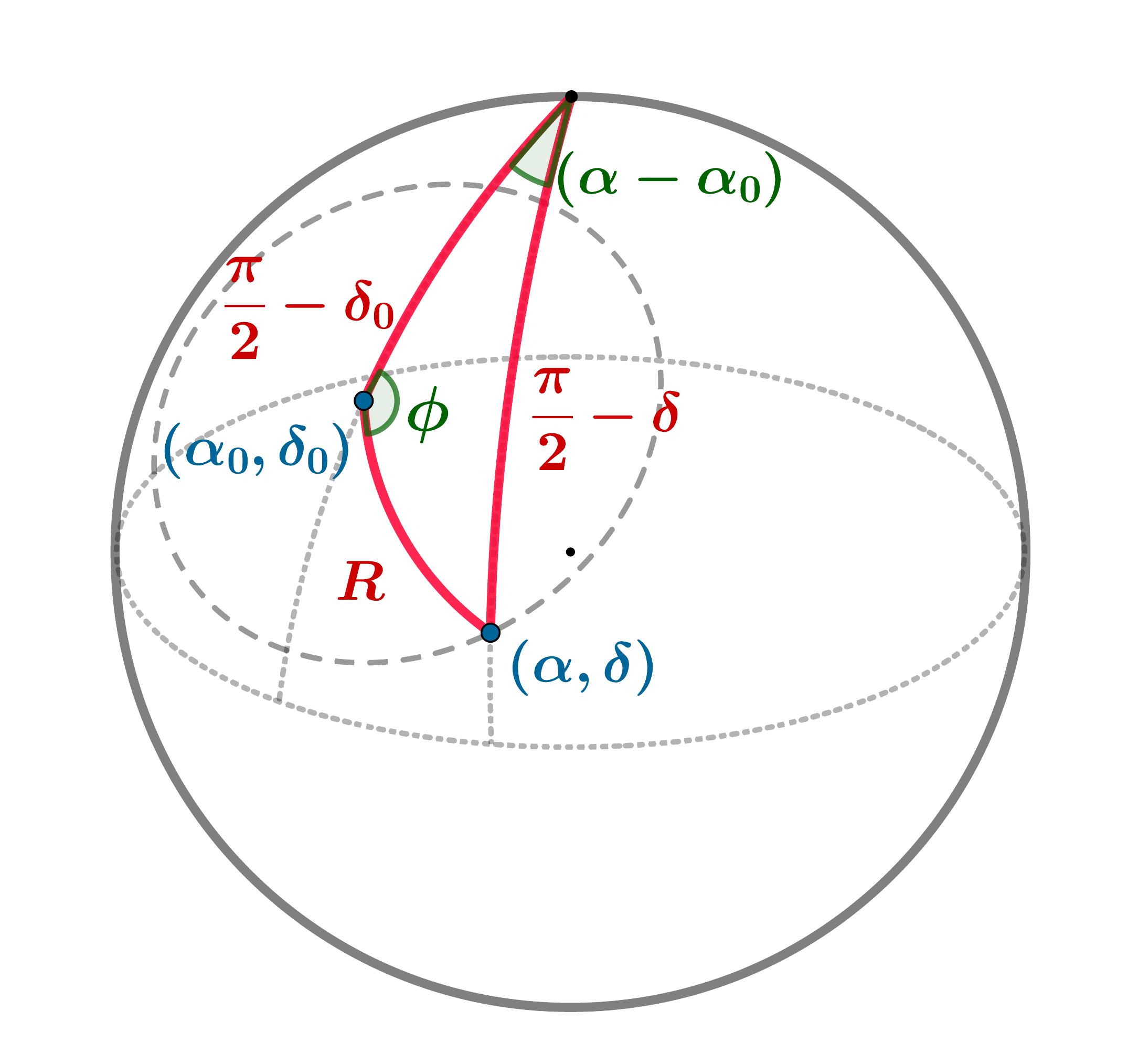}
\caption{{\it Spherical geometry:} Representation of the physical situation of a source projected in the plane of sky.}
\label{fig: sph-trig}
\end{figure}

In order to generate $n$ points uniformly distributed in a spherical cap, we shall first considerate Figure~\ref{fig: sph-trig}, with the geometry of the problem. From classical spherical trigonometry relations, it is straightforward to write:

\begin{subequations} \label{eq: sph-trig}
\begin{equation}
    \cos{R} = \sin{\delta} \sin{\delta_{0}} + \cos{\delta} \cos{\delta_{0}} \cos{(\alpha - \alpha_{0})} \ ,
\end{equation}
\begin{equation}
    \sin{\phi} = \displaystyle{\frac{\cos{\delta} \sin{(\alpha - \alpha_{0})}}{\sin{R}}} \ ,
\end{equation}
\begin{equation} 
    \sin{\delta} = \cos{R} \sin{\delta_{0}} + \sin{R} \cos{\delta_{0}} \cos{\phi} \ .
    \label{eq: sindelta}
\end{equation}
\end{subequations}

We wish to generate an uniform distribution of points in a spherical cap of radius $R_{\mathrm{lim}}$, so the probability of having a radius smaller than $R$ can be written as:

\begin{equation}
    \Pr \{r < R\} = \displaystyle{\frac{\text{Surface}(R)}{\text{Surface}(R_{\mathrm{lim}})}} \ ,
\end{equation} 
or, more precisely:

\begin{equation}
    \Pr \{r < R\} = \displaystyle{\frac{\int_{0}^{R} \int_{0}^{2 \pi} \rho^{2} \sin{\theta} \diff \theta \diff \varphi}{\int_{0}^{R_{\mathrm{lim}}} \int_{0}^{2 \pi} \rho^{2} \sin{\theta} \diff \theta \diff \varphi}} = \displaystyle{\frac{1-\cos{R}}{1-\cos{R_{\mathrm{lim}}}}} \ ,
    \label{eq: prob-r}
\end{equation}
where $\rho$ is the radius of the sphere, $\theta$ and $\varphi$ are the longitudinal and latitudinal angles of the sphere, that in our description have their origin at $(\alpha_{0},\delta_{0})$. Similarly, the probability that $\phi$ is smaller than an angle $\Theta$ is:

\begin{equation}
    \Pr \{\phi < \Theta\} = \Theta / 2 \pi \ .
    \label{eq: prob-theta}
\end{equation}

Therefore, in order to derive equations~(\ref{eq: random-positions}), one just needs to invert the relations~(\ref{eq: prob-r}) and (\ref{eq: prob-theta}), with respect to $R$ and $\Theta$, respectively. The following step, which is to convert the set of known $\alpha_{0}$, $\delta_{0}$, $R$ and $\phi$ into pairs of ($\alpha$, $\delta$) is done by first deriving $\delta$ with equation~(\ref{eq: sindelta}) and then $\alpha$ with:

\begin{equation} \label{eq: alphatrig}
    \alpha =
    \begin{cases}
       \alpha_{0} + \arccos{\left( \Lambda_{1} \right)} & \text{ , for } \Lambda_{2} > 0 \text{ and } 0 \leq \phi \leq \pi \\ \\
       
       \alpha_{0} + \arccos{\left( - \Lambda_{1} \right)}  & \text{ , for } \Lambda_{2} < 0 \text{ and } 0 \leq \phi \leq \pi \\
       
       \alpha_{0} - \arccos{\left( \Lambda_{1} \right)} & \text{ , for } \Lambda_{2} > 0 \text{ and } \pi < \phi < 2 \pi \\ \\
       
       \alpha_{0} -\arccos{\left( - \Lambda_{1} \right)}  & \text{ , for } \Lambda_{2} < 0 \text{ and } \pi < \phi < 2 \pi \\

    \end{cases}
\end{equation}
where we have the following correspondences:

\begin{subequations}
\begin{equation}
    \Lambda_{1} = \sqrt{1 - \displaystyle{\frac{\sin^{2}{\phi} \, \sin^{2}{R}}{1 - \sin^{2}{\delta}}}}
\end{equation}
\begin{equation}
    \Lambda_{2} = \displaystyle{\frac{\cos{R}-\sin{\delta} \sin{\delta_{0}}}{\cos{\delta} \cos{\delta_{0}}}}
\end{equation}
\end{subequations}

\subsection{Random proper motions}

For the field stars, we generated random variables that followed a symmetric Pearson VII distribution. The statistical approach to do so was to invert the probability that a field star proper motion modulus is smaller than M, i.e. the cumulative distribution function of M for a symmetric Pearson VII distribution of scale radius $a$ and characteristic slope $\tau$:

\begin{eqnarray}
    \mathrm{CDF}(\mathrm{M}) &\!\!\!\!\equiv\!\!\!\!& \int_{0}^{\mathrm{M}} f_{\mathrm{PM}}(\mu) \diff \mu \nonumber
    \\
    &\!\!\!\!=\!\!\!\!& - \int_{0}^{\mathrm{M}} \displaystyle{\frac{\tau+2}{a} \, \frac{\mu}{a} \, \left[ 1+ \left(\frac{\mu}{a} \right)^{2} \right]^{\tau/2}} \diff \mu \ .
    \label{eq: cdf-pm}
\end{eqnarray}
where $f_{\mathrm{PM}}$ is the distribution function of proper motions moduli for the Pearson VII symmetric distribution. The result of the integral above is:

\begin{equation}
    \mathrm{CDF}(\mathrm{M})  = 1 - \left[ 1+(\mathrm{M}/a)^{2} \right]^{1+\tau/2} \ .
\end{equation}

Thus, if U is a uniform random variable with boundaries [0, 1] (and thus, $1 - \mathrm{U} \equiv \mathrm{U}$, random PM variables can be generated with equation~(\ref{eq: random-pms}).

\section{Structural parameters of globular clusters and dwarf spheroidal galaxies}

We present the main results of our paper in Table~\ref{tab: results}, such as effective radii and bulk proper motions for over a hundred globular clusters and some of the main Local Group dwarf spheroidal galaxies. This table can be accessed in text format at the footnote link\footnote{\url{https://gitlab.com/eduardo-vitral/balrogo/-/raw/master/table\_gc\_dsph.dat}}.

\begin{table*}
\caption{Catalog of proper motions and other dynamical parameters of NGC globular clusters and Local Group dwarf spheroidal galaxies.}
\label{tab: results}
\centering
\renewcommand{\arraystretch}{1.2}
\tabcolsep=4.5pt
\begin{tabular}{llrrrrrrrcrr}
\hline\hline             
\multicolumn{1}{c}{Name} &
\multicolumn{1}{c}{Other ID} &
\multicolumn{1}{c}{$\alpha$} & 
\multicolumn{1}{c}{$\delta$} &
\multicolumn{1}{c}{$D$} &
\multicolumn{1}{c}{$\mu_{\alpha,*}$} & 
\multicolumn{1}{c}{$\mu_{\delta}$} & 
\multicolumn{1}{c}{$\epsilon_{\mu_{\alpha,*}}$} &
\multicolumn{1}{c}{$\epsilon_{\mu_{\delta}}$} &
\multicolumn{1}{c}{$\Sigma$} &
\multicolumn{1}{c}{$R_{1/2}$} &
\multicolumn{1}{c}{$\epsilon_{R_{1/2}}$} \\
\multicolumn{1}{c}{} &
\multicolumn{1}{c}{} &
\multicolumn{1}{c}{[deg]} &
\multicolumn{1}{c}{[deg]} &
\multicolumn{1}{c}{[kpc]} &
\multicolumn{1}{c}{[mas yr$^{-1}$]} &
\multicolumn{1}{c}{[mas yr$^{-1}$]} &
\multicolumn{1}{c}{[mas yr$^{-1}$]} &
\multicolumn{1}{c}{[mas yr$^{-1}$]} &
\multicolumn{1}{c}{} & 
\multicolumn{1}{c}{[arcmin]} &
\multicolumn{1}{c}{[arcmin]} \\ 
\multicolumn{1}{c}{(1)} &
\multicolumn{1}{c}{(2)} &
\multicolumn{1}{c}{(3)} &
\multicolumn{1}{c}{(4)} &
\multicolumn{1}{c}{(5)} &
\multicolumn{1}{c}{(6)} &
\multicolumn{1}{c}{(7)} &
\multicolumn{1}{c}{(8)} &
\multicolumn{1}{c}{(9)} &
\multicolumn{1}{c}{(10)} &
\multicolumn{1}{c}{(11)} &
\multicolumn{1}{c}{(12)} \\ 
\hline
NGC 104 & 47 Tuc & $  6.02242 $ & $ -72.08147 $ &  4.43 & $  5.250 $ & $ -2.565 $ & $  0.003 $ & $  0.003 $ & S & $  7.84 $ & $  0.02 $ \\
NGC 1261 & \multicolumn{1}{c}{--} & $  48.06694 $ & $ -55.21592 $ &  17.20 & $  1.602 $ & $ -2.065 $ & $  0.005 $ & $  0.005 $ & P & $  1.38 $ & $  0.01 $ \\
NGC 1851 & \multicolumn{1}{c}{--} & $  78.52821 $ & $ -40.04675 $ &  11.33 & $  2.140 $ & $ -0.664 $ & $  0.004 $ & $  0.005 $ & P & $  2.08 $ & $  0.01 $ \\
NGC 1904 & M 79 & $  81.04414 $ & $ -24.52423 $ &  13.27 & $  2.475 $ & $ -1.591 $ & $  0.005 $ & $  0.005 $ & P & $  1.56 $ & $  0.01 $ \\
NGC 2298 & \multicolumn{1}{c}{--} & $  102.24756 $ & $ -36.00532 $ &  10.80 & $  3.311 $ & $ -2.172 $ & $  0.006 $ & $  0.006 $ & S & $  1.18 $ & $  0.02 $ \\
NGC 2419 & \multicolumn{1}{c}{--} & $  114.53541 $ & $  38.88193 $ &  83.18 & $ -0.004 $ & $ -0.526 $ & $  0.011 $ & $  0.011 $ & S & $  0.91 $ & $  0.02 $ \\
NGC 2808 & \multicolumn{1}{c}{--} & $  138.01293 $ & $ -64.86349 $ &  10.21 & $  1.000 $ & $  0.276 $ & $  0.005 $ & $  0.005 $ & S & $  3.11 $ & $  0.01 $ \\
NGC 288 & \multicolumn{1}{c}{--} & $  13.18819 $ & $ -26.58238 $ &  9.00 & $  4.156 $ & $ -5.706 $ & $  0.004 $ & $  0.004 $ & S & $  2.57 $ & $  0.02 $ \\
NGC 3201 & \multicolumn{1}{c}{--} & $  154.40393 $ & $ -46.41249 $ &  4.60 & $  8.348 $ & $ -1.966 $ & $  0.002 $ & $  0.002 $ & S & $  4.50 $ & $  0.03 $ \\
NGC 362 & \multicolumn{1}{c}{--} & $  15.80943 $ & $ -70.84885 $ &  9.17 & $  6.683 $ & $ -2.543 $ & $  0.010 $ & $  0.010 $ & S & $  2.49 $ & $  0.01 $ \\
NGC 4147 & \multicolumn{1}{c}{--} & $  182.51853 $ & $  18.54192 $ &  18.20 & $ -1.716 $ & $ -2.103 $ & $  0.010 $ & $  0.010 $ & P & $  0.86 $ & $  0.02 $ \\
NGC 4372 & \multicolumn{1}{c}{--} & $  186.43919 $ & $ -72.65907 $ &  5.76 & $ -6.410 $ & $  3.298 $ & $  0.003 $ & $  0.003 $ & S & $  4.00 $ & $  0.03 $ \\
NGC 4590 & M 68 & $  189.86660 $ & $ -26.74405 $ &  10.13 & $ -2.736 $ & $  1.777 $ & $  0.004 $ & $  0.004 $ & P & $  2.11 $ & $  0.02 $ \\
NGC 4833 & \multicolumn{1}{c}{--} & $  194.89122 $ & $ -70.87650 $ &  6.56 & $ -8.377 $ & $ -0.962 $ & $  0.006 $ & $  0.006 $ & S & $  2.48 $ & $  0.02 $ \\
NGC 5024 & M 53 & $  198.23025 $ & $  18.16819 $ &  17.90 & $ -0.142 $ & $ -1.332 $ & $  0.004 $ & $  0.004 $ & S & $  2.47 $ & $  0.02 $ \\
NGC 5053 & \multicolumn{1}{c}{--} & $  199.11287 $ & $  17.69872 $ &  17.20 & $ -0.331 $ & $ -1.217 $ & $  0.004 $ & $  0.005 $ & S & $  2.29 $ & $  0.03 $ \\
NGC 5139 & $\omega$ Cen & $  201.69698 $ & $ -47.47950 $ &  5.24 & $ -3.253 $ & $ -6.757 $ & $  0.003 $ & $  0.003 $ & S & $  11.25 $ & $  0.03 $ \\
NGC 5272 & M 3 & $  205.54873 $ & $  28.37727 $ &  9.59 & $ -0.153 $ & $ -2.666 $ & $  0.004 $ & $  0.003 $ & P & $  3.85 $ & $  0.02 $ \\
NGC 5286 & \multicolumn{1}{c}{--} & $  206.61171 $ & $ -51.37425 $ &  11.45 & $  0.188 $ & $ -0.157 $ & $  0.006 $ & $  0.006 $ & S & $  1.90 $ & $  0.01 $ \\
NGC 5466 & \multicolumn{1}{c}{--} & $  211.36366 $ & $  28.53444 $ &  16.00 & $ -5.359 $ & $ -0.843 $ & $  0.008 $ & $  0.009 $ & P & $  2.30 $ & $  0.03 $ \\
NGC 5634 & \multicolumn{1}{c}{--} & $  217.40528 $ & $ -5.97638 $ &  27.20 & $ -1.689 $ & $ -1.473 $ & $  0.008 $ & $  0.007 $ & P & $  0.83 $ & $  0.02 $ \\
NGC 5694 & \multicolumn{1}{c}{--} & $  219.90217 $ & $ -26.53835 $ &  37.33 & $ -0.481 $ & $ -1.107 $ & $  0.033 $ & $  0.028 $ & P & $  0.63 $ & $  0.02 $ \\
NGC 5824 & \multicolumn{1}{c}{--} & $  225.99421 $ & $ -33.06854 $ &  30.90 & $ -1.204 $ & $ -2.228 $ & $  0.006 $ & $  0.006 $ & S & $  1.11 $ & $  0.04 $ \\
NGC 5897 & \multicolumn{1}{c}{--} & $  229.35163 $ & $ -21.01013 $ &  12.60 & $ -5.413 $ & $ -3.390 $ & $  0.006 $ & $  0.006 $ & S & $  2.29 $ & $  0.02 $ \\
NGC 5904 & M 5 & $  229.63841 $ & $  2.08097 $ &  7.57 & $  4.073 $ & $ -9.869 $ & $  0.004 $ & $  0.004 $ & S & $  4.26 $ & $  0.02 $ \\
NGC 5927 & \multicolumn{1}{c}{--} & $  232.00284 $ & $ -50.67305 $ &  9.08 & $ -5.051 $ & $ -3.214 $ & $  0.007 $ & $  0.007 $ & S & $  3.13 $ & $  0.06 $ \\
NGC 5946 & \multicolumn{1}{c}{--} & $  233.86904 $ & $ -50.65973 $ &  10.60 & $ -5.317 $ & $ -1.646 $ & $  0.012 $ & $  0.012 $ & S & $  1.05 $ & $  0.03 $ \\
NGC 5986 & \multicolumn{1}{c}{--} & $  236.51248 $ & $ -37.78644 $ &  10.56 & $ -4.193 $ & $ -4.555 $ & $  0.008 $ & $  0.008 $ & S & $  1.86 $ & $  0.01 $ \\
NGC 6093 & M 80 & $  244.26003 $ & $ -22.97611 $ &  8.86 & $ -2.930 $ & $ -5.588 $ & $  0.009 $ & $  0.009 $ & P & $  1.80 $ & $  0.02 $ \\
NGC 6101 & \multicolumn{1}{c}{--} & $  246.45171 $ & $ -72.20161 $ &  12.80 & $  1.761 $ & $ -0.258 $ & $  0.003 $ & $  0.003 $ & P & $  2.27 $ & $  0.03 $ \\
NGC 6121 & M 4 & $  245.89669 $ & $ -26.52584 $ &  1.93 & $ -12.515 $ & $ -19.011 $ & $  0.004 $ & $  0.004 $ & S & $  6.17 $ & $  0.09 $ \\
NGC 6139 & \multicolumn{1}{c}{--} & $  246.91658 $ & $ -38.84919 $ &  9.80 & $ -6.073 $ & $ -2.701 $ & $  0.010 $ & $  0.010 $ & S & $  1.40 $ & $  0.05 $ \\
NGC 6144 & \multicolumn{1}{c}{--} & $  246.80774 $ & $ -26.02352 $ &  8.90 & $ -1.746 $ & $ -2.614 $ & $  0.009 $ & $  0.009 $ & P & $  1.63 $ & $  0.02 $ \\
NGC 6171 & M 107 & $  248.13274 $ & $ -13.05381 $ &  5.70 & $ -1.945 $ & $ -5.973 $ & $  0.006 $ & $  0.005 $ & S & $  2.09 $ & $  0.02 $ \\
NGC 6205 & M 13 & $  250.42348 $ & $  36.46129 $ &  6.77 & $ -3.137 $ & $ -2.566 $ & $  0.003 $ & $  0.003 $ & S & $  3.87 $ & $  0.01 $ \\
NGC 6218 & M 12 & $  251.80908 $ & $ -1.94856 $ &  4.67 & $ -0.201 $ & $ -6.803 $ & $  0.004 $ & $  0.004 $ & S & $  2.88 $ & $  0.01 $ \\
NGC 6229 & \multicolumn{1}{c}{--} & $  251.74419 $ & $  47.52642 $ &  30.62 & $ -1.173 $ & $ -0.456 $ & $  0.009 $ & $  0.009 $ & P & $  0.71 $ & $  0.01 $ \\
NGC 6235 & \multicolumn{1}{c}{--} & $  253.35565 $ & $ -22.17748 $ &  13.52 & $ -3.942 $ & $ -7.590 $ & $  0.012 $ & $  0.012 $ & S & $  1.03 $ & $  0.02 $ \\
NGC 6254 & M 10 & $  254.28768 $ & $ -4.10033 $ &  4.96 & $ -4.760 $ & $ -6.609 $ & $  0.008 $ & $  0.008 $ & P & $  3.60 $ & $  0.02 $ \\
NGC 6256 & \multicolumn{1}{c}{--} & $  254.88617 $ & $ -37.12135 $ &  6.40 & $ -3.714 $ & $ -1.635 $ & $  0.018 $ & $  0.017 $ & K & $  1.17 $ & $  0.04 $ \\
NGC 6266 & M 62 & $  255.30247 $ & $ -30.11237 $ &  6.41 & $ -4.982 $ & $ -2.962 $ & $  0.011 $ & $  0.011 $ & S & $  3.06 $ & $  0.04 $ \\
NGC 6273 & M 19 & $  255.65702 $ & $ -26.26793 $ &  8.27 & $ -3.249 $ & $  1.656 $ & $  0.008 $ & $  0.008 $ & S & $  2.62 $ & $  0.02 $ \\
NGC 6284 & \multicolumn{1}{c}{--} & $  256.11976 $ & $ -24.76424 $ &  15.14 & $ -3.206 $ & $ -2.016 $ & $  0.011 $ & $  0.011 $ & S & $  1.00 $ & $  0.03 $ \\
NGC 6287 & \multicolumn{1}{c}{--} & $  256.28936 $ & $ -22.70776 $ &  9.40 & $ -5.002 $ & $ -1.875 $ & $  0.010 $ & $  0.009 $ & P & $  1.16 $ & $  0.02 $ \\
NGC 6293 & \multicolumn{1}{c}{--} & $  257.54342 $ & $ -26.58174 $ &  8.70 & $  0.878 $ & $ -4.322 $ & $  0.012 $ & $  0.011 $ & P & $  1.51 $ & $  0.03 $ \\
\hline
\end{tabular}
\parbox{\hsize}{\textit{Notes}: Columns are (1): Source name; (2): Alternative ID; (3): right ascension derived according to section~\ref{ssec: center}; (4): declination derived according to section~\ref{ssec: center}; (5): Distances in kpc from \protect\cite{Baumgardt+19}; (6): Bulk proper motion in right ascension (i.e., $[\diff \alpha / \diff t] \, \cos{\delta}$, in mas yr$^{-1}$); (7): Bulk proper motion in declination (i.e., $\diff \delta / \diff t$, in mas yr$^{-1}$); (8): uncertainty of $\mu_{\alpha,*}$, in mas yr$^{-1}$; (9): uncertainty of $\mu_{\delta}$, in mas yr$^{-1}$; (10): Surface density model preferred by AICc (see section~\ref{ssec: sd}). `P' stands for Plummer, `S' for S\'ersic and `K' for Kazantzidis; (11): Effective (two-dimensional) radius, in arcmin; (12): uncertainty on the effective radius, in arcmin.}
\end{table*}

\begin{table*}
\contcaption{}
\centering
\renewcommand{\arraystretch}{1.2}
\tabcolsep=4.5pt
\begin{tabular}{llrrrrrrrcrr}
\hline\hline             
\multicolumn{1}{c}{Name} &
\multicolumn{1}{c}{Other ID} &
\multicolumn{1}{c}{$\alpha$} & 
\multicolumn{1}{c}{$\delta$} &
\multicolumn{1}{c}{$D$} &
\multicolumn{1}{c}{$\mu_{\alpha,*}$} & 
\multicolumn{1}{c}{$\mu_{\delta}$} & 
\multicolumn{1}{c}{$\epsilon_{\mu_{\alpha,*}}$} &
\multicolumn{1}{c}{$\epsilon_{\mu_{\delta}}$} &
\multicolumn{1}{c}{$\Sigma$} &
\multicolumn{1}{c}{$R_{1/2}$} &
\multicolumn{1}{c}{$\epsilon_{R_{1/2}}$} \\
\multicolumn{1}{c}{} &
\multicolumn{1}{c}{} &
\multicolumn{1}{c}{[deg]} &
\multicolumn{1}{c}{[deg]} &
\multicolumn{1}{c}{[kpc]} &
\multicolumn{1}{c}{[mas yr$^{-1}$]} &
\multicolumn{1}{c}{[mas yr$^{-1}$]} &
\multicolumn{1}{c}{[mas yr$^{-1}$]} &
\multicolumn{1}{c}{[mas yr$^{-1}$]} &
\multicolumn{1}{c}{} & 
\multicolumn{1}{c}{[arcmin]} &
\multicolumn{1}{c}{[arcmin]} \\ 
\multicolumn{1}{c}{(1)} &
\multicolumn{1}{c}{(2)} &
\multicolumn{1}{c}{(3)} &
\multicolumn{1}{c}{(4)} &
\multicolumn{1}{c}{(5)} &
\multicolumn{1}{c}{(6)} &
\multicolumn{1}{c}{(7)} &
\multicolumn{1}{c}{(8)} &
\multicolumn{1}{c}{(9)} &
\multicolumn{1}{c}{(10)} &
\multicolumn{1}{c}{(11)} &
\multicolumn{1}{c}{(12)} \\ 
\hline
NGC 6304 & \multicolumn{1}{c}{--} & $  258.63435 $ & $ -29.46203 $ &  5.77 & $ -4.083 $ & $ -1.092 $ & $  0.017 $ & $  0.016 $ & S & $  0.95 $ & $  0.07 $ \\
NGC 6316 & \multicolumn{1}{c}{--} & $  259.15589 $ & $ -28.14002 $ &  11.60 & $ -4.975 $ & $ -4.611 $ & $  0.016 $ & $  0.016 $ & P & $  0.96 $ & $  0.08 $ \\
NGC 6325 & \multicolumn{1}{c}{--} & $  259.49692 $ & $ -23.76607 $ &  7.80 & $ -8.295 $ & $ -9.008 $ & $  0.013 $ & $  0.012 $ & S & $  0.78 $ & $  0.03 $ \\
NGC 6333 & M 9 & $  259.79907 $ & $ -18.51627 $ &  8.40 & $ -2.174 $ & $ -3.223 $ & $  0.010 $ & $  0.010 $ & S & $  2.30 $ & $  0.03 $ \\
NGC 6341 & M 92 & $  259.28081 $ & $  43.13600 $ &  8.44 & $ -4.935 $ & $ -0.627 $ & $  0.004 $ & $  0.004 $ & S & $  2.66 $ & $  0.01 $ \\
NGC 6342 & \multicolumn{1}{c}{--} & $  260.29224 $ & $ -19.58745 $ &  8.43 & $ -2.904 $ & $ -7.122 $ & $  0.010 $ & $  0.010 $ & S & $  2.16 $ & $  0.16 $ \\
NGC 6352 & \multicolumn{1}{c}{--} & $  261.37069 $ & $ -48.42210 $ &  5.30 & $ -2.167 $ & $ -4.436 $ & $  0.006 $ & $  0.005 $ & S & $  3.00 $ & $  0.08 $ \\
NGC 6355 & \multicolumn{1}{c}{--} & $  260.99435 $ & $ -26.35342 $ &  8.70 & $ -4.758 $ & $ -0.570 $ & $  0.017 $ & $  0.015 $ & S & $  1.13 $ & $  0.16 $ \\
NGC 6356 & \multicolumn{1}{c}{--} & $  260.89577 $ & $ -17.81304 $ &  15.10 & $ -3.765 $ & $ -3.400 $ & $  0.008 $ & $  0.008 $ & S & $  2.01 $ & $  0.07 $ \\
NGC 6362 & \multicolumn{1}{c}{--} & $  262.97906 $ & $ -67.04835 $ &  7.36 & $ -5.509 $ & $ -4.769 $ & $  0.002 $ & $  0.003 $ & S & $  2.74 $ & $  0.02 $ \\
NGC 6366 & \multicolumn{1}{c}{--} & $  261.93433 $ & $ -5.07988 $ &  3.79 & $ -0.335 $ & $ -5.161 $ & $  0.004 $ & $  0.004 $ & S & $  3.38 $ & $  0.04 $ \\
NGC 6380 & \multicolumn{1}{c}{--} & $  263.61666 $ & $ -39.06918 $ &  9.80 & $ -2.162 $ & $ -3.233 $ & $  0.018 $ & $  0.018 $ & S & $  0.90 $ & $  0.05 $ \\
NGC 6388 & \multicolumn{1}{c}{--} & $  264.07275 $ & $ -44.73566 $ &  10.74 & $ -1.313 $ & $ -2.712 $ & $  0.008 $ & $  0.008 $ & S & $  2.48 $ & $  0.02 $ \\
NGC 6397 & \multicolumn{1}{c}{--} & $  265.17540 $ & $ -53.67441 $ &  2.44 & $  3.254 $ & $ -17.653 $ & $  0.003 $ & $  0.003 $ & S & $  5.77 $ & $  0.05 $ \\
NGC 6401 & \multicolumn{1}{c}{--} & $  264.65386 $ & $ -23.90874 $ &  7.70 & $ -2.765 $ & $  1.438 $ & $  0.021 $ & $  0.019 $ & S & $  0.69 $ & $  0.05 $ \\
NGC 6402 & M 14 & $  264.40062 $ & $ -3.24594 $ &  9.31 & $ -3.575 $ & $ -5.057 $ & $  0.007 $ & $  0.007 $ & S & $  2.21 $ & $  0.01 $ \\
NGC 6426 & \multicolumn{1}{c}{--} & $  266.22795 $ & $  3.17013 $ &  19.80 & $ -1.805 $ & $ -2.981 $ & $  0.012 $ & $  0.012 $ & P & $  0.83 $ & $  0.03 $ \\
NGC 6440 & \multicolumn{1}{c}{--} & $  267.21945 $ & $ -20.35960 $ &  8.24 & $ -1.188 $ & $ -3.973 $ & $  0.021 $ & $  0.020 $ & S & $  1.07 $ & $  0.06 $ \\
NGC 6441 & \multicolumn{1}{c}{--} & $  267.55440 $ & $ -37.05147 $ &  11.83 & $ -2.541 $ & $ -5.372 $ & $  0.010 $ & $  0.010 $ & P & $  1.05 $ & $  0.03 $ \\
NGC 6453 & \multicolumn{1}{c}{--} & $  267.71571 $ & $ -34.59866 $ &  11.60 & $  0.205 $ & $ -5.982 $ & $  0.033 $ & $  0.032 $ & K & $  0.41 $ & $  0.04 $ \\
NGC 6496 & \multicolumn{1}{c}{--} & $  269.76531 $ & $ -44.26599 $ &  9.12 & $ -3.064 $ & $ -9.259 $ & $  0.005 $ & $  0.005 $ & P & $  1.43 $ & $  0.03 $ \\
NGC 6517 & \multicolumn{1}{c}{--} & $  270.45989 $ & $ -8.95957 $ &  10.60 & $ -1.564 $ & $ -4.470 $ & $  0.011 $ & $  0.015 $ & S & $  1.15 $ & $  0.06 $ \\
NGC 6522 & \multicolumn{1}{c}{--} & $  270.89201 $ & $ -30.03400 $ &  8.00 & $  2.583 $ & $ -6.491 $ & $  0.034 $ & $  0.030 $ & S & $  0.46 $ & $  0.03 $ \\
NGC 6528 & \multicolumn{1}{c}{--} & $  271.20670 $ & $ -30.05580 $ &  7.45 & $ -2.128 $ & $ -5.659 $ & $  0.036 $ & $  0.029 $ & K & $  0.57 $ & $  0.04 $ \\
NGC 6535 & \multicolumn{1}{c}{--} & $  270.96044 $ & $ -0.29765 $ &  6.50 & $ -4.219 $ & $ -2.938 $ & $  0.009 $ & $  0.009 $ & K & $  1.06 $ & $  0.03 $ \\
NGC 6539 & \multicolumn{1}{c}{--} & $  271.20722 $ & $ -7.58588 $ &  7.85 & $ -6.893 $ & $ -3.539 $ & $  0.010 $ & $  0.010 $ & S & $  1.91 $ & $  0.08 $ \\
NGC 6540 & \multicolumn{1}{c}{--} & $  271.53581 $ & $ -27.76529 $ &  5.20 & $ -3.749 $ & $ -2.819 $ & $  0.035 $ & $  0.034 $ & P & $  0.40 $ & $  0.06 $ \\
NGC 6541 & \multicolumn{1}{c}{--} & $  272.00984 $ & $ -43.71493 $ &  7.95 & $  0.285 $ & $ -8.843 $ & $  0.005 $ & $  0.005 $ & S & $  2.71 $ & $  0.02 $ \\
NGC 6544 & \multicolumn{1}{c}{--} & $  271.83574 $ & $ -24.99742 $ &  2.60 & $ -2.309 $ & $ -18.610 $ & $  0.011 $ & $  0.010 $ & K & $  4.57 $ & $  0.10 $ \\
NGC 6553 & \multicolumn{1}{c}{--} & $  272.31533 $ & $ -25.90775 $ &  6.75 & $  0.358 $ & $ -0.441 $ & $  0.021 $ & $  0.019 $ & P & $  1.29 $ & $  0.09 $ \\
NGC 6558 & \multicolumn{1}{c}{--} & $  272.57642 $ & $ -31.76354 $ &  7.20 & $ -1.744 $ & $ -4.152 $ & $  0.021 $ & $  0.020 $ & K & $  0.61 $ & $  0.05 $ \\
NGC 6569 & \multicolumn{1}{c}{--} & $  273.41198 $ & $ -31.82648 $ &  10.59 & $ -4.135 $ & $ -7.353 $ & $  0.011 $ & $  0.011 $ & P & $  0.77 $ & $  0.04 $ \\
NGC 6584 & \multicolumn{1}{c}{--} & $  274.65667 $ & $ -52.21581 $ &  13.18 & $ -0.093 $ & $ -7.204 $ & $  0.007 $ & $  0.006 $ & S & $  1.32 $ & $  0.02 $ \\
NGC 6624 & \multicolumn{1}{c}{--} & $  275.91879 $ & $ -30.36106 $ &  7.40 & $  0.127 $ & $ -6.947 $ & $  0.014 $ & $  0.014 $ & P & $  1.24 $ & $  0.04 $ \\
NGC 6626 & M 28 & $  276.13704 $ & $ -24.86987 $ &  5.43 & $ -0.302 $ & $ -8.930 $ & $  0.016 $ & $  0.016 $ & S & $  1.70 $ & $  0.12 $ \\
NGC 6637 & M 69 & $  277.84622 $ & $ -32.34811 $ &  8.80 & $ -5.063 $ & $ -5.834 $ & $  0.011 $ & $  0.011 $ & P & $  1.73 $ & $  0.03 $ \\
NGC 6638 & \multicolumn{1}{c}{--} & $  277.73436 $ & $ -25.49643 $ &  10.32 & $ -2.497 $ & $ -4.079 $ & $  0.014 $ & $  0.014 $ & S & $  0.76 $ & $  0.04 $ \\
NGC 6642 & \multicolumn{1}{c}{--} & $  277.97596 $ & $ -23.47616 $ &  8.05 & $ -0.176 $ & $ -3.898 $ & $  0.016 $ & $  0.014 $ & P & $  0.70 $ & $  0.04 $ \\
NGC 6652 & \multicolumn{1}{c}{--} & $  278.94010 $ & $ -32.99074 $ &  10.00 & $ -5.488 $ & $ -4.261 $ & $  0.008 $ & $  0.008 $ & S & $  0.93 $ & $  0.02 $ \\
NGC 6656 & M 22 & $  279.09980 $ & $ -23.90477 $ &  3.23 & $  9.840 $ & $ -5.618 $ & $  0.007 $ & $  0.007 $ & S & $  8.02 $ & $  0.15 $ \\
NGC 6681 & M 70 & $  280.80317 $ & $ -32.29214 $ &  9.31 & $  1.440 $ & $ -4.738 $ & $  0.008 $ & $  0.006 $ & P & $  1.67 $ & $  0.03 $ \\
NGC 6712 & \multicolumn{1}{c}{--} & $  283.26804 $ & $ -8.70596 $ &  6.95 & $  3.356 $ & $ -4.443 $ & $  0.009 $ & $  0.009 $ & P & $  1.98 $ & $  0.05 $ \\
NGC 6715 & M 54 & $  283.76386 $ & $ -30.47986 $ &  24.13 & $ -2.682 $ & $ -1.381 $ & $  0.003 $ & $  0.003 $ & S & $  2.73 $ & $  0.13 $ \\
NGC 6717 & Pal 9 & $  283.77515 $ & $ -22.70150 $ &  7.10 & $ -3.147 $ & $ -5.013 $ & $  0.010 $ & $  0.010 $ & S & $  0.94 $ & $  0.04 $ \\
NGC 6723 & \multicolumn{1}{c}{--} & $  284.88813 $ & $ -36.63226 $ &  8.30 & $  1.024 $ & $ -2.417 $ & $  0.005 $ & $  0.005 $ & S & $  2.33 $ & $  0.02 $ \\
NGC 6749 & \multicolumn{1}{c}{--} & $  286.31399 $ & $  1.89998 $ &  7.80 & $ -2.844 $ & $ -5.997 $ & $  0.012 $ & $  0.012 $ & P & $  1.38 $ & $  0.05 $ \\
NGC 6752 & \multicolumn{1}{c}{--} & $  287.71710 $ & $ -59.98458 $ &  4.25 & $ -3.163 $ & $ -4.034 $ & $  0.003 $ & $  0.003 $ & P & $  5.17 $ & $  0.02 $ \\
NGC 6760 & \multicolumn{1}{c}{--} & $  287.80024 $ & $  1.03046 $ &  7.95 & $ -1.091 $ & $ -3.610 $ & $  0.008 $ & $  0.008 $ & S & $  2.32 $ & $  0.10 $ \\
NGC 6779 & M 56 & $  289.14820 $ & $  30.18348 $ &  9.68 & $ -2.010 $ & $  1.612 $ & $  0.005 $ & $  0.005 $ & S & $  1.60 $ & $  0.01 $ \\
\hline
\end{tabular}
\end{table*}

\begin{table*}
\contcaption{}
\centering
\renewcommand{\arraystretch}{1.2}
\tabcolsep=4.5pt
\begin{tabular}{llrrrrrrrcrr}
\hline\hline             
\multicolumn{1}{c}{Name} &
\multicolumn{1}{c}{Other ID} &
\multicolumn{1}{c}{$\alpha$} & 
\multicolumn{1}{c}{$\delta$} &
\multicolumn{1}{c}{$D$} &
\multicolumn{1}{c}{$\mu_{\alpha,*}$} & 
\multicolumn{1}{c}{$\mu_{\delta}$} & 
\multicolumn{1}{c}{$\epsilon_{\mu_{\alpha,*}}$} &
\multicolumn{1}{c}{$\epsilon_{\mu_{\delta}}$} &
\multicolumn{1}{c}{$\Sigma$} &
\multicolumn{1}{c}{$R_{1/2}$} &
\multicolumn{1}{c}{$\epsilon_{R_{1/2}}$} \\
\multicolumn{1}{c}{} &
\multicolumn{1}{c}{} &
\multicolumn{1}{c}{[deg]} &
\multicolumn{1}{c}{[deg]} &
\multicolumn{1}{c}{[kpc]} &
\multicolumn{1}{c}{[mas yr$^{-1}$]} &
\multicolumn{1}{c}{[mas yr$^{-1}$]} &
\multicolumn{1}{c}{[mas yr$^{-1}$]} &
\multicolumn{1}{c}{[mas yr$^{-1}$]} &
\multicolumn{1}{c}{} & 
\multicolumn{1}{c}{[arcmin]} &
\multicolumn{1}{c}{[arcmin]} \\ 
\multicolumn{1}{c}{(1)} &
\multicolumn{1}{c}{(2)} &
\multicolumn{1}{c}{(3)} &
\multicolumn{1}{c}{(4)} &
\multicolumn{1}{c}{(5)} &
\multicolumn{1}{c}{(6)} &
\multicolumn{1}{c}{(7)} &
\multicolumn{1}{c}{(8)} &
\multicolumn{1}{c}{(9)} &
\multicolumn{1}{c}{(10)} &
\multicolumn{1}{c}{(11)} &
\multicolumn{1}{c}{(12)} \\ 
\hline
NGC 6809 & M 55 & $  294.99878 $ & $ -30.96479 $ &  5.30 & $ -3.430 $ & $ -9.314 $ & $  0.003 $ & $  0.003 $ & S & $  3.94 $ & $  0.02 $ \\
NGC 6838 & M 71 & $  298.44369 $ & $  18.77918 $ &  3.99 & $ -3.414 $ & $ -2.656 $ & $  0.004 $ & $  0.004 $ & S & $  3.37 $ & $  0.12 $ \\
NGC 6864 & M 75 & $  301.52017 $ & $ -21.92228 $ &  21.61 & $ -0.591 $ & $ -2.794 $ & $  0.009 $ & $  0.009 $ & P & $  1.08 $ & $  0.01 $ \\
NGC 6934 & \multicolumn{1}{c}{--} & $  308.54736 $ & $  7.40445 $ &  15.40 & $ -2.653 $ & $ -4.691 $ & $  0.015 $ & $  0.016 $ & P & $  1.24 $ & $  0.02 $ \\
NGC 6981 & M 72 & $  313.36541 $ & $ -12.53734 $ &  17.00 & $ -1.266 $ & $ -3.360 $ & $  0.006 $ & $  0.005 $ & P & $  1.13 $ & $  0.01 $ \\
NGC 7006 & \multicolumn{1}{c}{--} & $  315.37278 $ & $  16.18790 $ &  40.10 & $ -0.144 $ & $ -0.645 $ & $  0.021 $ & $  0.023 $ & P & $  0.44 $ & $  0.01 $ \\
NGC 7078 & M 15 & $  322.49304 $ & $  12.16699 $ &  10.22 & $ -0.652 $ & $ -3.808 $ & $  0.005 $ & $  0.005 $ & P & $  3.12 $ & $  0.01 $ \\
NGC 7089 & M 2 & $  323.36257 $ & $ -0.82332 $ &  10.51 & $  3.447 $ & $ -2.174 $ & $  0.010 $ & $  0.010 $ & S & $  2.54 $ & $  0.01 $ \\
NGC 7099 & M 30 & $  325.09221 $ & $ -23.17988 $ &  8.00 & $ -0.742 $ & $ -7.301 $ & $  0.006 $ & $  0.005 $ & S & $  2.16 $ & $  0.01 $ \\
NGC 7492 & \multicolumn{1}{c}{--} & $  347.11116 $ & $ -15.61142 $ &  26.55 & $  0.763 $ & $ -2.319 $ & $  0.008 $ & $  0.009 $ & S & $  1.02 $ & $  0.03 $ \\
Bootes dSph & Boo dSph & $  210.00004 $ & $  14.49996 $ & \multicolumn{1}{c}{--} & $ -0.399 $ & $ -1.065 $ & $  0.018 $ & $  0.016 $ & P & $  12.45 $ & $  2.86 $ \\
Carina dSph & PGC 19441 & $  100.40293 $ & $ -50.96613 $ & \multicolumn{1}{c}{--} & $  0.528 $ & $  0.118 $ & $  0.008 $ & $  0.008 $ & P & $  8.71 $ & $  0.30 $ \\
Draco dSph & UGC 10822 & $  260.05981 $ & $  57.92121 $ & \multicolumn{1}{c}{--} & $  0.032 $ & $ -0.183 $ & $  0.008 $ & $  0.009 $ & S & $  6.98 $ & $  0.24 $ \\
Fornax dSph & ESO 356-4 & $  39.99709 $ & $ -34.44920 $ & \multicolumn{1}{c}{--} & $  0.379 $ & $ -0.358 $ & $  0.002 $ & $  0.002 $ & S & $  14.80 $ & $  0.07 $ \\
Leo I dSph & PGC 29488 & $  152.11721 $ & $  12.30651 $ & \multicolumn{1}{c}{--} & $ -0.078 $ & $ -0.090 $ & $  0.014 $ & $  0.014 $ & S & $  2.93 $ & $  0.05 $ \\
Leo II dSph & PGC 34176 & $  168.36717 $ & $  22.15282 $ & \multicolumn{1}{c}{--} & $ -0.092 $ & $ -0.143 $ & $  0.027 $ & $  0.026 $ & S & $  2.41 $ & $  0.09 $ \\
Sculptor dSph & PGC 3589 & $  15.03901 $ & $ -33.70893 $ & \multicolumn{1}{c}{--} & $  0.103 $ & $ -0.149 $ & $  0.003 $ & $  0.003 $ & P & $  9.71 $ & $  0.11 $ \\
Sextans dSph & LEDA 88608 & $  153.26204 $ & $ -1.61463 $ & \multicolumn{1}{c}{--} & $ -0.360 $ & $  0.031 $ & $  0.015 $ & $  0.013 $ & P & $  21.39 $ & $  1.28 $ \\
Ursa Minor dSph & PGC 54074 & $  227.29745 $ & $  67.21439 $ & \multicolumn{1}{c}{--} & $ -0.133 $ & $  0.045 $ & $  0.010 $ & $  0.010 $ & S & $  13.97 $ & $  0.74 $ \\
\hline
\end{tabular}
\end{table*}


\bsp	
\label{lastpage}
\end{document}